\RequirePackage{fix-cm}
\documentclass[natbib,smallextended]{svjour3}       
\smartqed  
\usepackage{graphicx}
\usepackage{amsmath}
\usepackage{amssymb}
\usepackage{slashed}
\usepackage{color}
\usepackage{soul}
\usepackage{upgreek}
\usepackage{blindtext}
\usepackage{mathrsfs}
\usepackage{dsfont}
\usepackage[normalem]{ulem}
\usepackage{boxhandler}
\usepackage{mathtools} 
\usepackage{braket} 
\usepackage{relsize}
\usepackage[UKenglish]{babel} 
\usepackage{anyfontsize}
\usepackage[colorlinks=true,citecolor=blue,urlcolor=blue,breaklinks]{hyperref}

\newcommand{\av}[1]{\langle {#1} \rangle}
\newcommand{\be}{\begin{equation}}
\newcommand{\ee}{\end{equation}}
\newcommand{\bea}{\begin{eqnarray}}
\newcommand{\eea}{\end{eqnarrray}}

\newcommand{\cP}{\mathcal P} 
 
\newcommand{\vd}{\vec d}
\newcommand{\vD}{{\vec D}}
\newcommand{\re}{\mathbb R}

\newcommand{\cA}{\mathcal A}
\newcommand{\cT}{\mathcal T}
\newcommand{\cH}{\mathcal H} 
\newcommand{\fut}{\mathrm{Fut}}
\newcommand{\past}{\mathrm{Past}}
\newcommand{\ifut}{\mathrm{IFut}}
\newcommand{\ipast}{\mathrm{IPast}}

\newcommand{\cyl}{\mathrm{cyl}}
\newcommand{\stem}{\mathrm{stem}}
\newcommand{\hD}{\widehat \Delta} 
\newcommand{\hW}{\widehat W} 

\newcommand{\hP}{\widehat{\Phi}}

\newcommand{\vkpj}{v_{\mathbf k}}
\newcommand{\vkpjp}{v_{\mathbf k}^+}
\newcommand{\vkpjm}{v_{\mathbf k}^-}

\newcommand{\hB}{\widehat{\Box}}

\newcommand{\lk}{\lambda_{\mathbf{k}}}

\newcommand{\bC}{\mathbf C}
  
\newcommand{\bB}{\mathbf B} 
\newcommand{\bK}{{\mathbf K}} 
\newcommand{\mink}{\mathbb M}
\newcommand{\bL}{\mathbf L}
 \newcommand{\bl}{\mathbf l}

\newcommand{\twod}{\mathrm{2d}}

\newcommand{\pprec}{\mathsmaller{\prec\! \prec}}

\newcommand{\vol}{\mathrm{vol}}

\newcommand{\cO}{\mathcal O} 
\newcommand{\cG}{\mathcal G} 
\newcommand{\cN}{\mathcal N}

\newcommand{\Alex}{\mathbf A}

\newcommand{\negs}{\!\!\!}
\newcommand{\cR}{\mathcal R} 
\newcommand{\cW}{\mathcal W}
\newcommand{\cF}{\mathcal F} 
\newcommand{\tF}{\widetilde F} 
\newcommand{\btF}{\mathbf{\widetilde F}}

\newcommand{\cS}{\mathcal S}
\newcommand{\bS}{\mathbf{S}}

\newcommand{\tbS}{\textbf{S}}
\newcommand{\F}[1]{F_{#1}}
\renewcommand{\P}[1]{P_{#1}}

\newcommand{\bCk}{\mathbf C_k}

\newcommand{\nno}{\nonumber}
\newcommand{\ff}{{ f}_0}

\newcommand{\wzeta}{\zeta_0}

\newcommand{\G}[1]{\Gamma\left(#1 \right)}
\newcommand{\mFm}[4]{\, _{#1}F_{#1}\left( \genfrac{}{}{0pt}{}{#2}{ #3  }  \bigg| #4 \right) }
\newcommand{\Poch}[2]{\left( #1 \right)_{#2}}

\newcommand{\dm}{d}
\newcommand{\bbN}{\mathbb N}
\newcommand{\contg}{\mathcal H}
\newcommand{\bn}{\mathbf n} 
\newcommand{\bN}{\mathbf N} 
\newcommand{\ca}{\mathcal A} 
\newcommand{\rc}{\rho_c} 
\newcommand{\rp}{\rho_p} 

\newcommand{\rk}{\rho_\kappa} 
\newcommand{\cC}{\mathcal C}
\newcommand{\cM}{\mathcal M}
\newcommand{\bO}{\mathbf O}
\newcommand{\bbn}{\mathbf n} 
 
\newcommand{\bR}{\mathbf R} 
\newcommand{\cAlex}{\mathbf I} 
\newcommand{\dns}{d_s}

\newcommand{\lc}{\ell_c}
\newcommand{\llk}{\ell_\kappa}
\newcommand{\lp}{\ell_p}
\newcommand{\postcau}{\mathcal P}
\newcommand{\csgtree}{\mathcal T}

\newcommand{\Og}{\Omega_g}
\newcommand{\tOg}{\tilde {\Og} }
\newcommand{\tD}{\tilde D} 
\newcommand{\tc}{\tilde c}
\newcommand{\fA}{\mathfrak A} 
\newcommand{\tfA}{\tilde \fA}
\newcommand{\tmu}{\tilde \mu}
\newcommand{\fS}{\mathfrak S} 
\newcommand{\tfS}{\tilde \fS}
\newcommand{\hmu}{\hat \mu}
\newcommand{\dlor}{\mathcal G_{Lor}}
\newcommand{\dca}{\mathcal G_{A}}
\newcommand{\fca}{f_a}
\newcommand{\fcb}{f_b} 
\newcommand{\complex}{\mathbb C}
\newcommand{\tth}{\mathrm th}
\newcommand{\cB}{\mathcal B}
\newcommand{\vs}{\mathcal{V}}

\def\nn{\nonumber}
\journalname{Living Reviews in Relativity}

\begin{document}

\title{The causal set approach to quantum gravity}  
 

\author{Sumati Surya}
\institute{S. Surya \at
  Raman Research Institute, \\
  CV Raman Ave, Sadashivanagar, \\ 
  Bangalore 560080, \\
  India \\
  \email{ssurya@rri.res.in}
}


\date{Received: date / Accepted: date}

\maketitle

\begin{abstract}
The causal set theory (CST) approach to quantum gravity  postulates that at the most  fundamental level, spacetime is
discrete, with the spacetime continuum replaced by locally finite posets or ``causal sets''.  The partial
order on a causal set represents a  proto-causality relation while local finiteness
  encodes an intrinsic  discreteness. In the continuum approximation the former corresponds to the spacetime  causality
  relation   and the latter to  a fundamental spacetime atomicity, so that 
  finite volume regions in  the continuum contain only  a finite number of causal set elements.  CST is deeply rooted in the Lorentzian character of spacetime, where a  primary role is  played by
  the causal structure poset.  Importantly, the assumption of a fundamental discreteness in 
 CST  does not violate local  Lorentz invariance in the continuum approximation.   On the other hand, the combination of discreteness and
 Lorentz invariance gives rise to a characteristic non-locality which distinguishes CST  from most other approaches to
 quantum gravity.

 In this review we  give a broad, semi-pedagogical
introduction to CST,  highlighting key results  as well as some of the key open questions. This review is intended both for the
  beginner student in quantum gravity as well as more seasoned researchers in the field.
\end{abstract}

\setcounter{tocdepth}{3}
\tableofcontents

\section{Overview} 

In this review,  causal set theory (CST) refers to the
specific proposal made by
Bombelli, Lee, Meyer and Sorkin (BLMS) in their 1987 paper
\citep{blms}.  In CST, the space of Lorentzian geometries is  replaced by the set of  locally finite posets, or
\emph{causal sets}.   These causal sets encode  the
twin principles of causality and discreteness. In the continuum approximation 
of CST, where elements of the causal set 
set represent spacetime events, the order relation on the
causal set   corresponds to the spacetime causal order and the  cardinality of an ``order interval''  to the spacetime volume 
of the associated causal interval.

This review is intended as a semi-pedagogical 
introduction to CST. The aim  is to give a broad survey of the
main results and open questions and  to direct the reader
to some of the many interesting open research problems in CST, some of which are
accessible even to the beginner.

We begin in
Sect.~\ref{sec:history} with a historical perspective 
on the ideas behind  CST. The
twin principles of discreteness and causality at the heart of
CST have both been proposed -- sometimes independently and sometimes together -- starting with
\cite{Riemann} and \cite{robbone,robbtwo},  and somewhat later by \cite{zeeman,kp,finkelstein,hemion} and \cite{myrheim}, 
culminating in the CST proposal of BLMS \citep{blms}. 
 The continuum approximation of CST is an implementation of a deep
result in Lorentzian geometry due to \cite{hkm}
and its generalisation by \cite{malament}, which states that the causal structure determines the conformal geometry of a future and past distinguishing
causal spacetime. In following this
history, the discussion will be necessarily somewhat technical. For those unfamiliar with the terminology of causal
structure we point to standard texts \citep{HE,BE,Wald,penrose}.

In Sect.~\ref{sec:cst}, we state the 
 CST proposal and describe its continuum approximation, in which spacetime causality is equivalent to the order relation
 and finite spacetime volumes to cardinality. Not all causal sets have a continuum approximation -- in fact we will see
 that most do not. Those that do are referred to as  \emph{manifold-like}. Important to CST is its
 ``Hauptvermutung'' or fundamental conjecture, which roughly states that a 
 manifold-like causal set is equivalent to the  continuum spacetime, modulo differences up to the discreteness scale. Much of the discussion on the Hauptvermutung is centered on
 the question of how to estimate the 
 closeness of Lorentzian manifolds or more generally,  causal sets. While there is no full proof of the  conjecture, there
 is growing body of  evidence in its favour as we will see in Sect.~\ref{sec:kinematics}. An important outcome of CST
 discreteness in the continuum approximation 
 is that it does not violate Lorentz invariance as shown in an elegant theorem by \cite{bomhensor}. Because of the centrality of this result we review this construction in some
 detail. The combination of discreteness and Lorentz invariance moreover gives rise to an inherent and characteristic
 non-locality, which distinguishes CST from other discrete approaches. Following \cite{lambdatwo}, we then discuss
 how the twin principles behind CST force us to take  certain ``forks in the road'' to quantum gravity.

We present some of the key developments in  CST in Sects.~\ref{sec:kinematics}, \ref{sec:matter} and \ref{sec:dynamics}. We
begin with the kinematical structure of theory and the program of ``geometric reconstruction'' in Sect.~\ref{sec:kinematics}. Here, the aim
is to reconstruct manifold invariants from  \emph{order invariants}  in a manifold-like causal set. These are functions on
the causal set that are independent of the labelling or ordering of the elements in the causal set.    
Finding the appropriate order invariants that correspond to manifold invariants can be challenging, since there is little in the mathematics
literature which correlates  order theory to  Lorentzian geometry via the CST continuum approximation. 
Extracting such invariants requires new technical tools and insights sometimes requiring a rethink of familiar aspects
of continuum Lorentzian geometry. We  will describe some of the progress made in this direction over the years 
\citep{myrheim,bg,meyer,bommeyer,bomthesis,reid,homology,rw,sorkinnonlocal,bd,dionthesis,gaussbonnet,intervals,rss,bdjs,tlbdry,bomemad,esv}. The 
correlation between order invariants and manifold invariants  in the continuum approximation lends support for the
Hauptvermutung and simultaneously establishes weaker, observable-dependent versions of the conjecture.

Somewhere between dynamics and kinematics is the study of quantum fields on
manifold-like causal sets, which we describe in Sect.~\ref{sec:matter}. The simplest system is free
scalar field theory on a causal set approximated by $d$-dimensional Minkowski spacetime $\mink^d$. Because causal sets do not admit a natural Hamiltonian
framework, a fully covariant construction is required to obtain the quantum field theory vacuum. A natural starting
point is the advanced and retarded Green functions
for a free scalar field theory since  it is defined using the causal structure of the spacetime. The explicit form for
these Green functions were found for causal sets approximated by $\mink^d$ for $d=2,4$ 
\citep{daughton,johnston,johnstonthesis} as well as  de~Sitter spacetime \citep{dsx}.  In trying to find a quantisation scheme on the causal
set without reference to the continuum, 
\cite{johnstontwo} found a novel covariant definition of the discrete scalar field vacuum, starting from the covariantly
defined Peierls' bracket formulation of quantum field theory. Subsequently 
\cite{sorkinsj} showed that the construction is also valid in the continuum, and can be used to give an alternative
definition of the quantum field theory vacuum. This  \emph{Sorkin--Johnston (SJ) vacuum} provides a new 
insight into quantum field theory and  has stimulated the interest
of the algebraic field theory community \citep{fv12,bf14,fewsterart}. The SJ vacuum has also been used to calculate Sorkin's spacetime
entanglement entropy (SSEE) \citep{bklsEE,sorkinEE} in a causal set \citep{yasamaneecont,causetee}. The calculation in $d=2$ is 
surprising since it gives rise to a volume law rather than an area law. What this means for causal set entanglement
entropy is still an open question. 

In Sect.~\ref{sec:dynamics}, we describe the CST approach to quantum
dynamics, which roughly follows two directions.
The first, is based on ``first principles'', where one starts with a general set of
axioms which respect microscopic covariance and causality. An important
class of such theories is the set of Markovian \emph{classical sequential growth} (CSG)  models of Rideout and Sorkin
\citep{csg,csgtwo,csgrg,davidthesis,rv}, which we will describe in some detail. The dynamical framework finds a natural
expression in terms of measure theory,
with the classical covariant observables represented by a covariant event algebra $\fA$ over the sample space $\Omega_g$ of past finite
causal sets \citep{observables,observablesds}. One of the main questions in CST dynamics is whether the overwhelming entropic
presence of the \emph{Kleitman--Rothschild (KR)} posets in $\Omega_g$ can be overcome by the dynamics \citep{kr}. These KR posets are
highly  non-manifold-like and ``static'', with just three ``moments of time''. Hence, if the continuum approximation is
to emerge in the classical limit of the theory, then the entropic contribution  from the KR posets should be suppressed
by the dynamics in this limit.  In the CSG models, 
the typical causal sets generated are  very ``tall'' with countable rather than finite moments  of time and, though not
quite manifold-like, are very
unlike the KR posets or even the subleading entropic contributions from non-manifold-like causal sets
\cite{dharone,dhartwo}.
The CSG models have generated some interest in the mathematics community, and new mathematical tools are now being used
to study the asymptotic structure of the theory \citep{grnick,gmone,gmtwo,grahammalwina}.

In CST, the appropriate route to quantisation is via the quantum measure or decoherence functional defined
in the double-path integral formulation \citep{qmeasureone,qmeasuretwo,sorkinqmeasure}. In the quantum versions of the
CSG (\emph{quantum sequential growth} or QSG) models the transition probabilities of CSG are replaced
by the decoherence functional. While
covariance can be easily imposed, a quantum version of microscopic causality is still missing \citep{joecausality}.
Another indication of the non-triviality of quantisation comes from a prosaic 
generalisation of transitive percolation, which is the simplest of the CSG models. In this ``complex percolation''  
dynamics the quantum measure does not
extend to the full algebra of observables which is an impediment to the construction of 
covariant quantum observables \citep{djs}. This can  be somewhat alleviated by taking a physically motivated approach to
measure theory \citep{ec}. An important future direction is to
construct  covariant observables in a wider class of quantum dynamics and look for a quantum version of coupling
constant renormalisation.   

Whatever the ultimate quantum dynamics  however, little sense can be
made of the theory without a fully developed quantum interpretation for closed systems, essential to quantum gravity. Sorkin's co-event interpretation
\citep{sorkinalogic,kochenspecker} provides a promising avenue based on the quantum measure approach. While a discussion
of this is outside of the scope of the present work, one can use the broader ``principle of preclusion'', i.e., that sets of zero quantum measure do not
occur\citep{sorkinalogic,kochenspecker}, to make a limited set of  predictions in complex percolation ({\it Sorkin and Surya, work in progress}).

The second approach to quantisation is more pragmatic, and uses the continuum inspired path integral formulation of
quantum gravity for causal sets. 
Here, the path integral is replaced by a sum over the sample space $\Omega$ of causal sets, using the \emph{Benincasa--Dowker} (BD) action, which limits to  the  Einstein--Hilbert action \citep{bd}. This can be viewed as an
effective, continuum-like dynamics, arising from the more fundamental dynamics described above. A recent
analytic calculation in \cite{carliploomis} showed that a sub-dominant 
class of non-manifold-like causal sets, the bilayer posets, are suppressed in the path integral when using the BD
action, under certain dimension dependent conditions satisfied by the parameter space. This gives hope that an 
effective dynamics might be able to overcome the entropy of the non-manifold-like causal sets.

In \cite{2dqg}, \cite{2dhh}, and \cite{fss}, Markov Chain Monte Carlo (MCMC) methods were used for a dimensionally restricted sample space
$\Omega_{2d}$ of 2-orders, which corresponds to topologically trivial $d=2$ causal set quantum gravity. 
The quantum partition function over causal sets can be rendered into a statistical partition function via an analytic
continuation of a  ``temperature'' parameter, while retaining  the Lorentzian character of the theory. This theory
exhibits a first order phase transition \citep{2dqg,fss} between a manifold-like phase and a layered, non-manifold-like
one. MCMC methods have also been used to examine the sample space $\Omega_n$ of $n$-element causal sets and to estimate the
onset of asymptotia, characterised by 
the dominance of the KR posets \citep{onset}. These techniques have recently  been extended to topologically non-trivial
$d=2$ and $d=3$ CST ({\it Cunningham and Surya, work in progress}). While this approach gives us expectation values of covariant observables
which allows for a straightforward interpretation, relating it to the complex or quantum partition function is
non-trivial and an open problem. 

In Sect.~\ref{sec:phen}, we describe in brief some of the exciting
phenomenology that comes out of the \emph{kinematical} structure of
causal sets. This includes the momentum space diffusion coming from CST
discreteness (``swerves'') \citep{swerves} and the effects of non-locality on quantum field theory \citep{sorkinnonlocal}, which includes a 
recent proposal for dark matter \citep{darkmatter}. Of these, the most striking is the 1987 prediction of
Sorkin for the value of the cosmological constant
$\Lambda$ \citep{lambdaone,lambdatwo}. While the original argument was a kinematic estimate, subsequently dynamical
models of fluctuating $\Lambda$ were examined \citep{lambdathree,eptwo,recentlambda} and have been compared with recent observations
\citep{recentlambda}. This is an exciting future direction of research in CST which interfaces intimately with
observation. We conclude with a brief outlook for CST in Sect.~\ref{sec:outlook}.

Finally, since this is an extensive review, to assist the reader  we have made a  list of some of the key definitions,
as well as the abbreviations  in Appendix \ref{sec:aone}.  

As is true of all other approaches to quantum gravity, CST
is not as yet a complete theory. Some of the challenges faced are universal to quantum gravity as a whole, while
others are specific to the approach. Although we have developed a reasonably good grasp of the 
kinematical structure of CST and some progress has been made in the construction of effective quantum dynamics, CST still lacks a
satisfactory quantum dynamics built from first principles. Progress in this direction is therefore very important for
the future of the program.  From a broader perspective, it is the opinion of this author that a deeper understanding of CST will help provide key insights into the nature of quantum
gravity from a fully Lorentzian, causal perspective, whatever ultimate shape the final theory takes. 

It is not possible for this review to be truly complete. The hope is that the interested reader will use it 
as a springboard to the existing literature. Several older reviews  exist with differing emphasis  
\citep{lambdaone,valdivia,joereviewone,fayreview, lightlinks, fayreview, joereviewtwo,walldenreview}, some of which 
have an in depth discussion of the conceptual underpinnings of CST.  The
focus of the current review is to provide as cohesive an account of the program as possible, so as to be
useful to a starting researcher in the field. For more technical details, the reader is urged to look at the original
references.

 \section{A historical perspective}
\label{sec:history} 

One of the most important conceptual realisations that arose from the
special and general theories of relativity in the early part of the 20th century, was that space and
time are part of a single construct, that of spacetime.  At a
fundamental level, one does not exist without the other. Unlike
Riemannian spaces, spacetime has a Lorentzian signature $(-, +,+,+)$ which
gives rise to local lightcones and an associated  global causal
structure. The causal structure $(M,\prec)$ of a  causal spacetime\footnote{Henceforth, we will assume that
  spacetime is causal, i.e., without any closed causal curves.} $(M,g)$ is a partially ordered set or poset, with
$\prec$ denoting the causal ordering on the ``event-set'' $M$. 

\begin{figure}[ht]
\centering \resizebox{4in}{!}{\includegraphics{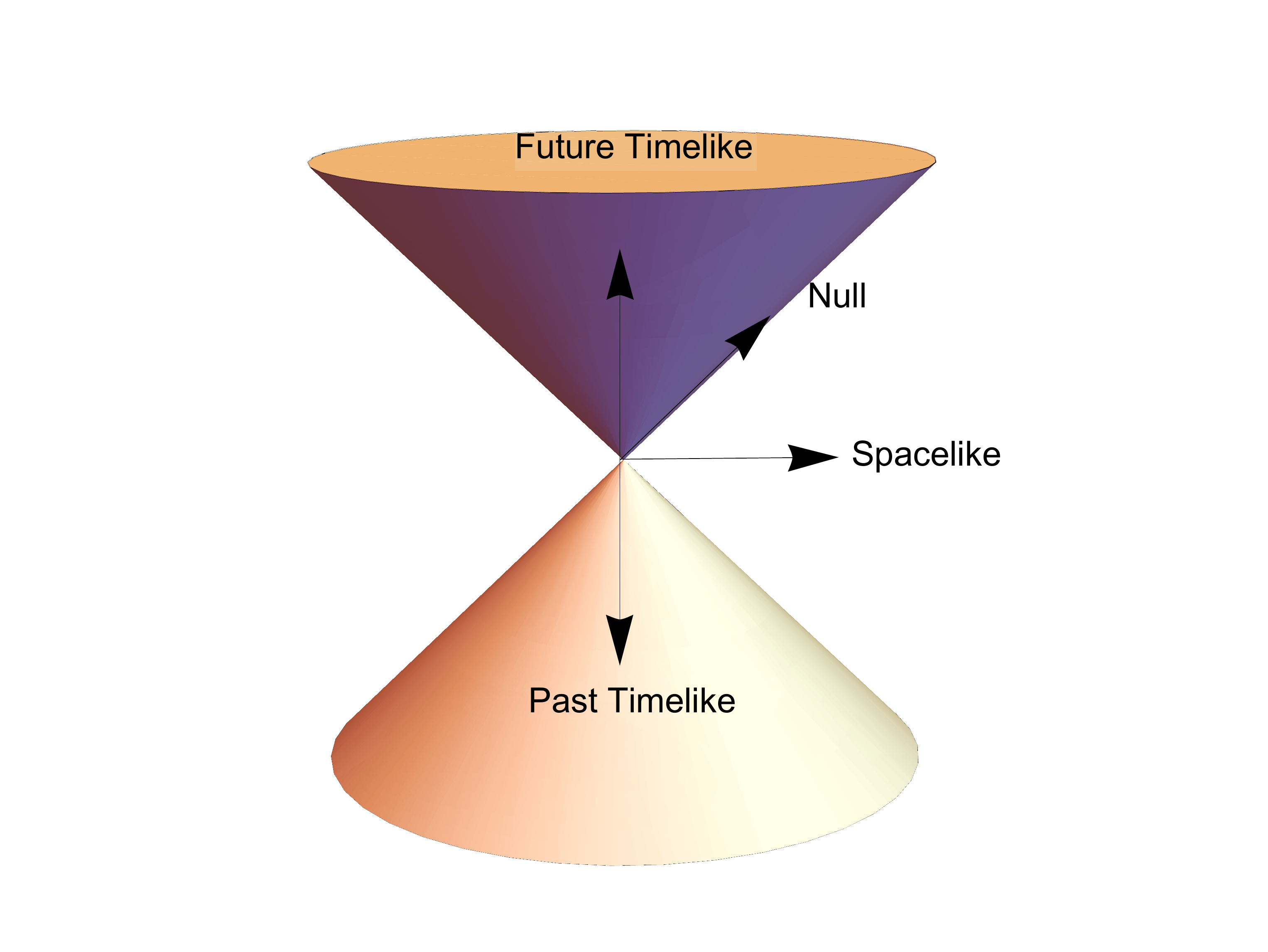}}
\caption{The local lightcone of a Lorentzian spacetime.}
\end{figure}

Causal set theory (CST) as proposed in \cite{blms}, takes the Lorentzian
character of spacetime and the causal structure poset in particular, as a crucial starting point to
quantisation. It is inspired by a long but sporadic history of investigations into
Lorentzian geometry, in which the connections between $(M,\prec)$
and the conformal geometry were eventually established. This history, while not a part of the
standard narrative of General Relativity, is relevant to the sequence
of ideas that led to CST. In looking for a quantum theory of spacetime, $(M,\prec)$ has also been paired with
discreteness, though the earliest ideas on discreteness go back to pre-quantum and pre-relativistic physics. We now give a
broad review of this history. 

The first few decades after the formulation of General Relativity were dedicated to
understanding the physical implications of the theory
and to finding solutions to the field equations. The attitude  towards Lorentzian geometry was mostly 
practical:  it was seen as a simple, though odd,  generalisation
of Riemannian geometry.\footnote{Hence the term
  ``pseudo-Riemannian''.}  There were however early attempts to understand this new geometry and to use causality as a
starting point. Weyl
and Lorentz  (see \citealt{weyl}) used light rays to attempt a
reconstruction of $d $ dimensional Minkowski spacetime {$\mink^d$}, while 
\cite{robbone,robbtwo} suggested an axiomatic framework for
spacetime where the causal precedence on the collection of events was seen to play a critical role.
It was only several decades later, however,  that the mathematical structure of Lorentzian geometry began to be explored more vigorously.

In a seminal paper titled ``Causality Implies the Lorentz Group'', \cite{zeeman} identified the {chronological poset}
$(\mink^d,\pprec)$ in $\mink^d$, where  {{$\pprec$}} denotes  the  {chronological relation} on the event-set $\mink^d$. Defining a {chronological automorphism}\footnote{Zeeman used the term ``causal'' instead of
 ``chronological'', but we will follow the more modern usage of these terms \citep{HE,Wald}.} $\fca$ of
$\mink^d$ as the chronological poset-preserving bijection
\begin{equation}
 \fca: \mink^d \rightarrow \mink^d, \quad x \, \,\pprec \, \, y \Leftrightarrow \fca(x) \, \,\pprec \, \,\fca(y), \, \, \forall \, \, x,y
 \in \mink^d, 
 \end{equation} 
Zeeman showed that the group of chronological automorphisms $\dca$ is
isomorphic to  the group $\dlor$ 
of inhomogeneous Lorentz transformations and dilations on
$\mink^d$ when $\dm >2$.  While it is simple to see that
the generators of $\dlor$ preserve the chronological structure so that $\dlor \subseteq \dca$, the
converse is not obvious. In his proof Zeeman showed that every $\fca \in \dca$ maps light rays to light rays, such that
parallel light rays remain parallel and moreover that the map is  linear.  In Minkowski spacetime every chronological
automorphism is also a causal automorphism, so a Corollary to Zeeman's theorem is that the group of causal
automorphisms is isomorphic to $\dlor$. This is a remarkable result, since it states that the physical invariants
associated with $\mink^d$ follow naturally from its \emph{causal structure poset} $(\mink^d,\prec)$ where $\prec$
denotes the causal relation on the event-set $\mink^d$.

\cite{kp} subsequently generalised
Zeeman's ideas to an arbitrary causal spacetime $(M,g)$ where they identified both $(M,\prec)$ and $(M,\pprec)$ with
the event-set $M$,  devoid of the differential and topological structures associated with a
spacetime. They defined an abstract causal space axiomatically, using both 
$(M,\prec)$ and $(M,\pprec)$ along with a mixed
transitivity condition between the relations $\prec$ and $\pprec$, which mimics that in a causal spacetime.

Zeeman's result in $\mink^d$ was then generalised to a larger class of spacetimes by
\cite{hkm} and \cite{malament}. A chronological bijection generalises Zeeman's
chronological automorphism between two 
spacetimes $(M_1,g_1)$ and $(M_2,g_2)$, and is a chronological order preserving bijection, 
\begin{equation}
 \fcb :M_1 \rightarrow M_2, \quad x \, \,\pprec_1 \, \, y \Leftrightarrow \fcb(x) \, \,\pprec_2 \, \,\fcb(y), \, \, \forall \, \, x,y \in
 M_1, 
\end{equation}
where $\pprec_{1,2}$ refer to the chronology relations on $M_{1,2}$, respectively. 
The existence of a chronological bijection between two strongly causal spacetimes\footnote{A point $p$ in a spacetime is
  said to be strongly causal if every neighbourhood of $p$ contains a subneighbourhood such that no causal curve
  intersects it more than once. All the events in a  strongly causal spacetime are strongly causal.} was equated by  \cite{hkm} to the
existence of a {conformal isometry}, which is a bijection $f:M_1\rightarrow M_2 $ such that $f, f^{-1}$ are smooth (with
respect to the manifold topology and differentiable structure) and $f_*g_1=\lambda g_2$ for a real, smooth, strictly positive function
$\lambda $ on $M_2$. \cite{malament}  then generalised this result to the larger class of \emph{future and past
 distinguishing}  spacetimes.\footnote{These are spacetimes in which the
chronological past and future $I^\pm(p)$ of each event $p$ is unique,
i.e., $I^\pm(p)=I^\pm(q) \Rightarrow p=q$.} 
We refer to these results collectively as the Hawking--King--McCarthy--Malament theorem or \emph{HKMM} theorem, summarised as 

\begin{theorem} {\bf Hawking--King--McCarthy--Malament (HKMM)} \\
If a chronological bijection $\fcb$ exists between two $\dm$-dimensional spacetimes which are both future and
past distinguishing, then these spacetimes are conformally isometric when $\dm>2$. \label{HKMM}
\end{theorem}

It was shown by  \cite{levichev} that a causal bijection implies a
chronological bijection and hence the above theorem can be generalised
by replacing ``chronological'' with ``causal''. Subsequently \cite{ps} showed that the causal structure poset
$(M,\prec)$ of these spacetimes also contains information about the spacetime dimension. 

Thus, the causal structure poset $(M,\prec)$ of a future and past distinguishing spacetime is
equivalent its conformal geometry.  This means that $(M,\prec)$
is equivalent to the spacetime, except for the local volume  element encoded in the conformal
factor $\lambda$, which is a single scalar. As phrased by \cite{finkelstein}, the causal structure in $d=4$ 
is therefore $\left(9/10\right)^{\mathrm{th}}$ of the metric! 

\begin{figure}[ht]
\centering \resizebox{4in}{!}{\includegraphics{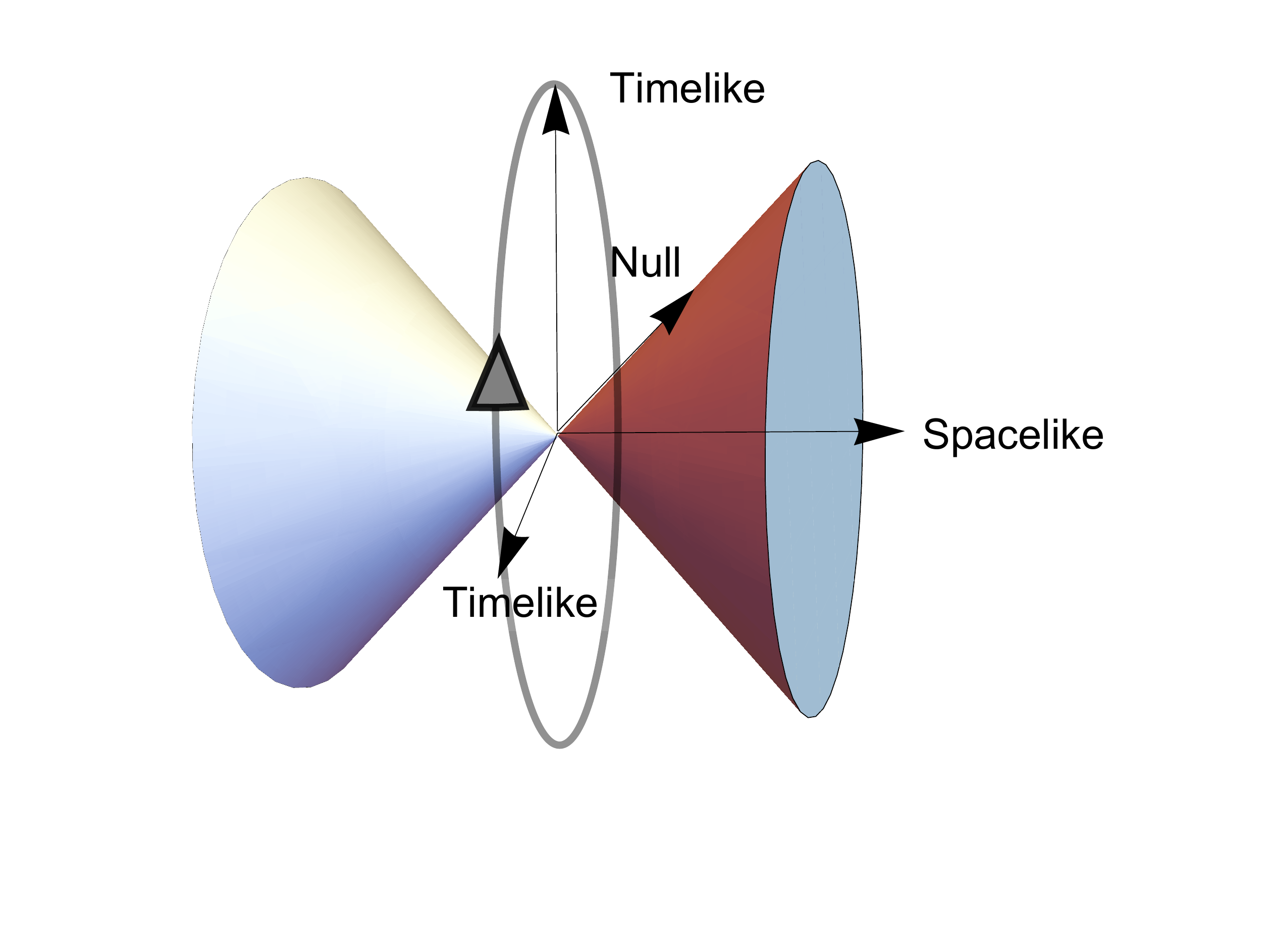}}
\caption{An example of a signature $(-,-,+,+)$ spacetime with one spatial dimension suppressed. It is
    not possible to distinguish a  past from a  future timelike direction and hence order events,  even locally.}
\label{nonlor.fig}
\end{figure}

En route to a theory of quantum gravity one must pause to ask:
what ``natural''  structure of spacetime  should be quantised? Is it the metric  or is it the causal structure poset?  The
former can be defined for all signatures,  but the latter is  an
exclusive  embodiment  of a causal Lorentzian spacetime. In Fig.~\ref{nonlor.fig}, we show a 3d projection of a  non-Lorentzian and
non-Riemannian $d=4$ ``space-time'' with  signature $(-,-,+,+)$. The fact that a time-like direction can be
continuously transformed into any other while still remaining time-like means that there is no order  relation in the space and
hence no associated causal structure poset. We can thus view  the  causal structure poset as an essential  embodiment of  Lorentzian spacetime.  

Perhaps the first explicit  statement of intent to quantise the {causal structure} of spacetime, rather than the
spacetime geometry was  by \cite{kp},
who listed, as one of their motivations for axiomatising the causal structure:
\begin{quote}
\textit{``To admit structures which can be very different from a
  manifold. The possibility arises, for example, of a locally
  countable or discrete event-space equipped with causal relations macroscopically similar to those of a space-time continuum.''
}
\end{quote}

This brings to focus  another historical thread of ideas 
important to CST, namely that  of spacetime discreteness.   
The idea  that  the  continuum is a  mathematical
construct which approximates an underlying physical discreteness was already
present in the writings of Riemann as he ruminated on the physicality of the continuum \citep{Riemann}:
\begin{quote}
\textit{``Now it seems that the empirical notions on which the metric determinations of  Space are based, the concept of a
  solid body and that of a light ray; lose their validity in the infinitely small; it is therefore quite definitely
  conceivable that the metric relations of Space in the infinitely small do not conform to the hypotheses of  geometry;
  and in fact one ought to assume this as soon as it permits a simpler way of explaining phenomena.''} 
\end{quote}

Many years later, in their explorations of spacetime and quantum theory, Einstein and Feynman
each questioned the physicality of the continuum \citep{Einstein,Feynman}.  
These ideas were also expressed in Finkelstein's  ``spacetime code'' \citep{finkelstein}, and most relevant to CST, in
Hemion's use of \emph{local finiteness},   to obtain discreteness in the causal
structure poset 
\citep{hemion}.  This last condition is the  requirement there are only a finite number of fundamental spacetime 
elements  in any finite volume  \emph{Alexandrov interval} $\Alex[p,q]\equiv I^+(p)\cap I^-(q)$.

Although these ideas of spacetime discreteness resonate with the appearance of discreteness in quantum theory, 
the latter  typically manifests itself as  a discrete spectrum of  a continuum observable. The discreteness
proposed above is different: one is replacing the fundamental degrees of
freedom, \emph{before} quantisation, already  at the kinematical
level of the theory.

The most immediate motivation for discreteness however comes from the HKMM theorem itself.  The missing
$\left(1/10\right)^{\mathrm{th}}$ of the $d=4$ metric is the volume element. A discrete causal set can supply this volume
element by substituting the continuum volume with cardinality. 
This idea was already present in Myrheim's remarkable (unpublished)  CERN preprint \citep{myrheim}, which contains many  of the main  ideas of CST. Here he states:
\begin{quote}
\textit{``It seems more natural to regard the metric as a statistical property of discrete spacetime. Instead we
  want to suggest that the concept of absolute time ordering, or causal ordering of, space-time points, events, might
  serve as the one and only fundamental concept of a discrete space-time geometry. In this view space-time is nothing
  but the causal ordering of events.''}  
\end{quote}

The statistical nature of the poset is a key proposal that survives into CST with the  spacetime continuum emerging 
via a random Poisson sprinkling. We will see this explicitly in Sect.~\ref{sec:cst}. Another key concept which plays a
role in the dynamics is that the
order relation replaces coordinate time and any evolution of spacetime takes meaning only in this intrinsic sense
\citep{lambdatwo}.

There are of course many other motivations for spacetime discreteness. One of the expectations from a theory of quantum gravity is
that the Planck scale will introduce a natural cut-off which cures both  the UV divergences of quantum field theory and
regulates black hole entropy. The realisation of this hope lies in the details of a given discrete theory, and CST
provides us a concrete way to study this question, as we will discuss in Sect.~\ref{sec:matter}.  

It has been 31 years  since the original CST proposal of BLMS \citep{blms}. The early work shed
considerable  light on key aspects of the theory \citep{blms,bommeyer,bg} and resulted in  Sorkin's prediction of the
cosmological constant $\Lambda$ 
\citep{lambdaone}.  There was a seeming  hiatus in the 1990s, which ended  in the early 2000s
with  exciting  results from the Rideout--Sorkin  classical sequential growth  models \citep{csgone,csgtwo,csgrg,davidthesis}.
There have been several non-trivial results  in CST in the intervening 19 odd years.  In the following sections we will
make a broad sketch of the theory and its key  results,  with this historical perspective in
mind.

\section{The causal set hypothesis}
\label{sec:cst}

We begin with the definition of a causal set: \\

\noindent {\bf Definition:}  A set $C$ with an order relation $\prec$ is a \emph{causal set}  if it is 
\begin{enumerate}  
\item \emph{Acyclic}: $x\prec y$ and $y \prec x $ $\Rightarrow x=y$, $\forall
  x,y \in C$ 
\item \emph{Transitive}: $x\prec y$ and $y \prec z $ $\Rightarrow x \prec z$, $\forall
  x,y,z \in C$ 
\item \emph{Locally finite}:  $\forall x,y \in C$, $|\cAlex[x,y]|< \infty$, where $\cAlex[x,y]\equiv \fut(x) \cap \past(y)$ \,,  
\end{enumerate} 
where $|.|$ denotes the cardinality of the set, and\footnote{These are  the \emph{exclusive} future and past sets since
  they do not include the element itself.} 
\begin{eqnarray}  
\fut(x) &\equiv&\{w \in C| x \prec w, x \neq w\} \nn \\ 
  \past(x) &\equiv&\{w \in C| w \prec x, x \neq w \}.
 \label{eq:futpast}                    
\end{eqnarray}
We refer to $\cAlex[x,y]$ as an \emph{order interval}, in analogy with the  Alexandrov interval in the continuum. 
The acyclic and transitive conditions together define a partially
ordered set or poset, while the  condition of local finiteness
encodes discreteness.  
\begin{figure}[ht]
\centering \resizebox{3in}{!}{\includegraphics{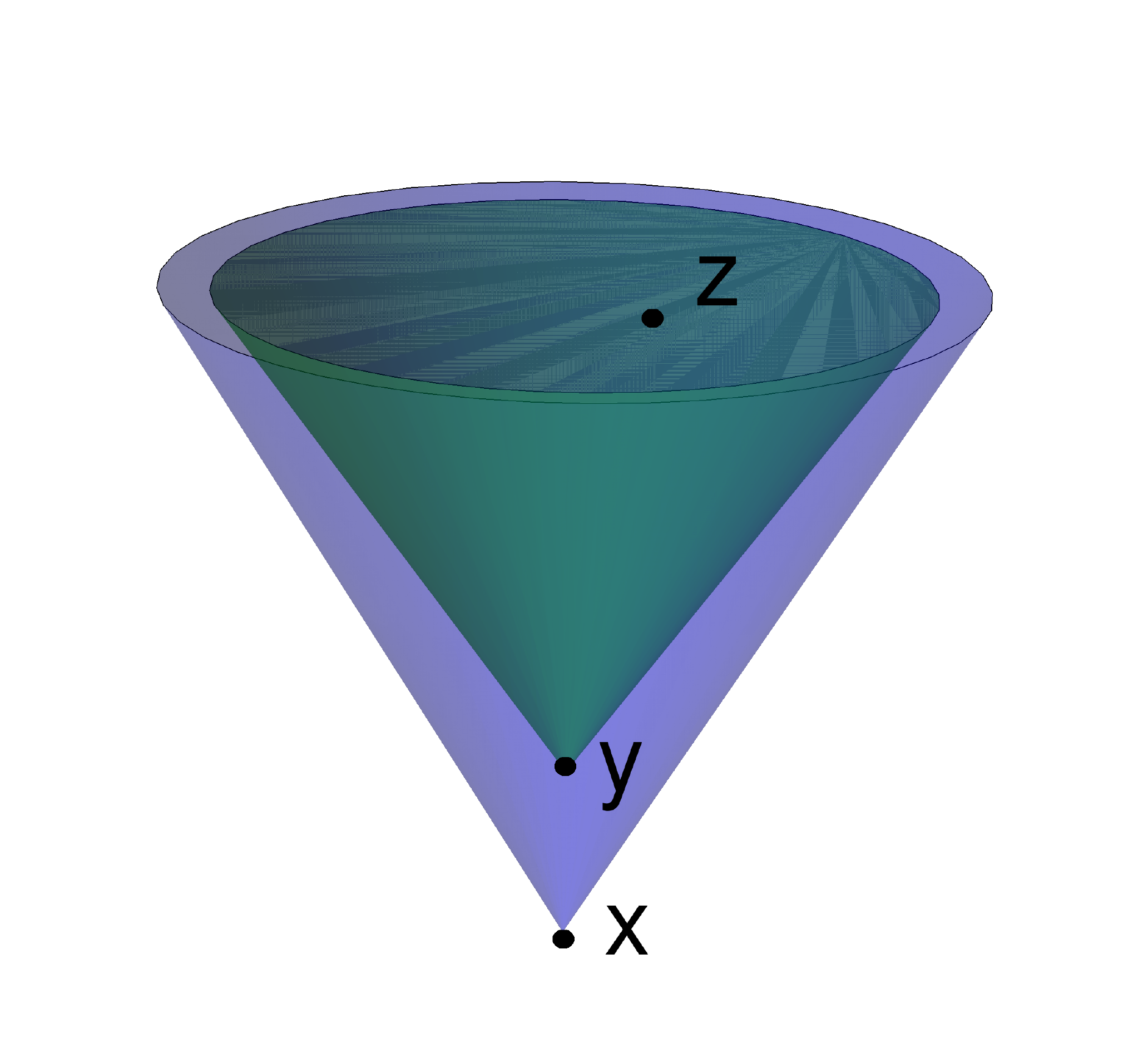}}
\caption{The transitivity condition $x \prec y, y \prec z \Rightarrow x\prec z$ is satisfied by the causality
    relation $\prec$ in any Lorentzian  spacetime.}
\label{transitive.fig}
\end{figure}

\begin{figure}[ht]
\centering \resizebox{3.5in}{!}{\includegraphics{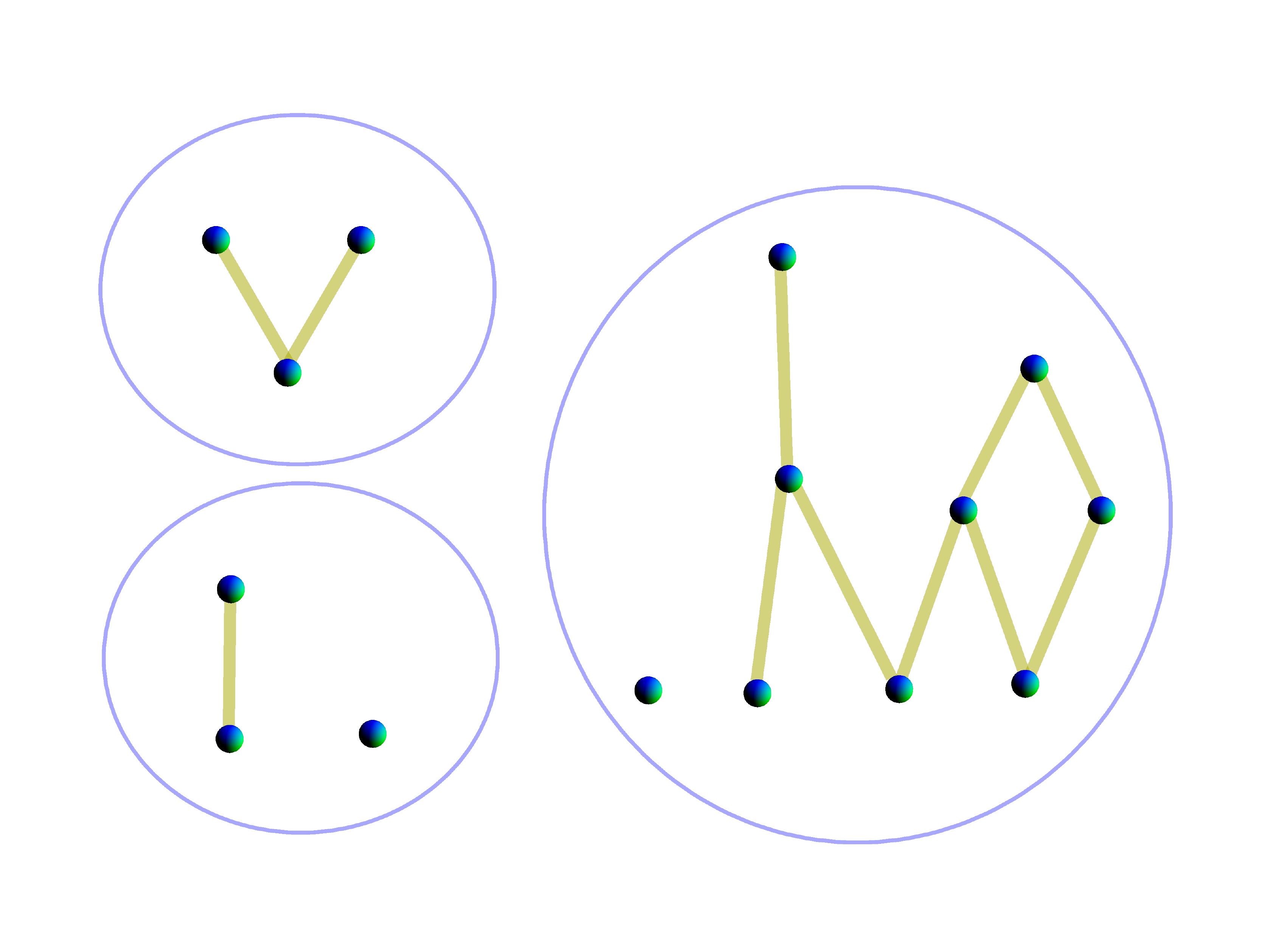}}
\caption{The \emph{Hasse diagrams} of some simple finite cardinality causal sets. Only the nearest neighbour
    relations or \emph{links} are depicted. The remaining relations are deduced from transitivity.}
\label{transitivity.fig}
\end{figure}

The content of the HKMM theorem can be summarised in the statement:
\begin{equation} 
\mathrm{Causal\, \, Structure + Volume\,\, Element = Lorentzian \, \, Geometry},     
\end{equation}
which lends itself to a discrete rendition, dubbed  
the ``CST slogan'': 
\begin{equation} 
\mathrm{Order + Number \sim Lorentzian\,\, Geometry }.   
\label{ordernumber} 
\end{equation}
One  therefore assumes a fundamental correspondence between the number of elements in a region of the
  causal set and the continuum volume element that it represents. The condition of local finiteness means that all
  order intervals  in the causal set are of finite cardinality and hence correspond in the continuum to
  finite volume. This CST slogan captures  the essence of the (yet to be specified)  continuum approximation of a {manifold-like }   causal set, which we denote by $C \sim (M,g)$. 
While the continuum causal structure gives the continuum conformal geometry  via the HKMM theorem, the  discrete  causal
structure represented by the underlying causal set is conjectured to approximate  the entire spacetime geometry. Thus,  discreteness supplies the missing conformal factor, or the missing $\left(1/10\right)^{\mathrm{th}}$ of the metric, in $d=4$.

Motivated thus, CST makes the following  radical proposal \citep{blms}:  
\begin{enumerate}    
\item \label{qgcst} Quantum gravity is a quantum theory of causal sets. 
\item  \label{ca} 
A continuum spacetime $(M,g)$ is an approximation of an underlying causal set $C \sim (M,g)$, where  
\begin{enumerate}  
\item  \label{order} Order $\sim $ Causal Order 
\item  \label{number} Number $\sim $ Spacetime Volume 
\end{enumerate} 
\end{enumerate}

In CST, the {kinematical} space of $d=4$ continuum spacetime geometries or histories is replaced with a \emph{
  sample space} $\Omega$ of causal sets.  Thus,  discreteness is viewed not only as a tool for regulating the
continuum, but as a fundamental feature of quantum spacetime. $\Omega$ includes causal sets that
have no continuum counterpart, i.e., they cannot be related via Conditions (\ref{order})  and
(\ref{number}) to \emph{any}  continuum spacetime in any dimension. These non-manifold-like causal sets are expected to
play an important role in the deep quantum regime. In order to make this precise we need to define what it means for a
causal set to be manifold-like, i.e., to make precise the relation ``$C \sim (M,g)$''.

Before doing so, it is important to understand the need for a continuum approximation at all.  Without it, Condition
(\ref{qgcst}) yields  \emph{any}  quantum theory of locally finite posets: one then has the full freedom of choosing any poset calculus to construct
a quantum dynamics, without need to connect with the continuum.   Examples of such poset approaches to quantum gravity
include those by  \cite{finkelstein} and \cite{hemion}, and more recently \cite{lee}. What distinguishes CST from
these approaches is the critical role played by both causality and discrete covariance which informs the choice of the
dynamics as well the physical observables.  In particular, condition (\ref{ca}) is the requirement  that in the continuum approximation these observables should
correspond to appropriate continuum topological and geometric covariant observables.

What do we mean by the \emph{continuum approximation}   Condition (\ref{ca})?  We begin to answer this  by looking
for the underlying causal set of a  causal spacetime $(M,g)$.  A useful analogy to keep in mind  is that of a
macroscopic  fluid, for example a glass of water.  Here, there are a multitude of  molecular-level configurations
corresponding to the same macroscopic state.  Similarly, we expect there to be a multitude of  causal sets  approximated by the same
spacetime $(M,g)$. And,  just as the set of allowed microstates of the glass of
water depends on the molecular size, the causal set microstate depends on the \emph{discreteness scale $V_c$}, which is
a fundamental spacetime volume cut-off.\footnote{The most obvious choice for $V_c$ is the Planck volume, but we will not
  require it at this stage.}

Since the causal set $C$ approximating $(M,g)$ is locally finite,  it represents a proper subset of the event-set $M$.
An   \emph{  embedding} is  the injective map
\begin{equation} \Phi:C \hookrightarrow (M,g), \quad x \prec_C y \Leftrightarrow \Phi(x) \prec_M \Phi(y),
\end{equation} 
 where $\prec_{C} $ and $\prec_M$ denote  the order relations in $C $ and $ M$ respectively.  Not every causal set can
 be embedded into a given spacetime $(M,g)$. Moreover, even if an embedding exists, this is not sufficient to ensure
 that $C \sim (M,g)$ since only Condition (\ref{order})  is satisfied.  In addition to correlate the cardinality
 of the causal set with the spacetime volume element, Condition (\ref{number}),   the embeddings must also be
 \emph{uniform}  with respect
 to the spacetime volume measure of   $(M,g)$.  A causal set  is said to approximate a spacetime $C \sim (M,g)$ at \emph{density}
 $\rc=V_c^{-1}$ if there exists    
 a \emph{faithful embedding} 
 \begin{equation}
   \Phi: C \hookrightarrow M, \quad \Phi(C) \mathrm{\ is \ a \ uniform \ distribution \ in \ }(M,g) \mathrm{\ at \
     density \ } \rho_c, 
 \end{equation}
 where by \emph{uniform} we mean with respect to the spacetime volume measure of   $(M,g)$.

\begin{figure}
\centerline{
  \includegraphics[width=0.5\textwidth]{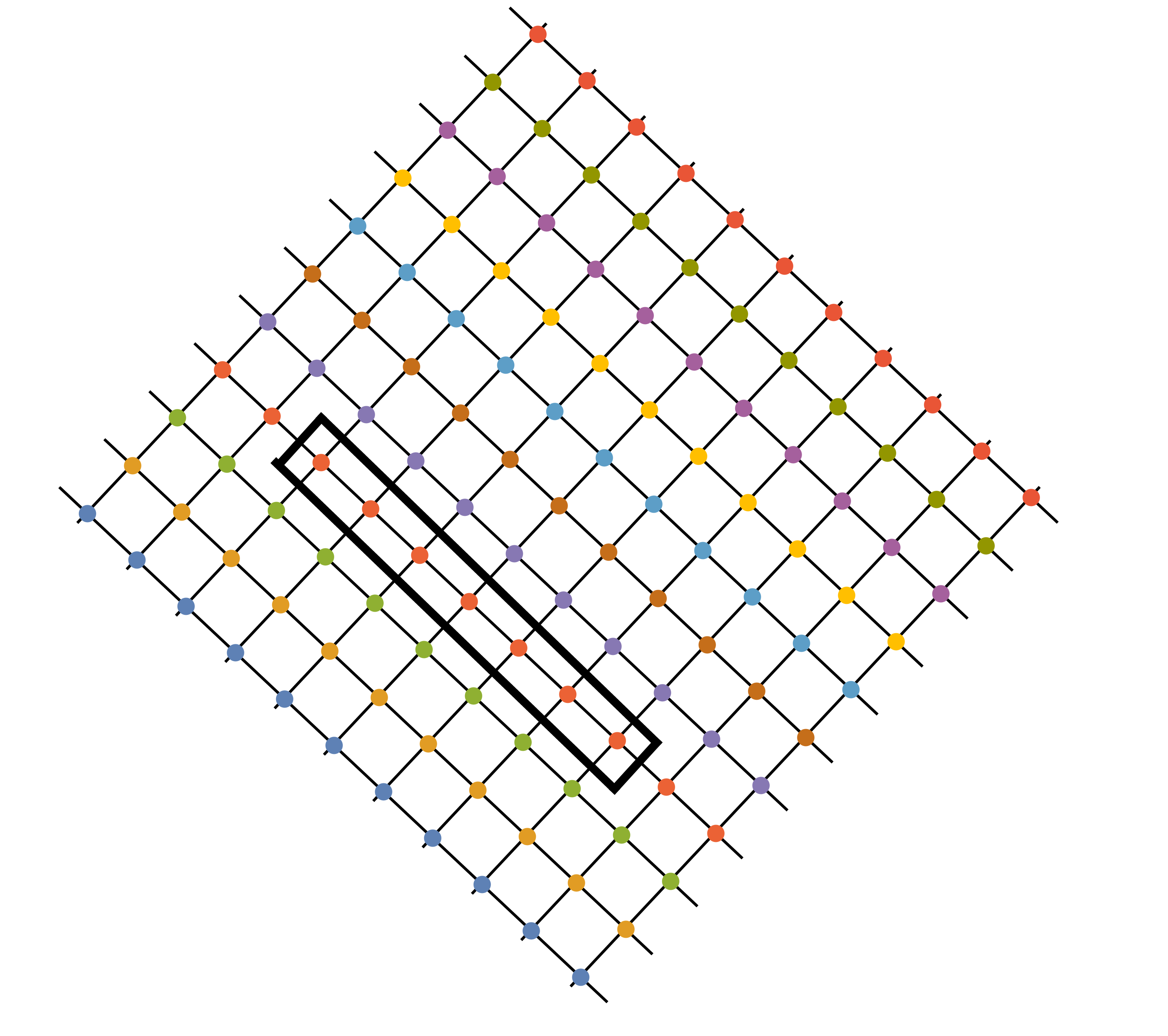}
  \includegraphics[width=0.5\textwidth]{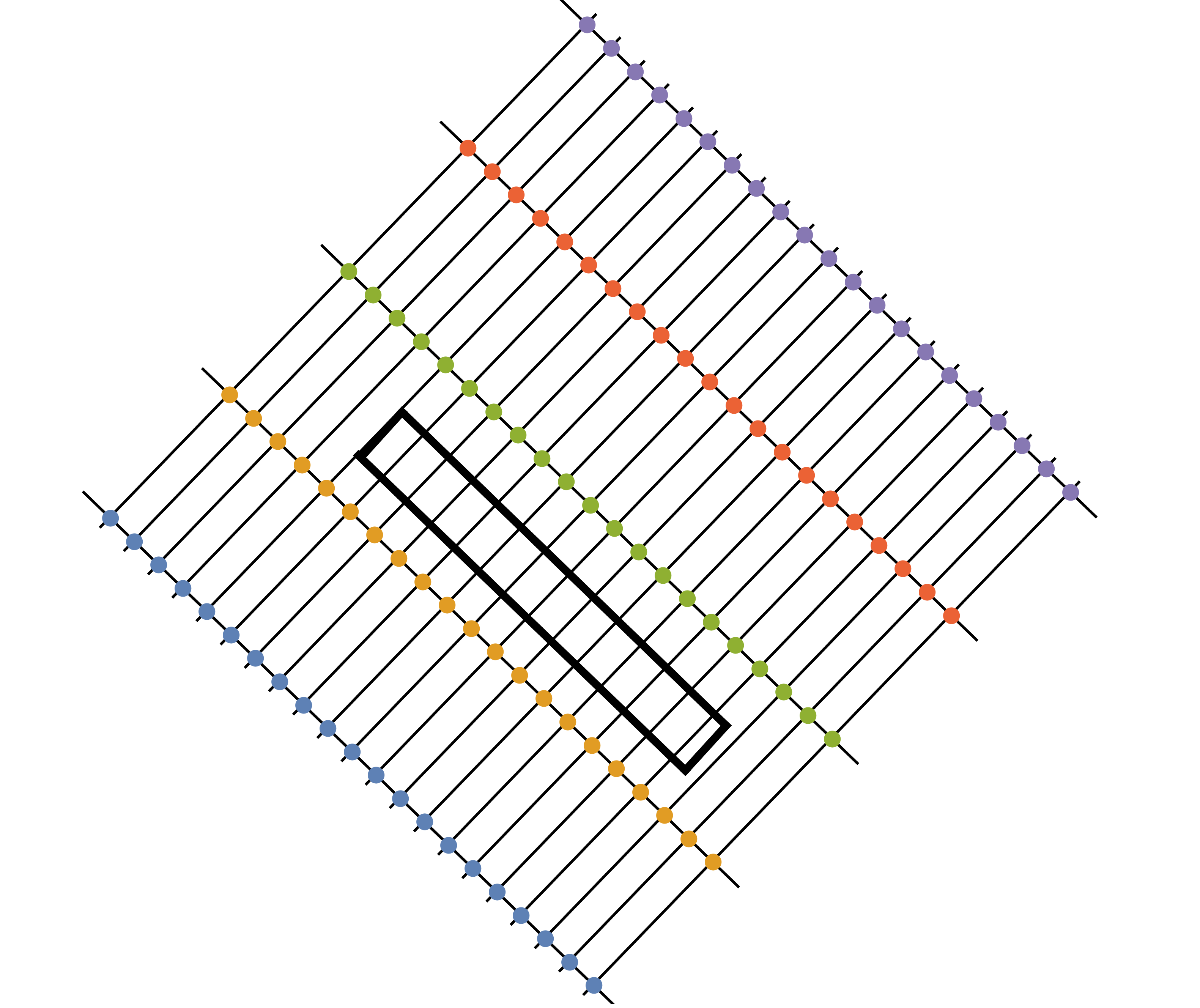}
}
\caption{The lightcone lattice in $d=2$. The lattice on the left looks ``regular'' in a fixed frame but 
  transforms into the ``stretched''  lattice on the right under a boost. The $n \sim \rho_cV$ correspondence cannot be
  implemented as seen from  the example of the Alexandrov interval, which  contains $n=7$ lattice points in the lattice
  in the left but is empty after a boost.}
\label{lattice.fig}
  \end{figure}

The uniform distribution at density $\rho_c$ ensures that  every finite spacetime volume $V$ is represented
by  a finite number of elements $n \sim \rho_c V $ in the causal set.  It is natural to make these finite spacetime regions
causally convex, so that they can be constructed from unions of Alexandrov intervals $\Alex[p,q] $ in $(M,g)$. 
 However,  we must
ensure covariance, since the goal is to be able to recover the approximate covariant spacetime geometry.  This is why 
$\Phi(C)$ is required to be uniformly distributed in $(M,g)$ with respect to the spacetime volume measure. It is obvious
that a ``regular'' lattice cannot do the job since it is  not  regular in all frames or coordinate systems. Hence  it is not
possible to consistently  assign $n \sim \rho_c V$ to such lattices (see Fig.~{\ref{lattice.fig}}). 

The issue of symmetry breaking is of course obvious even in Euclidean space. Any regular discretisation breaks the rotational
and translational  symmetry of the continuum. In the  lattice calculations for 
QCD, these symmetries are restored only  in the continuum limit, but 
are broken as long as the discreteness persists.  In \cite{tdlee} it was suggested
that symmetry can be restored in a  randomly generated  lattice where there lattice points are uniformly distributed via
a Poisson process. This has the advantage of not picking any preferred direction and hence not explicitly breaking
symmetry, at least on average. We will discuss this point in greater detail further on.  

Set in the context of spacetime, the Poisson distribution is a natural choice for $\Phi(C)$, with the probability of finding $n$ elements in a
spacetime region of volume $v$  given by 
\begin{equation} 
P_v(n)=\frac{(\rc v)^n}{n!}\exp^{-\rc v}.  
\label{eq:poisson} 
\end{equation} 
This means that on the \emph{average}
\begin{equation} 
\av{\bbn}=\rc v, 
\end{equation}  
where $\bn$ is the random variable associated with the random causal set $\Phi(C)$.   
This distribution then gives us the covariantly defined $n \sim  \rc V$ correspondence we seek.\footnote{Since $\Phi(C)$ is a random
  causal set, any function of ${\mathbf F}: C \rightarrow \re $ is therefore a random variable.} 

In a \emph{Poisson sprinkling} into a
spacetime $(M,g)$ at density $\rho_c$ one selects points in $(M,g)$ uniformly at random and imposes a partial ordering on these elements via the induced spacetime causality relation.  Starting from $(M,g)$, we can then
obtain an ensemble of ``microstates'' or   causal sets, which we denote by   $\cC(M,\rc)$,  via  the Poisson sprinkling.\footnote{$\cC(M,\rc)$
  explicitly depends on the spacetime metric $g$, which we have suppressed for brevity of notation.} Each causal set thus obtained is a
\emph{realisation}, while any ``averaging'' is done over the whole ensemble.

\begin{figure}[ht]
\centering \resizebox{3in}{!}{\includegraphics{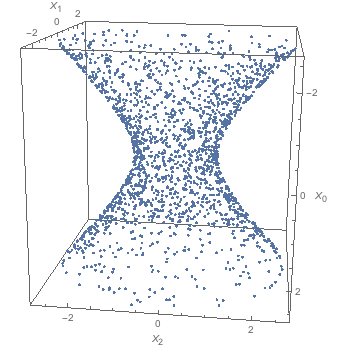}}
\caption{A Poisson sprinkling into a portion of 2d de Sitter spacetime embedded in $\mink^3$. The  relations on the elements are deduced from the causal structure of  $\mink^3$.}
\label{causalset.fig}
\end{figure}

We say that a causal set $C$ is approximated by a spacetime $(M,g)$ if $C$ can be
obtained from  $(M,g)$ via  a high probability Poisson sprinkling. Conversely, for every  $C \in \cC(M,\rc)$  there is a natural embedding map
\begin{equation}
\Phi: C \hookrightarrow  M \,,  
\end{equation}
where $\Phi(C)$ is a particular realisation in $\cC(M,\rc)$. In Fig.~\ref{causalset.fig}, we show a causal set 
obtained by Poisson sprinkling into $d=2$ de Sitter spacetime.

That there is a fundamental discrete randomness even \emph{kinematically}  is not always easy for a newcomer to CST to come to terms with.  Not only
does CST posit a fundamental  discreteness,  it also requires it to be probabilistic. Thus, even before coming to 
quantum probabilities, CST makes us work with a  classical, stochastic discrete geometry.  

Let us state some obvious, but important aspects of Eq.~(\ref{eq:poisson}). Let $\Phi: C \hookrightarrow  (M,g)$ be a faithful embedding at density $\rho_c$.  While the set of all finite
volume regions\footnote{We assume that these are always causally convex.}   $v$ possess {on average}  $\av{\bn}=\rho_c
v$ elements of $C$,\footnote{Henceforth we will
  identify $\Phi(C)$ with $C$, whenever $\Phi$ is a faithful embedding.} the  Poisson fluctuations are given by  $\delta n =\sqrt{n}$. 
Thus, it is possible that  the region contains no elements at all, i.e., there is a ``void''.   An important
question to ask is how large a void is allowed, since  a  sufficiently large void
would have an obvious  effect on our macroscopic perception of a
continuum. If  spacetime is unbounded, as it is in  Minkowski spacetime,
the probability for the existence of a void of \emph{any} size is one.  Can this be compatible at all with the idea of an
emergent continuum in which the classical world can exist,  unperturbed by the vagaries of quantum gravity? 

The presence of a macroscopic void means that the continuum approximation is not realised in this region. A prediction
of CST is then that  the emergent continuum regions of spacetime are 
bounded both spatially and temporally, even if the underlying causal set is itself ``unbounded'' or countable. Thus, a continuum universe is not
viable forever. However, since the current phase of the observable universe \emph{does} have a continuum realisation one
has to ask whether this is compatible with CST discretisation.  In \cite{swerves} the
probability for there to be  at least one  nuclear size void  $ \sim 10^{-60}m^4 $ was calculated in a region of Minkowski spacetime
which is  the size of our present universe.  Using general considerations they found that the probability is of order $ 10^{84}
\times 10^{168} \times e^{-{10^{72}}}$, which is an absurdly small number!  Thus, CST poses no phenomenological inconsistency in
this regard. 

\begin{figure}[ht]
\centering \resizebox{3in}{!}  {\includegraphics[angle=45]{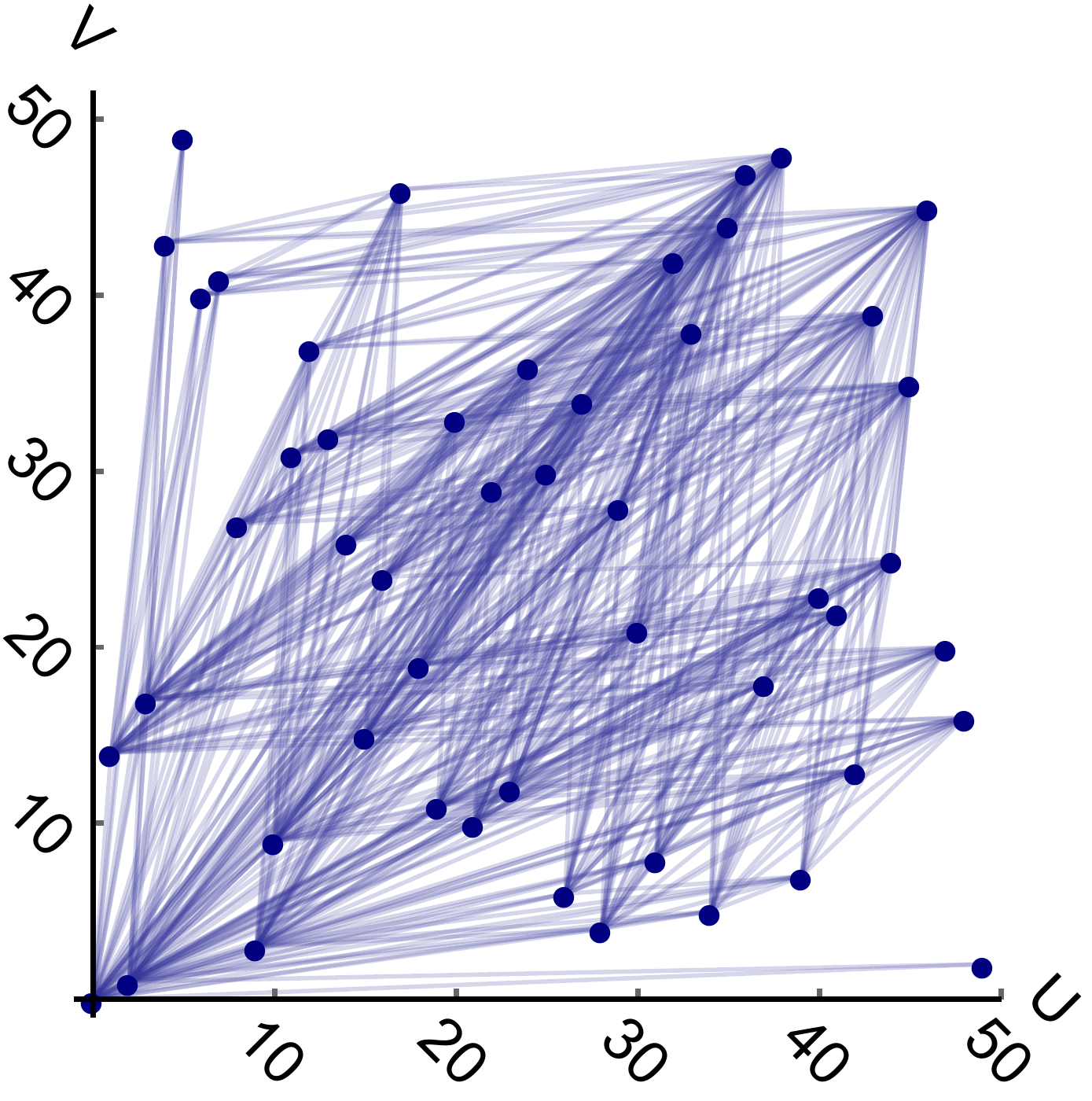}}
\caption{A \emph{Hasse diagram}  of a causal set that faithfully embeds into a causal diamond in $\mink^2$. In a Hasse diagram
  only the nearest neighbour relations or \emph{links} are shown. The remaining relations follow by transitivity.}
\label{2drandom.fig}
\end{figure}

\begin{figure}[ht]
\centering {\resizebox{3.5in}{!}{\includegraphics{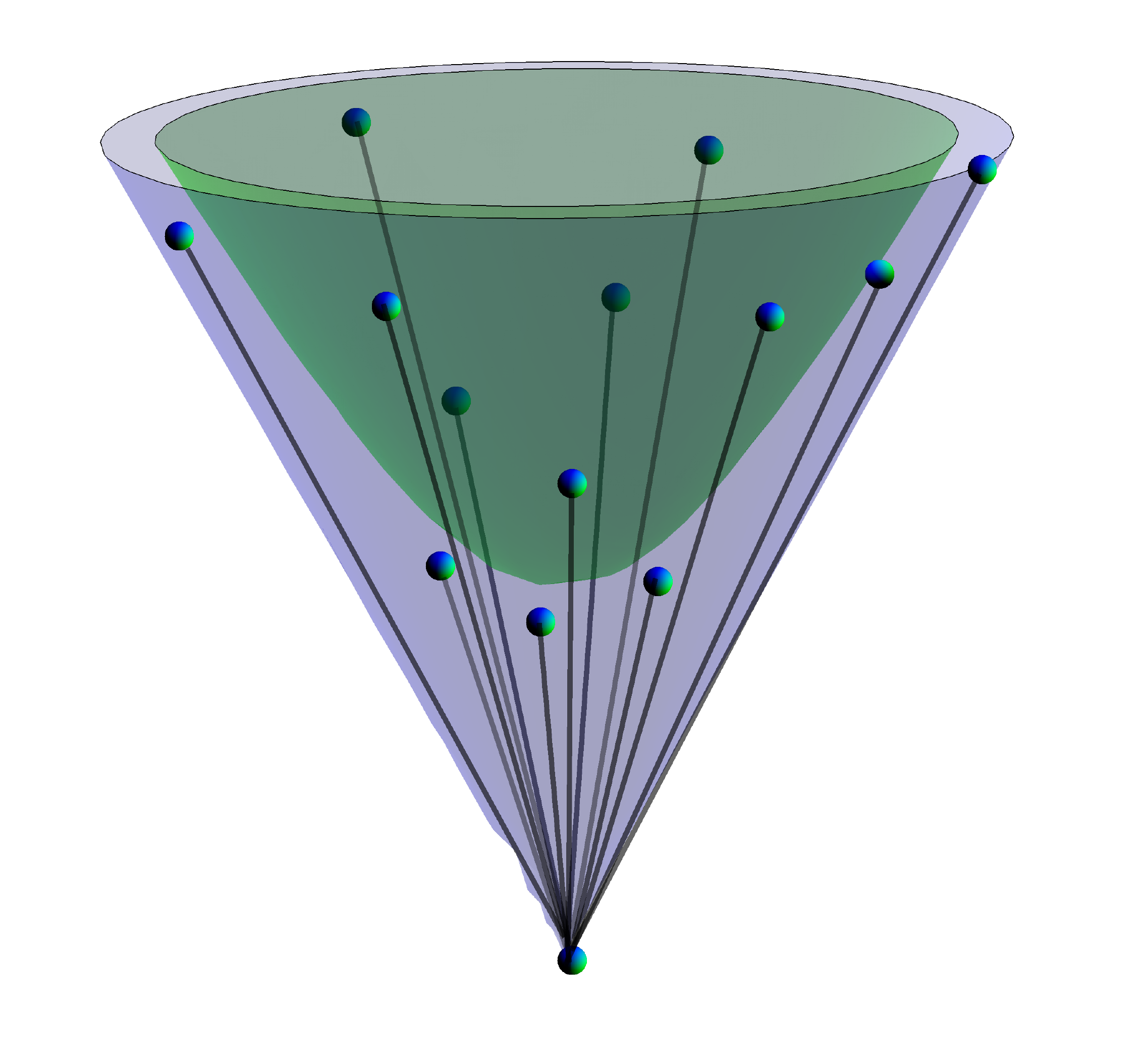}}}
\caption{The \emph{valency} or number of nearest neighbours  of an element in a causal set obtained from a Poisson sprinkling into $\mink^2$ is infinite.}
\label{valency.fig}
\end{figure}

An example of a manifold-like causal set $C$ which is obtained via a Poisson sprinkling into a
2d causal diamond is shown in Fig.~\ref{2drandom.fig}.
A striking feature of the resulting graph is that there is a high degree of connectivity. In the
 \emph{Hasse diagram} of Fig.~\ref{2drandom.fig} only the nearest neighbour relations or {links} are depicted with the remaining relations following from
 transitivity.  $e \prec e' \in C$  is said to be  a  \emph{link} if   $\nexists \, \, e''\in C$  
such 
 that $e''\neq e,e'$ and $e\prec e''\prec e'$.    In 
 a  causal set that is obtained from a Poisson sprinkling, the \emph{valency}, i.e., the number of nearest neighbours or
 links from any given element  is typically very large. This is an important feature of continuum
 like causal sets and results from the fact that the elements of $C$ are uniformly distributed in $(M,g)$.  For a given
 element $e \in C$, the
 probability of an event $x \succ e $ to 
be a link is equal to the probability that the Alexandrov interval $\Alex[e,x]$ does not contain any
elements of $C$. Since 
\begin{equation} 
P_V(0)=e^{-\rho_c V},  
\end{equation}  
the probability is significant only when $V \sim V_c$. As shown in  Fig.~\ref{valency.fig}, in  $\mink^d$, the set of events within a
proper time $\propto (V)^{1/d}$ to the future (or past) of a point $p$ lies in the region between the future light cone and the
hyperboloid $-t^2+\Sigma_i x_i^2 \propto (V)^{2/d}$, with $t>0$. 
Up to fluctuations, therefore,  most of the future links to $e$ lie within the hyperboloid with $V=V_c \pm \sqrt{V_c}$. This is a  non-compact, infinite volume
region and hence the number of future links to $e$ is (almost surely) infinite.   Since linked elements  are the nearest
neighbours of $e$, this means the valency of 
 the graph $C$ is infinite. It is this feature of manifold-like causal sets which gives rise to a characteristic  
``non-locality'',  and   plays a critical  role in the continuum approximation of CST, time and again.

The Poisson distribution is not the only choice for a uniform distribution.
A pertinent question is whether a different choice of distribution is possible, which would 
lead to a different manifestation of the continuum approximation.   In \cite{aspoisson}, this question was addressed in
some detail.  Let $C \sim (M,g)$  at density $\rc$.  Consider $k$ non-overlapping Alexandrov intervals of
volume $V$ in $(M,g)$. Since $C$ is uniformly  distributed,  $\av{\bn} = \rho_c V$. The most  optimal choice of
distribution, is also  one in which the fluctuations $\delta \bn/\av{\bn}=\sqrt{\av{(\bn-\av{\bn})^2}}/{\av{\bn}}$ are minimised. This ensures
that $C$ is as close to the continuum as possible. For the Poisson distribution $\delta \bn/\av{\bn} = 1/\sqrt{\av{\bn}} = 1/\sqrt{\rho_c V}$.  Is this as good as it
gets? It was shown that for $d>2$,  and under certain further technical assumptions, the Poisson distribution indeed
does the best job. Strengthening these results is important as it can improve our understanding of  the continuum approximation.

\subsection{The Hauptvermutung or fundamental conjecture of CST} 
\label{ssec:haupt} 

An important question is the uniqueness of  the continuum approximation associated to  a causal set $C$.  Can a given
$C$ be faithfully embedded at density $\rho_c$  into two different spacetimes,
$(M,g)$ and $(M',g')$?  We expect that this is the case  if $(M,g) $ and $ (M',g')$ differ
on scales  smaller than 
$\rho_c$, or that they are, in an appropriate sense, ``close'' $(M,g) \sim (M',g')$.   Let us assume that a causal set \emph{can} be identified with two macroscopically distinct spacetimes at the same density
$\rho_c$.  Should this be interpreted as a hidden duality between these spacetimes, as is the case for example for isospectral manifolds
or mirror manifolds in string theory \citep{string}?  The answer is clearly in the negative, since the aim of the CST
continuum approximation is to ensure that $C$ contains \emph{all} the information
in $(M,g)$ at scales above $\rho_c^{-1}$. Macroscopic non-uniqueness would therefore mean that the intent of
the CST  continuum approximation is \emph{not}  satisfied. 

We thus state the fundamental conjecture of CST: 

\textit{\textbf{The Hauptvermutung of CST:}  $C$ can  be faithfully embedded at density $\rho_c$  into two distinct spacetimes,
$(M,g)$ and $(M',g')$  iff  they are approximately isometric.}    

By an \emph{approximate  isometry }, $(M,g) \sim (M',g')$ at density $\rc$, we mean that $(M,g)$ and $ (M',g')$ differ only at scales smaller
 than $\rho_c$.  Defining such an isometry  rigorously is challenging, but concrete proposals have been made by 
 \cite{bomclose,noldusone,noldustwo,bomnoldus,bomnoldustwo},  en route to a full proof of the conjecture. Because of the
 technical nature of these results, we  will discuss it only very briefly in the next
 section, and  instead use the above intuitive and functional definition of closeness. 

 Condition (\ref{qgcst}) tells us that the kinematic space of Lorentzian geometries must be replaced by a \emph{sample
   space }  $\Omega$ of causal sets.  Let $\Omega$ be the set of all 
countable causal sets and $\contg$ the set of all possible Lorentzian geometries,  in all dimensions.  If $\sim$ denotes
the approximate isometry at a given $\rc$, as  discussed above, the quotient
space $\contg/\!\!\sim$ corresponds to the set of all continuum-like causal sets $\Omega_{cont}\subset \Omega$ at that $\rc$.  Thus,
causal sets in 
$\Omega$  correspond to Lorentzian geometries of  \emph{all} dimensions!  Couched this way, we see that   CST dynamics
has the daunting task of not only obtaining manifold-like causal sets in the classical limit, but also ones that have dimension $d=4$.  

\begin{figure}[htb]
\centering
\includegraphics[width=\textwidth]{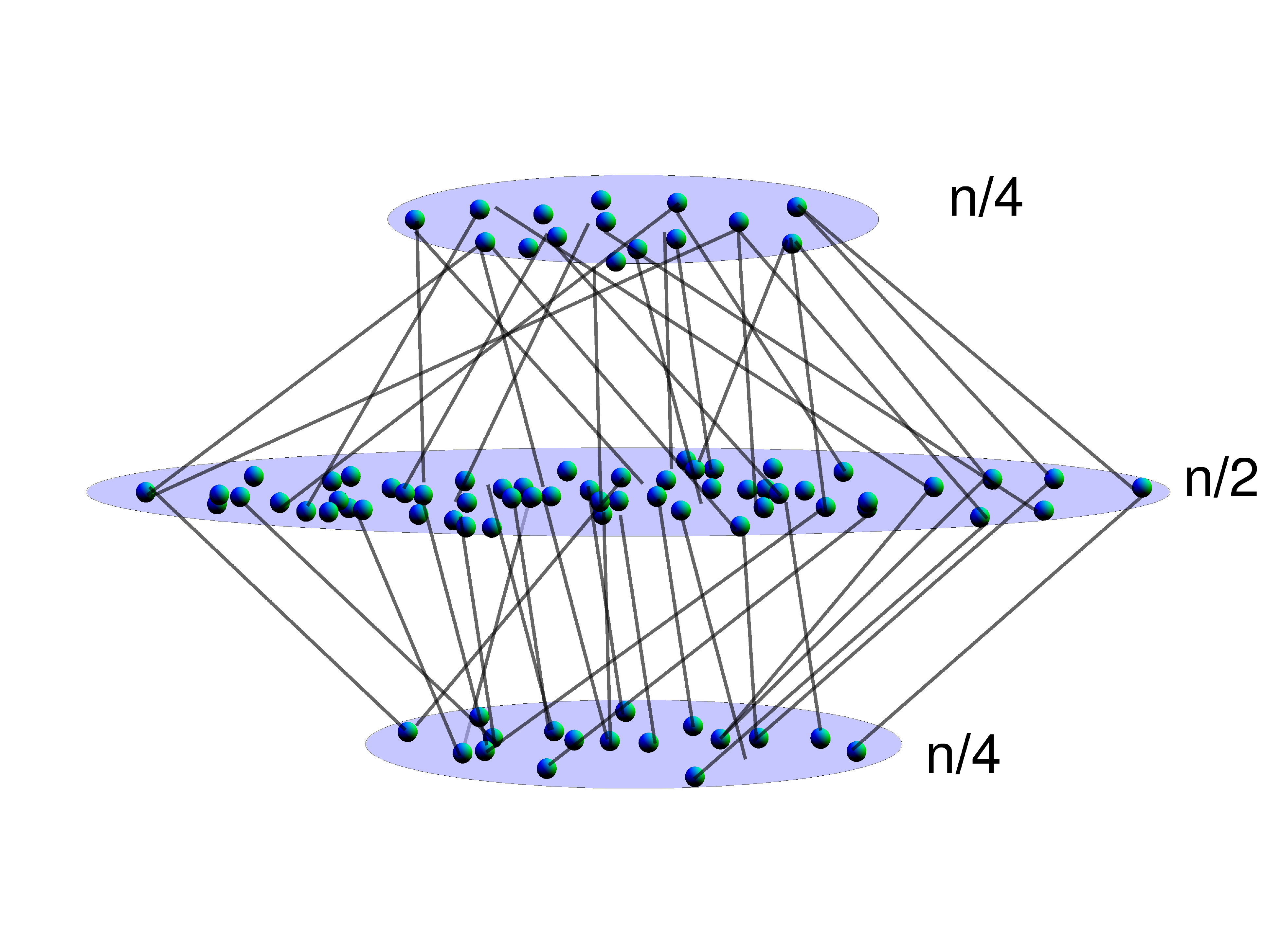}
\caption{A Kleitman--Rothschild or KR poset.}
\label{kr.fig}
\end{figure}

As mentioned in the  introduction, the sample space of $n$ element causal sets $\Omega_n$ is dominated by the KR posets
depicted in Fig.~\ref{kr.fig} and are hence very non-manifold-like    \citep{kr}. A KR poset has   three ``layers'' (or abstract ``moments of
time''),  with roughly $n/4$ elements in the bottom and top layer and such that each  element  in the bottom
layer is related to roughly half those in the  middle layer, and similarly each element  in the top layer is related to
roughly half those in the middle layer.  The number of  KR posets grows as $\sim 2^{{n^2}/{4}}$ and hence must play a role in the
deep quantum regime. Since they are non-manifold-like  they pose  a challenge to the dynamics, which must overcome
their entropic dominance in the classical  limit of the theory.  Even if the entropy from these KR  posets is
  suppressed by an appropriate choice of dynamics, however, there is a sub-dominant hierarchy of non-manifold-like posets (also layered)
  which also need to be 
  reckoned with  \citep{dharone,dhartwo,pst}. 

Closely tied to the continuum approximation  is  the notion of ``coarse graining''. Given a spacetime $(M,g)$ the
set $\cC(M,\rc)$ can be obtained for different values of $\rc$.  Given a causal set $C$ which faithfully embeds into $(M,g)$ at $\rc$, 
one can then \emph{coarse grain} it to a smaller subcausal set  $C' \subset C$  which faithfully embeds into $(M,g)$  at
$\rc' <\rc$.   A natural coarse graining would be via a random selection of elements in $C$ such that for every $n$ elements
of $C$ roughly $n'=(\rc'/\rc) n$ elements are chosen. Even if $C$ itself does not faithfully embed into $(M,g)$ at $\rc$, it is
possible that a coarse graining of $C$ can be   embedded. This would be in keeping with our sense in CST that the deep
quantum regime need not be manifold-like.   One
can also envisage manifold-like causal sets with   a regular  fixed lattice-like structure  attached to each element similar to a 
``fibration'',  in the spirit of Kaluza--Klein theories. Instead of the coarse graining procedure, it would be more
appropriate to  take the  quotient with respect to this fibre to obtain the continuum like causal
set. Recently, the implications of coarse graining in CST, both
dynamically  and
kinematically, were considered in \cite{astridcg}  based on renormalisation techniques. 

\subsection{Discreteness without Lorentz breaking}

It is often assumed that a fundamental discreteness is incompatible with continuous symmetries.  As was pointed
out  in \cite{tdlee},  in the Euclidean context, symmetry can be preserved on average in a random lattice.
In \cite{bomhensor}, it was shown that a causal set in $\cC(\mink^d,\rc)$ not only preserves Lorentz invariance on
average, but in \emph{every} realisation, with respect to the Poisson distribution.  Thus,  in a very specific sense a manifold-like   causal set does not break Lorentz invariance.  In
order to see the contrast between the Lorentzian and Euclidean cases we present the arguments of \cite{bomhensor}
starting with the easier  Euclidean case.

Consider the Euclidean plane $\cP = (\re^2,\delta_{ab})$, and let   $\Phi: \cC(\cP,\rc) \hookrightarrow \cP$ be the
natural  embedding map, where $\cC(\cP,\rc)$ denotes the ensemble of Poisson sprinklings into $\cP$ at density $\rc$. A rotation $r \in SO(2)$ about a point $p \in \cP$, induces a map
$r^* : \cC(\cP,\rc) \rightarrow \cC(\cP,\rc)$, where $r^*=\Phi^{-1}\circ r \circ \Phi$ and similarly a 
translation $t$ in $\cP$ induces the map $t^*: \cC(\cP,\rc)
\rightarrow \cC(\cP,\rc)$. The action of the  Euclidean group  is clearly not transitive on $\cC(\cP,\rc)$ but has non-trivial orbits which provide a fibration of $\cC(\cP,\rc)$. Thus 
the ensemble $\cC(\cP,\rc)$ preserves the Euclidean group on average. This is
the sense in which the discussion of \cite{tdlee} states that the random discretisation preserves the Euclidean group.

The situation is however different for a \emph{given} realisation $ P \in \cC(\cP,\rc)$. Fixing an 
element $e \in \Phi(P)$,
we  define a {direction} $\vd \in S^1$, the space of unit vectors in
$\cP$ centred at $e$.  Under  a rotation $r$ about $e$, $\vd \rightarrow r^*(\vd)\in S^1$.  In general, we want a rule that assigns a
natural direction to every $P \in \cC(\cP,\rc)$. One simple choice is to find the 
closest element to $e$  in $\Phi(P)$, which \emph{is}  well defined in this    Euclidean context.  Moreover, this element is almost surely unique, since the probability of  two elements being at the same radius from $e$ is zero in a Poisson distribution.
Thus we can  define  a  ``direction map''  $\vD_e: \cC(\cP,\rc) \rightarrow S^1$ for a fixed $e \in \Phi(P)$ consistent with the rotation map, i.e.,  $\vD_e$ commutes with any $r\in SO(2)$, or is \emph{equivariant}. 

Associated with  $\cC(\cP,\rc)$, is a probability distribution $\mu$ arising  from the Poisson sprinkling  which associates with
every measurable set $\alpha$ in $\cC(\cP,\rc)$ a probability $\mu(\alpha) \in [0,1]$. The Poisson distribution being
volume preserving  
\citep{stoyan},  the measure on $\cC(\cP,\rc)$  moreover must be independent of the action of the  Euclidean group on $\cC(\cP,\rc)$, i.e.:  $\mu \circ r =\mu$.

In analogy with a continuous map, a measurable map is one whose 
preimage from  a measurable set is itself a measurable set. The natural map $\vD$ we have defined is a measurable map, and we can use it to define a measure on $S^1$: $\mu_\vD \equiv \mu \circ
\vD^{-1}$.   
Using the invariance of $\mu$ under rotations  and the equivariance of $\vD$ under rotations 
\begin{equation} 
\mu_\vD=\mu\circ r \circ \vD^{-1} = \mu \circ \vD^{-1} \circ r
=\mu_\vD \circ r \, \, \, \forall \, \, r \in SO(2),  
\end{equation} 
we see that $\mu_\vD$ is also invariant under rotations. Because 
$S^1$ is compact, this does not lead to a contradiction. In analogy with the construction used in \cite{bomhensor} for the Lorentzian case, we choose a  measurable set  $s\equiv (0,2\pi/n) \in S^1$. A  
rotation by  $r(2\pi/n)$, takes $s \rightarrow s'$ which is non-overlapping, so that after  $n$ successive rotations,
$r^n(2\pi/n)\circ s = s$. Since  each rotation does not change
$\mu_\vD$ and  $\mu_\vD(S^1)=1$, this means that $\mu_\vD(s)=1/n$.   
Thus,  it \emph{is} possible to assign a consistent direction for a given realisation $P\in \cC(\cP,\rc)$ and hence break
Euclidean symmetry. 

However, this is not the case for the space of  sprinklings $\cC({\mink^d},\rc)$ into $\mink^d$, where the hyperboloid  $\cH^{d-1}$ now
denotes  the space of future directed unit vectors and is invariant under the Lorentz group $SO(n-1,1)$ about a fixed
point $p\in \mink^{d-1}$.  To begin with, there is no  ``natural'' direction map. Let $C \in \cC(\mink^d,\rc)$.  To find an element  which is
closest to some fixed $e \in \Phi(C)$, one has to take the infimum over $J^+(e)$ , or some suitable Lorentz invariant
subset of it, which being non-compact, does not exist.  Assume that \emph{some} measurable direction map $D: \Omega_{\mink^d} \rightarrow \cH^{d-1}$,  does exist.  Then the above arguments imply  that $\mu_D$ must be invariant under Lorentz boosts. The
action of successive Lorentz transformations $\Lambda$ can take a given
measurable set $h \in \cH^{d-1}$ to an infinite number of copies that are
non-overlapping, and \emph{of the same measure}. Since $\cH^{d-1}$ is non-compact,  this is not  possible  unless each set is of measure zero, but since this is true for any measurable set $h$ and we require $\mu_D(\cH^{d-1})=1$, this is a contradiction.  This proves the following theorem \citep{bomhensor}:   

\begin{theorem} In dimensions $n>1$ there exists no equivariant
measurable map $\vD: \cC(\mink^d,\rc) \rightarrow \cH$, i.e., 
\begin{equation} 
\vD \circ \Lambda = \Lambda \circ \vD \, \, \forall \, \Lambda \in
SO(n-1,1).  
\end{equation}   
\end{theorem}

\begin{figure}[ht]
\centering \resizebox{2in}{!}  {\includegraphics[width=\textwidth]{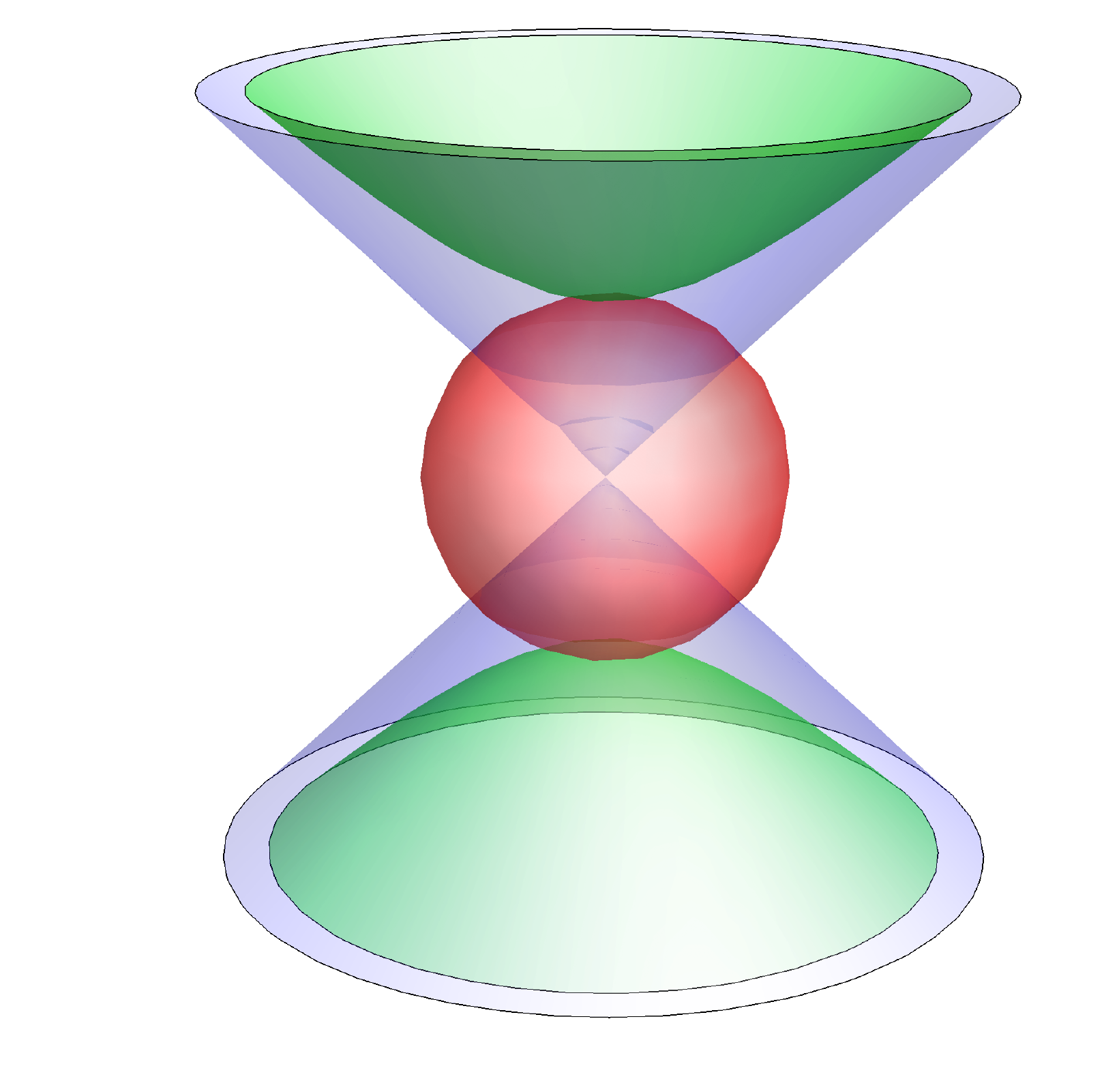}}
\caption{The space of unit directions in $\re^d$ is $S^{d-1}$, while the space of unit timelike vectors in $\mink^d
  $ is $\cH^{d-1}$.}
\label{Directions.fig}
\end{figure}

In other words, even for a given sprinkling $\omega \in \Omega_{\mink^d}$ it is not possible to consistently pick a
direction  in $\cH^{d-1}$. Consistency means that under a boost $\Lambda: \omega \rightarrow \Lambda\circ w$, and
hence $D(\omega) \rightarrow \Lambda \circ D(\omega) \in \cH^{d-1}$. Crucial to this  argument is the use of the Poisson
distribution.\footnote{It is interesting to ask if other choices of uniform distribution satisfy the above theorem. If so, then
our criterion for a uniform distribution could not only include ones that minimise the fluctuations but also those that
respect Lorentz invariance.}  Thus, an important prediction of CST is local Lorentz invariance.  Tests of Lorentz invariance over the last couple of decades have produced an ever-tightening
bound, which is therefore consistent with CST \citep{li}.

\subsection{Forks in the road: What makes CST so ``different''?} 
\label{ssec:forks}

In many ways CST  doesn't  fit  the  standard paradigms adopted by other approaches to quantum
gravity and it is worthwhile trying to understand the source of this difference.  The program is minimalist but
also 
rigidly constrained by its continuum approximation. The ensuing non-locality  means that the apparatus of local
physics is not readily available to CST.

\cite{lambdaone}  describes the route to quantum gravity and the various forks at which one has to make choices.
Different routes  may lead to the same destination: for example
(barring interpretational questions),   simple quantum systems  can be
described equally well by the path integral and the canonical approach. However, this
need not be the case in gravity: a set of consistent choices may lead  you down a
unique path, unreachable from another route. Starting from broad principles,  Sorkin argued that  certain
choices at a  fork are preferable to others for a theory quantum gravity. These include the choice of  Lorentzian over Euclidean,
the path integral over canonical quantisation and discreteness over the continuum.   This  set of choices leads to a
CST-like theory, while choosing the Lorentzian-Hamiltonian-continuum route  leads to a canonical approach like Loop
Quantum Gravity.

Starting with CST as the final destination, we can work backward to retrace our steps to see what forks had
to be taken and why other routes are impossible to take. The choice  at the discreteness versus continuum fork and the
Lorentzian versus Euclidean fork are obvious from our earlier discussions. As we  explain below, the other essential fork that \emph{has} to be taken
in CST is the histories approach to quantisation.

One of the standard routes to quantisation is via  the canonical approach.  Starting with the phase space of a
classical system, with or without constraints,  quantisation rules  give rise to the 
familiar apparatus of Hilbert spaces and self adjoint operators.   In quantum gravity,  apart
from interpretational issues, this route has difficult technical hurdles, some of which have been partially
overcome \citep{ashtekar}. Essential to the canonical formulation is the $3+1$ split of a spacetime $M=\Sigma \times \re$,
where $\Sigma$ is a Cauchy hypersurface, on which are defined the canonical phase space variables which capture the intrinsic and
extrinsic geometry of $\Sigma$. 

\begin{figure}[!htb]
\centering \resizebox{3in}{!}  {\includegraphics[width=\textwidth]{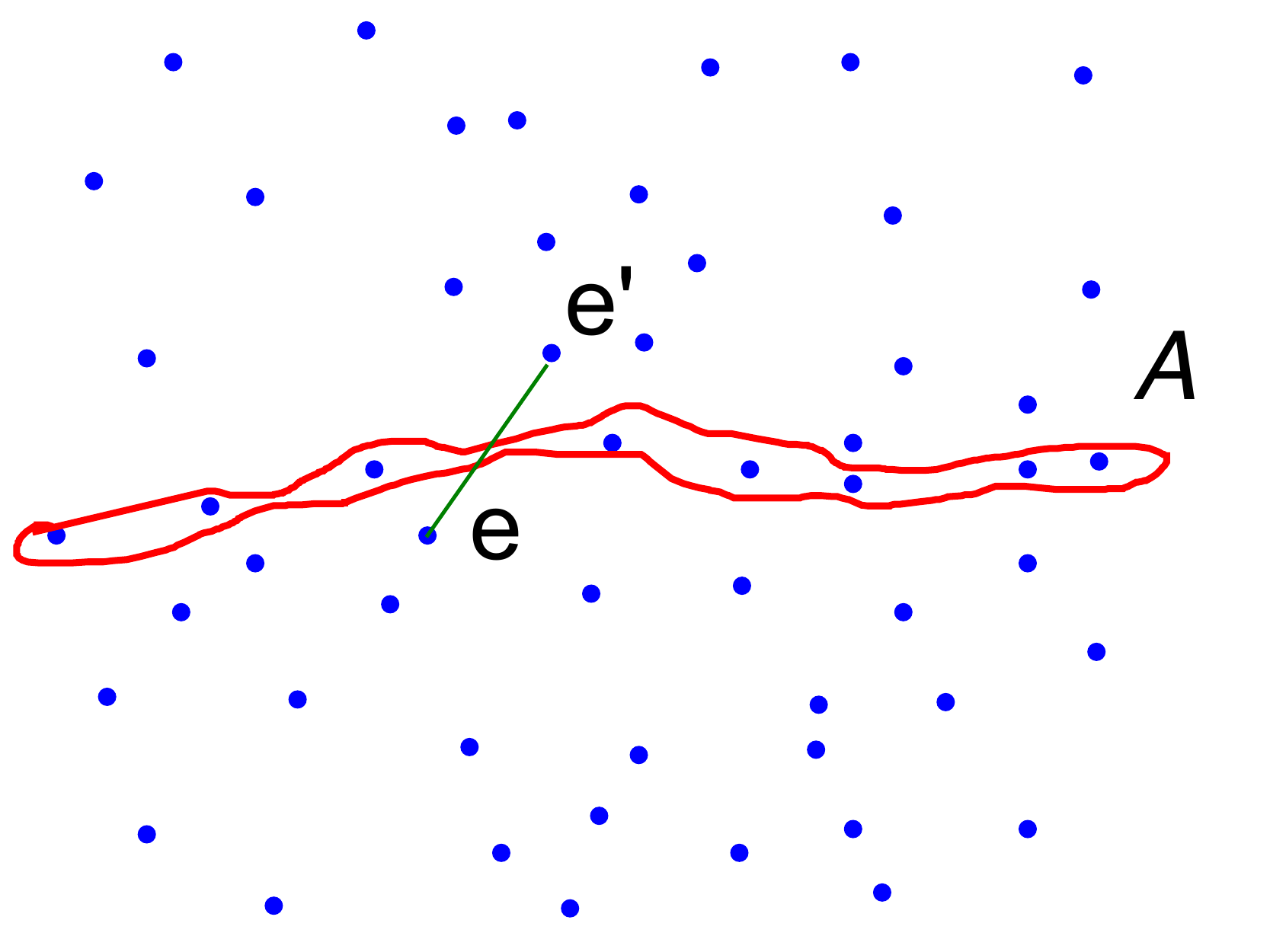}}
\caption{A ``missing link'' from $e$ to $e'$ which ``bypasses'' the inextendible antichain $\ca$.}
\label{Missinglinks.fig}
\end{figure}

The continuum approximation of CST however, does not allow a meaningful definition of a Cauchy hypersurface, because of 
the `` graphical non-locality'' inherent in a continuum like  causal set,
as we will now show.  We begin by defining an \emph{antichain} to be a set of unrelated elements in
  $C$, and an \emph{inextendible antichain} to be an antichain $\cA \subset C$ such that every element $e \in C
  \backslash \cA$ is related to an element of $\cA$.  The   natural choice for a   discrete analog  of a
Cauchy  hypersurface is therefore an {inextendible antichain} $\ca$, which separates the \emph{set} $C$ into its future and past, so that we can
express $C=\fut(\ca) \sqcup
\past(\ca) \sqcup \ca$,  with  $\sqcup$ denoting disjoint union. However, an element  in $\past(\ca)$ can be related via
a link to an 
element in $\fut(\ca)$ thus ``bypassing'' $\ca$. An example of a  ``missing link'' is  depicted in Fig
\ref{Missinglinks.fig}.   This means that unlike a Cauchy hypersurface, $\ca$ is not a summary of its
past, and hence a canonical decomposition using Cauchy hypersurfaces is not viable \citep{antichain}.  On the other
hand, each causal
set is a  ``history'',  and since the sample space of causal sets is countable, one can construct a path integral or path-sum as
over causal sets. We will describe the dynamics of causal sets in more detail in Sect.~\ref{sec:dynamics}. 

Before moving on, we comment on the condition of local finiteness which, as we have pointed out,   provides an intrinsic
  definition of spacetime discreteness. This does not need a continuum approximation. An alternative definition would be
  for the causal set to be  countable, which along with the continuum approximation is sufficient to ensure the number
  to volume correspondence. This includes causal sets with  order intervals of infinite cardinality. This
  allows us to extend  causal set discretisation to more general spacetimes, like anti de Sitter spacetimes, where there
exist  events $p,q$ in the spacetime for which  $\vol(\Alex[p,q])$ is not finite. However, what is ultimately of
interest is the dynamics and in particular, the sample space $\Omega$ of causal sets. In the growth models we will encounter in 
Sect.~\ref{ssec:csg},\ref{ssec:beable} and \ref{ssec:qsg} the sample space consists of past finite posets, while in the
continuum-inspired dynamics of Sect.~\ref{ssec:partn} it consists of finite element posets. Thus, while countable posets may be relevant to a broader
framework in which to study the  dynamics of causal sets, it suffices for the present to focus on locally finite
posets.

\section{Kinematics or geometric reconstruction}
\label{sec:kinematics} 

In this section we discuss the program of geometric reconstruction in which topological and geometric
invariants of a continuum spacetime $(M,g)$ are ``reconstructed'' from the  underlying ensemble of causal sets. The assumption that such a reconstruction
exists for any covariant observable in $(M,g)$ comes from the Hauptvermutung of CST discussed in Sect.~\ref{sec:cst}.

In the statement of the Hauptvermutung,  we used the phrase ``approximately
isometric'', with the promise of an explanation in this section.  A rigorous definition  requires the
notion of  closeness of two Lorentzian spacetimes. In Riemannian geometry, one has the Gromov--Hausdorff distance \citep{petersen}, but there is no simple extension to
Lorentzian geometry, in part because of the indefinite signature. In  \cite{bommeyer}  a measure of closeness of two Lorentzian manifolds was  given in terms of a pseudo distance
function, which however is neither symmetric nor satisfies the triangle inequality. 
Subsequently, in a series of papers,  a true distance function was defined on the space of Lorentzian
geometries, dubbed the Lorentzian Gromov--Hausdorff distance \citep{bomclose,noldusone,noldustwo,bomnoldus,bomnoldustwo}.  While this makes
the statement of the Hauptvermutung precise, there is as yet no complete proof. Recently, 
a purely order theoretic criterion has been used to determine the closeness of causal sets and  prove a version  of the
Hauptvermutung ({\it Sorkin and Zwane, work in progress}).

Apart from these more formal constructions, as we will describe below, a large body of evidence has accumulated in
favour of the Hauptvermutung.  In the program of geometric reconstruction, we look for order invariants in continuum
like causal sets which correspond to manifold (either topological or geometric) invariants of the spacetime.  These
\emph{manifold invariants} include dimension, spatial topology, distance functions between fixed elements in the spacetime,
scalar curvature, the discrete Einstein--Hilbert action, the Gibbons-Hawking-York boundary terms, Green functions for
scalar fields, and the d'Alembertian operator for scalar fields. The identification of the order invariant $\cO$ with
the manifold invariant $\cG$ then ensures that a causal set $C$ that faithfully embeds into $(M,g)$ cannot faithfully
embed into a spacetime with a different manifold invariant $\cG'$.\footnote{This is in the sense of an ensemble,
  since the faithful embedding is defined statistically.}  Thus, in this sense two manifolds can be defined to be  close with respect to
their specific manifold invariants. We can then state the  limited,  order-invariant version of the Hauptvermutung: 

\textit{\textbf{$\cO$-Hauptvermutung:}  If $C$ faithfully
  embeds into $(M,g)$ and $(M',g')$ then $(M,g)$ and $(M',g')$ have the same manifold invariant $\cG$ associated with 
  $\cO$.}

The longer our list of correspondences between order invariants and manifold invariants, the closer we are
to proving the full Hauptvermutung.

In order to correlate a manifold invariant $\cG$ with an order invariant  $\cO$, we must   recast geometry in purely
order theoretic terms.  Note that since locally finite posets appear in a wide range of
contexts,  the poset literature contains several order invariants, but these are typically not related to  the manifold
invariants of interest to us.  The challenge is to choose the appropriate invariants that  correspond to manifold  invariants. Guessing and verifying this using
both analytic and numerical tools is the art of geometric
reconstruction. 

A \emph{labelling} of a causal set $C$ is an injective map: $C \rightarrow \mathbb N$, which is the 
analogue of a choice of coordinate system in the continuum.  By an \emph{order invariant} in a finite causal set $C$ we mean a function $\cO: C \rightarrow \re$ such that $\cO$ is
independent of the {labelling} of $C$. For a manifold-like causal set\footnote{We remind the reader
  that the ensemble depends on the spacetime $(M,g)$ but we suppress the dependence on $g$ for the sake of brevity.} $C \in \cC(M,\rc)$, associated to every order
invariant $\cO$ is the random variable $\bO$ whose expectation value $\av{\bO} $ in the ensemble $\cC(M,\rc)$ is either equal to or limits (in
the large $\rc$ limit) to a manifold invariant $\cG$ of $(M,g)$.  We will typically  restrict to compact regions of $(M,g)$ in order to deal
with finite values of $\bO$.

The first candidates for geometric order invariants were defined for $\cC(\Alex[p,q],\rc)$ where  $\Alex[p,q]$
is an  Alexandrov interval in $\mink^d$. Some of these have 
been later generalised to Alexandrov intervals (or causal diamonds)  in \emph{Riemann Normal Neighbourhoods (RNN)} in
curved spacetime. These manifold invariants are in this sense ``local''. 
In order to find \emph{spatial}  global invariants, the relevant spacetime region is  
a \emph{Gaussian Normal Neighbourhood (GNN)} of a compact Cauchy hypersurface in a globally hyperbolic spacetime. As
discussed in Sect.~\ref{sec:cst}  compactness is
necessary for manifold-likeness since otherwise  there is a finite probability for there to be  arbitrarily large voids which negates the
discrete-continuum correspondence.   


Before proceeding, we remind the reader that we are restricting ourselves to  manifold-like causal sets in this section
only because of the focus on  CST kinematics and the continuum approximation. All the order invariants, however, can be calculated
for \emph{any}  causal set, manifold-like  or not. These order invariants give us an important class of covariant
observables,  essential to constructing a 
quantum theory of causal sets. As we will see in Sect.~\ref{sec:dynamics} they play an important role in the
quantum dynamics. 

The analytic results in this section are typically found in the continuum limit,  $\rc \rightarrow \infty$.  Strictly
speaking, this limit is unphysical in CST because of the assumption of a fundamental discreteness. There are 
fluctuations at  finite $\rc $ which give important deviations from the continuum with potential phenomenological consequences. These are 
however  not always easy to calculate analytically and hence require simulations 
to assess  the size of fluctuations at finite $\rc$. As we will see below, CST kinematics therefore needs a combination
of analytical and numerical tools.

\subsection{Spacetime dimension estimators} 
\label{ssec:dimn} 

The earliest result in CST is a dimension estimator for Minkowski spacetime due to \cite{myrheim}\footnote{This remarkable preprint also
  contains the first expression, again without  detailed proof, of the volume of a small causal diamond in an arbitrary
  spacetime.} and predates BLMS \citep{blms}.  A closely related dimension estimator 
was given by \cite{meyer}, which is now collectively known as the \emph{Myrheim--Meyer}
dimension estimator.  

The {number of relations}  $R$ in a finite $n$ element causal set $C$ is the number
of ordered 
pairs $e_i, e_j \in C$ such that $e_i \prec e_j$.  Since the maximum number of possible  relations on $n$ elements  is
$\binom{n}{2}$,  the  {ordering fraction} is defined as 
\begin{equation} 
r =\frac{2R}{n(n-1)}. 
\end{equation}
It was shown by \cite{myrheim}  that $r$ is dependent only on the dimension
when $C$ faithfully embeds into $\mink^d$.    

We now describe the construction of a closely related dimension estimator by \cite{meyer}. Consider an Alexandrov interval $\Alex_d[p,q] \subset \mink^d$ of volume $V>> \rc^{-1}$.   
We are interested in calculating the expectation value of the random variable $\bR$ associated with $R$ for the ensemble
$\cC(\Alex_d,\rc)$.  This is the probability that a pair of elements $e_1,e_2 \in \Alex_d[p,q]$ are related.    Given $e_1$, the
probability of  there being an $e_2$ in its future is given by the volume of the region $J^+(e_1) \cap J^-(p)$
in units of the discreteness scale, while the probability to pick $e_1$  is
given by the volume of $\Alex_d[p,q]$. This joint probability can be calculated as follows.

Without loss of generality, choose  $p=(-T/2,0,\ldots, 0)$ and
$q=(T/2, 0, \ldots, 0)$, so that the total volume
\begin{equation}
  V = \zeta_d T^d,  \quad  \zeta_d \equiv \frac{\vs_{d-2}}{2^{d-1} d(d-1)}
  \label{eq:zetad} 
\end{equation} 
 with  $\vs_{d-2}$  the volume of the unit ${d-2}$ sphere. 
For this choice,  
 \begin{equation} 
   \av{\bR} = \rc^2 \int\limits_{\Alex_d}\,\,\,  dx_1 \negs \negs\int\limits_{J^+(x_1) \cap J^-(q)}\negs\negs dx_2  =\,
   \rc^2 \, \,  \zeta_d \,  \int\limits_{\Alex_d}\,\,dx_1 T_1^d \,,
\end{equation} 
where $T_1$ is the proper time from $x_1$ to $q$, and $\Alex_d \equiv\Alex_d[p,q]$. Evaluating the integral, one finds 
\begin{equation} 
\label{ctwoflat} 
 \av{\bR} = \rc^2 V^2 \frac{\Gamma(d+1) \Gamma(\frac{d}{2})}{4 \Gamma(\frac{3d}{2})}. 
\end{equation}
Using $\av{\bbn} = \rc V$,  \cite{meyer} obtained a dimension estimator from $\av{\bR}$ by noting that the ratio
\begin{equation} 
\label{eq:dimension} 
\frac{\av{\bR}}{\av{\bbn}^2} = \frac{\Gamma(d+1) \Gamma(\frac{d}{2})}{4 \Gamma(\frac{3d}{2})} \equiv \ff(d) 
\end{equation} 
is a function only of $d$. In  the large $n$ limit 
this is is half of Myrheim's ordering fraction $r$.

However, the fluctuations in $\av{\bR}$ are
large and hence the right dimension cannot be obtained from  a single realisation $C \in \cC(\Alex_d,\rc)$, but rather
by  averaging over the ensemble.  
For large enough $\rc$, however, the relative
fluctuations should become smaller, and allow one to distinguish causal sets obtained from sprinkling into different
dimensional Alexandrov intervals. Such systematic tests have been carried out numerically using sprinklings into
different spacetimes by \cite{reid} and show a general convergence as $\rc$ is taken to be large, or
equivalently the interval size is taken to be large.

 How can we use this dimension estimator in practice?  Let $C$ be a causal set of sufficiently large cardinality
 $n$. If the dimension obtained from Eq.~(\ref{eq:dimension})  is approximately  an integer $d$, this means that $C$ cannot be distinguished
 from a causal set that  belongs to $\cC(\Alex_d,\rc)$ using {just} the dimension estimator, for $n \sim \rc \vol(\Alex_d)$. We denote this by   $C \sim_d
\Alex_d$. This also means that $C$ cannot be a typical member of $\cC(\Alex_{d'},1)$ for dimension $d'\neq
d$, so that $C \not\sim_{d'} \Alex_{d'}$. The equivalence $C\sim_d \Alex_d$ itself does not of course imply that   $C \sim \Alex_d$ or  even  that $C$ is manifold-like. Rather, it is the limited
statement that its dimension estimator is the same as that of a typical causal set in $ \cC(\Alex_d,\rc)$ for  $n \sim
\rc \vol(\Alex_d)$.

This is our first example of a $\cO$-Hauptvermutung, where the order invariant $\cO$ is
the  ordering fraction $r$ and the spacetime  dimension $d$  is the corresponding manifold invariant $\cG$. This example
provides a useful  template in the search for manifold-like order
  invariants some of which we will describe in the next few subsections.

Using simulations \cite{carlipdrone} recently obtained the Myrheim--Meyer
dimension  as function of interval size for nested intervals in a causal set in $\cC(\Alex_d,\rc)$ for
$d=3,4,5$. As the interval size decreases, they found that the resulting causal sets are likely to
be disconnected due to the large fluctuations at  small volumes. In the extreme case, there
is a single point with
no relations and hence the Myrheim--Meyer dimension goes to $\infty$ rather than $0$. Using a criterion to discard such
disconnected regions, it was shown that this dimension estimator gives a value of $2$ at small volumes, 
even when $d=3,4,5$, in support of the dimensional reduction conjecture in quantum gravity \citep{carlipdr} which we 
discuss briefly in Sect.~\ref{sec:matter}. 

\begin{figure}[ht]
\centering \resizebox{3in}{!}  {\includegraphics[width=\textwidth]{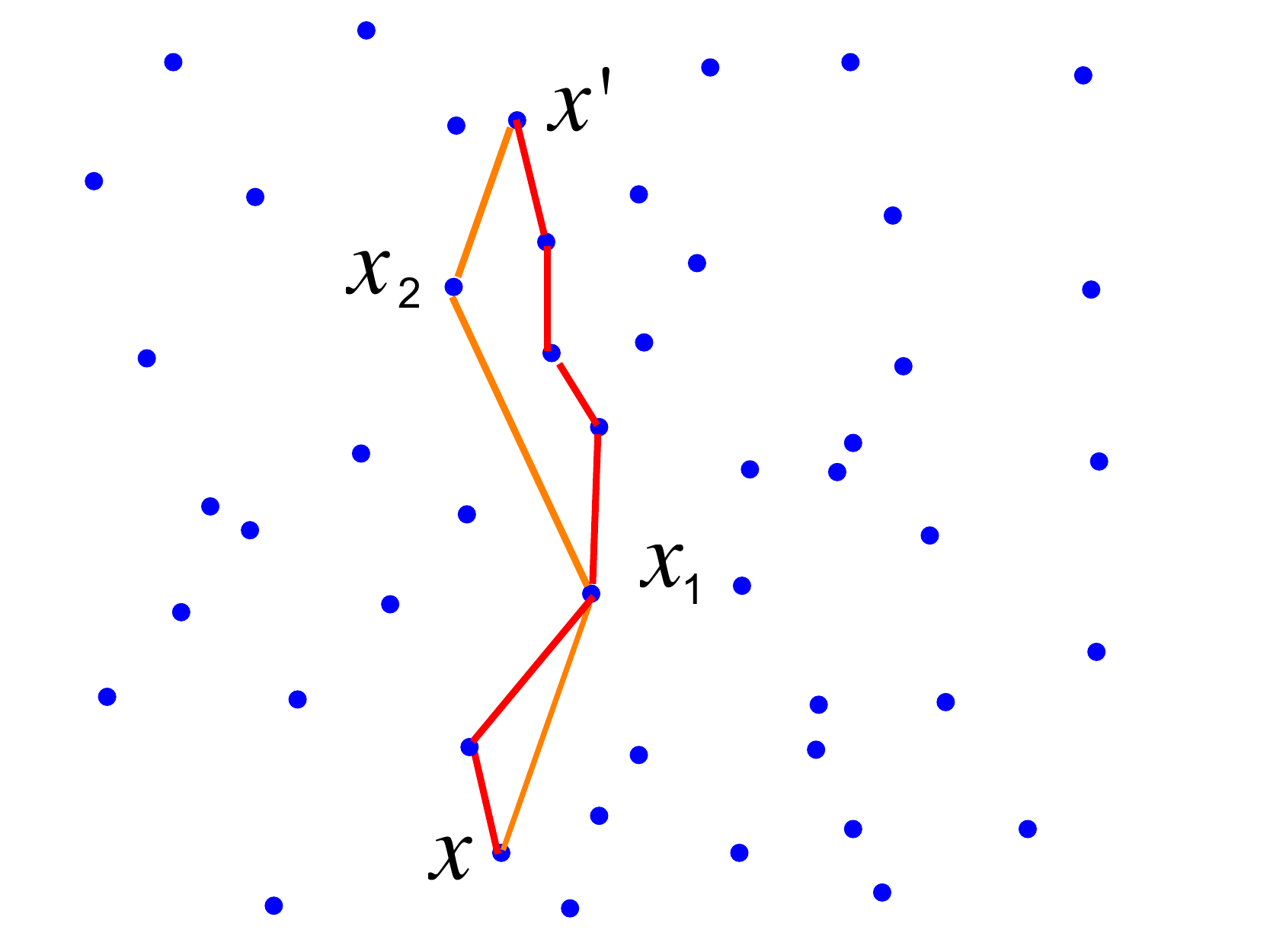}}
\caption{Two different chains between $x$ and $x'$. One is a $k=4$ chain and the other is a $k=7$ chain.}
\label{chain.fig}
\end{figure}

Meyer's construction is in fact
more general and yields a whole family of dimension estimators.  If we think of the relation  $e_1\prec e_2$ as a \emph{chain} $c_2$ of two elements, then a \emph{$k$-chain} $c_k$ is the causal sequence $e_1 \prec e_2 \ldots \prec e_{k-1} \prec
e_k$ (see Fig.~\ref{chain.fig}), where the \emph{length} of $c_k$ is defined as $k-2$.
We denote the \emph{abundance}, or number  of the $c_k$'s contained  in $C$, by $C_k$. Its expectation value in
$\cC(\Alex_d[p,q],\rc)$ is therefore  given by a sequence of $k$ nested integrals over a sequence of nested Alexandrov
intervals,  $\Alex_d[p,q] \supset  I(x_1,q) \supset I(x_2,q) \ldots I(x_k, q) $ which, as was shown by \cite{meyer},  can be
calculated inductively to give 
\begin{equation} \label{eq:ckflat}  
\av{\bCk} =\rc^k\chi_k V^k, \qquad 
\chi_k\equiv \frac{1}{k} \biggl(\frac{\Gamma(d+1)}{2}\biggr)^{k-1}
\frac{\Gamma(\frac{d}{2})\Gamma(d)}{\Gamma(\frac{kd}{2})\Gamma(\frac{(k+1)d}{2})}.
\end{equation} 
Thus for any $k,k'$, the ratio of $\av{\bC_k}^{1/k}$ to ${\av{\bC_{k'}}^{1/k'}}$ only depends on the dimension. This
gives a multitude of dimension estimators.

Meyer's calculation of $\av{\bC_k}$ was   generalised to a small causal diamond $\Alex_d[p,q]$ that lies in an  RNN of a general 
spacetime, i.e., one for which $RT^2 << 1$, where  $T$ is the proper time from $p$ to $q$ and $R$ denotes components of
the curvature  at the centre of the
diamond  \citep{rss}. In such a region the dimension  satisfies the more complicated equation 
\begin{eqnarray} 
\ff^2(d) \biggl( -\frac{1}{3} \frac{(d+2)}{(3d+2)} -
\frac{(4d+2)}{(2d+2)}\biggl(\frac{\av{\bC_3}}{\chi_3}\biggr) ^{\frac{4}{3}} \frac{1}{\av{\bC_1}^4}  &&  \nn \\ 
+ \frac{1}{3} \frac{(4d+2)(5d+2)}{(2d+2)(3d+2)} \frac{\av{\bC_4}}{\chi_4} \frac{1}{\av{\bC_1}^4} \biggr)
&=&- \frac{\av{\bC_2}^2}{\av{\bC_1}^4} \,,
\end{eqnarray} 
where $\ff(d)$ is given by Eq.~(\ref{eq:dimension}). It is straightforward to show that the expression above reduces to the Myrheim--Meyer dimension
estimator in $\mink^d$.  The calculation of \cite{rss} uses a  result of \cite{ks}, which makes explicit
earlier calculations of the volume of a causal diamond in an RNN \citep{myrheim,gs}. 
The $C_k$ themselves
are order invariants and hence are covariant observables for finite element causal sets.

This class of dimension estimators is just one among several that have appeared in the literature, including the {mid-point scaling estimator} \citep{bomthesis,reid}, and more
recent ones \citep{intervals,bomemad}. We refer the reader to the literature for more details. 

\subsection{Topological invariants}
\label{ssec:topology} 
 
The next step in our reconstruction is that of topology.  There are several poset topologies described in the
literature (see \citealt{stanley} as well as  \citealt{suryatop} for a review). However, our interest is in finding one that most closely resembles the ``coarse'' continuum topology. It is clear that the full manifold topology cannot be reproduced in a causal set since it
requires arbitrarily small open sets. However, according to the Hauptvermutung,  topological invariants like the homology groups and the fundamental
groups of $(M,g)$ should be encoded in the causal set. 

 A natural  choice for a topology in $C$ based on the order relation  is one generated by the order
intervals  $\cAlex[e_i,e_j] \equiv \fut(e_i) \cap \past(e_j)$. Indeed, in the continuum the topology generated by their
analogs, the Alexandrov intervals,  can be shown to be  equivalent to the manifold topology  in strongly causal
spacetimes \citep{penrose}.   However, even for a causal set approximated by
 a finite region of $\mink^d$,  this order-interval topology is roughly discrete or
 trivial.  This is because the intersection of any two intervals in the continuum can be  of order the discreteness scale
and hence  contain  just  a  single element of the causal set, thus trivialising the topology.  
A way forward is to use the causal structure to obtain a locally finite open covering of $C$ and construct the
  associated  ``nerve simplicial complex'' (see \citealt{munkres}). 

In \cite{homology,stablehomology},  a ``spatial'' homology of $C$ was obtained in this manner by considering an inextendible 
 antichain $\cA \subset C$ (see Sect.~\ref{ssec:forks}),    which is an (imperfect) analog of a Cauchy hypersurface.
 The natural topology on $\cA$ is the discrete topology since
there are no causal relations amongst the elements. In order to
provide a topology on $\cA$, one needs to ``borrow'' information from a neighbourhood of
$\cA$. The method devised was to consider
elements to the future of $\cA$ and ``thicken'' by a parameter $v$  to some collar
neighbourhood  $\cT_v(\cA) \equiv\{e | \, |\ifut(\cA)\cap \ipast(e)| \leq v \}$. Here $\ifut$ and $\ipast$ denote the \emph{inclusive future and
  past} respectively, where for any $S \subset C$, $\ifut(S)=\fut(S) \cup S$ and  $\ipast(S)=\past(S) \cup S$.

A topology can then be induced on $\cA$ from $\cT_v(\cA)$ by considering the open cover $\{\cO_v \equiv\past(e) \cap
\cA\}$ of $\cA$, for  $e \in {\cM}_v(\cA)$, the  set of future most elements of  $\cT_v(\cA)$.  The 
``nerve''  simplicial complex $\cN_v(\cA)$ can be constructed 
from $\{\cO_v\}$ for  every $v$.   For a spacetime $(M,g)$  with compact Cauchy hypersurface $\Sigma$, and for  $C \in \cC(M,\rc)$ 
it was shown in \cite{homology,stablehomology}  that there exists a range of
values of  $v$ such that  $\cN_v(\cA)$ is  homological to $\Sigma$ (up to the discreteness scale)  as long as there is a sufficient
separation  between the discreteness  scale $\ell_c\equiv V_c^{{1}/{d}}$ and $\ell_K$ the scale of extrinsic curvature of
$\Sigma$.  

One might also imagine a similar construction on $C$ using  the nerve simplicial complex  of 
  causal intervals of a given minimal cardinality  $v$ which cover $C$. However, in the continuum the intersection of such intervals may not
  only  be of order the discreteness scale, but also such that they ``straddle'' each other. As an example consider the
  equal volume intervals  $\Alex[p_1,q_1], \Alex[p_2,q_2]$ in $\mink^2$  where $p_1,q_1$ are at $x=0$ in a  frame
  $(x,t)$, with the $x$-coordinate of
$p_2$ being  $<0$ and that of $q_2$ being $>0$.  These two intervals not only intersect, but  straddle each other, i.e.,
the set difference 
$\Alex[p_1,q_1] \backslash \Alex[p_2,q_2]$ is disconnected as is  $\Alex[p_2,q_2]\backslash \Alex[p_1,q_1]$. By choosing $p_2,q_2$
appropriately, the intersection  region can be made very ``thin'', pushing most of the volume of $\Alex[p_2,q_2]$ out of
$\Alex[p_1,q_1]$. Thus, while they intersect in $\mink^2$ these intervals  would not intersect in the corresponding causal set
$C$. This results in a non-trivial cycle in the associated nerve simplicial complex for $C$, which is absent  in the
continuum. Such a  construction can be therefore made to work only in a sufficiently localised region within $C$.

An example of  a localised of subset of $C$ is the region   sandwiched between two inextendible and
non-overlapping antichains $\cA_1$ and $\cA_2$. The resulting homology constructed from the nerve
simplicial complex {of the order intervals of volume $\sim v$} is then is associated with 
a spacetime region rather than just space, and hence includes topology change. While preliminary investigations
  along these lines have been started, there is much that remains to be understood.  Another possibility for
characterising the spatial homology uses  chain complexes but this has only been partially
investigated.  A further open direction is to obtain  the causal set analogues of {other} topological invariants.

\subsection{Geodesic distance: timelike, spacelike and spatial} 
\label{ssec:distance}

In Minkowski spacetime, the proper time between two events is the 
\emph{longest} path between them; the shortest 
path between two time-like separated events  is of course any  zig-zag null path, which has zero length. In a causal set
$C$, if  $e_i \prec e_f$, one can construct different  chains from $e_i$ to $e_f$, of varying lengths. A  natural choice
for the discrete timelike geodesic
distance between $e_i$ and $e_f$ is  the length of the longest chain, which we denote by  $l(e_i,e_j)$, as was suggested
by \cite{myrheim}.  
It was shown  in  \cite{bg} that the expectation value of the associated random variable $\bl$  in the ensemble $\cC(\Alex_d,\rc)$  limits to a dimension dependent constant 
\begin{equation} 
\lim_{\rho_c \rightarrow \infty} \frac{\av{\mathbf{l}(x,x')}} {(\rc  V(x,x'))^{1/d}} =   m_d  
\end{equation} 
where 
\begin{equation} 
1.77\leq\frac{2^{1-\frac{1}{d}}}{\Gamma(1+\frac{1}{d})}\leq
m_d\leq \frac{2^{1-\frac{1}{d}}e\,(\Gamma(1+d))^{\frac{1}{d}}}{d}\leq
         2.62 
\end{equation} 
For a finite  $\rc$, the fluctuations in $l(e_i,e_j)$ are  very  large \citep{meyer,bachmat} and hence the correspondence
becomes meaningful only when averaged over a large ensemble. 

In \cite{rss}, an expression for the proper time $T$ of 
a small causal diamond $\Alex_d$ in an RNN of a  $d$ dimensional spacetime was obtained to leading order correction in terms of the random
variables $\bC_k$ associated to the abundance of $k$-chains,    
\begin{equation} 
\label{T}
T^{3d}=\frac{1}{2d^2 \rc^3}\biggl(J_1-2J_2+J_3 \biggr).   
\end{equation} 
where 
\begin{equation} 
\label{moreidentities}
J_k \equiv  (kd +2)  ((k+1)d+2) \frac{1}{\zeta_d^3}\biggl(\frac{\av{\bC_k}}{\chi_k }\biggr)^{3/k},  
\end{equation}
with $\av{\bC_k}$  the ensemble average in  $\cC(\Alex_d,\rc)$ and $\zeta_d, \chi_k$ defined in Eqs.~(\ref{eq:zetad}) and
(\ref{eq:ckflat}). This definition is not intrinsic to a single causal set
but requires the full ensemble. Nevertheless, it is of interest to study the intrinsic version of the expression by
replacing  $\av{\bC_k}$ by   $C_k$ for each causal set and then taking the  ensemble average to check for
convergence.   Recent simulations suggest that these expressions converge fairly rapidly to their continuum values. 

Spacelike distance is far less straightforward to compute from the
poset, because events that are spacelike to each other have no natural
relationship to each other. We saw this already in trying to find a topology on the inextendible
antichain. Thus, the relationship must be ``borrowed'' from the
elements in the causal past and future of the spacelike events.  \cite{bg} defined the following,  naive spatial distance function in
$\mink^d$. For a given spacelike pair $p,q \in \mink^d$, the 
{common future and past} are defined as 
$ J^+(p,q)\equiv J^+(p) \cap J^+(q)$ and  $ J^-(p,q)\equiv J^-(p) \cap J^-(q)$ respectively. For  every $r \in  J^+(p,q) $ and
$ s  \in J^-(p,q) $ let $\tau(s,r)$ be the timelike distance. Then the  naive distance function is given by 
\begin{equation} 
\label{eq:sdist} 
\dns(p,q) \equiv min_ {r,s} \tau(r,s). 
\end{equation} 
While this is a perfectly good continuum definition of the distance in $\mink^d$, it fails for the causal set when $d>2$
since the number of pairs $(r,s)$ which minimise $\tau(r,s)$ lies  in the region between a co-dimension $2$ hyperboloid
and the light cone $\tau=0$.  In the causal set we
can use the length of the maximal chain $\bl(r,s)$ to obtain $\tau(r,s)$, but in $d>2$ since there are an infinite
number of proper time minimising pairs $(r,s)$, there will almost surely be those for which $\bl(r,s)$ is drastically
underestimated. The minimisation in Eq.~(\ref{eq:sdist}) will then always give $2$ as the spatial distance!

\cite{rw}  generalised  the naive distance function using minimising pairs $(r,s)$ such that either $r$ or $s$ is
linked to both $p$ and $q$.  Instead of minimising over these pairs (again infinite), the \emph{2-link} distance can be 
calculated by averaging over the pairs. Numerical simulations for the naive distance and the 2-link distance for sprinklings
into a finite region of $\mink^3$ show that the latter stabilises as a function of $\rc$. The former underestimates the
spatial distance compared to the continuum, and the latter overestimates it. The spatial distance functions of both \cite{bg}
and \cite{rw} are however strictly
``predistance'' functions since they do not satisfy the triangle inequality.

Recently a one-parameter family of discrete  induced spatial distance functions was proposed for an inextendible antichain
in a causal set by \cite{esv}.   To begin with, a  one parameter family of continuum  induced distance functions
$d_\epsilon$ was constructed for a globally hyperbolic region $(M,g) $  of spacetime with Cauchy hypersurface
$\Sigma$ using  only the causal structure and the volume element. In  $\mink^d$ with $\Sigma$ a constant time slice in
an inertial reference frame, 
the volume of a past causal cone from $p \succ \Sigma$ has a simple relation to the diameter $D$ of the base of the
cone $J^-(p)\cap \Sigma$
\begin{equation}
  \label{zetadef} 
\vol(J^-(p)\cap J^+(\Sigma))=\zeta_d \biggl(\frac{D}{2}\biggr)^d. 
\end{equation}
Since $D$ is the distance between any two antipodal points on the $S^{d-2}\subset \Sigma$,  this simple
formula defines the induced distance on $\Sigma$. In a general spacetime this
formula can be used to extract an approximate induced distance function in a sufficiently
small region of $\Sigma$.  In order to define the distance function on all of $\Sigma$, a meso-scale $\epsilon$
must be introduced, and the full distance function  can then be obtained by minimising over all segmented paths, such that each
segment is bounded from above by $\epsilon$.   For  $\epsilon
<<  \ell_K$, the scale of the extrinsic curvature of $\Sigma$,   $d_\epsilon$  was shown in \cite{esv} to converge to 
the induced spatial distance function $d_h$ on  $(\Sigma,h)$.

Since the $d_\epsilon$ are constructed from the causal
structure and volume element they are readily defined on  an inextendible antichain on a causal set.  For causal sets
in $\cC(M,\rc)$ with $\Sigma \subset M$ the discrete distance function $d_\epsilon$ was shown to significantly overestimate the
continuum induced distance on $\Sigma$ when the latter is close to the  
discreteness scale $(V_c)^{{1}/{d}}$. This  discrete \emph{``asymptotic silence''} of \cite{ems} mimics the narrowing of
light cones in the UV, and can be traced to the large fluctuations expected around the discreteness scale.  At larger
distances, on the other hand,  $d_\epsilon$ is a good approximation of the continuum induced
distance when $(V_c)^{{1}/{d}}<< \epsilon << \ell_K$.  It was
shown moreover that the  continuum induced distance is slightly underestimated for positive curvature and slightly overestimated  for negative curvature, when restricted
to small regions of $\Sigma$. This was confirmed by extensive  numerical
simulations in $\mink^d$ for $d=2,3$ (see Fig.~\ref{distance.fig}).  This works paves the way to
recovering more spatial geometric invariants from the causal set, and is currently in progress (Eichhorn, Surya and Versteegen).  
\begin{figure}[ht]
  \includegraphics[width=0.5\linewidth, keepaspectratio]{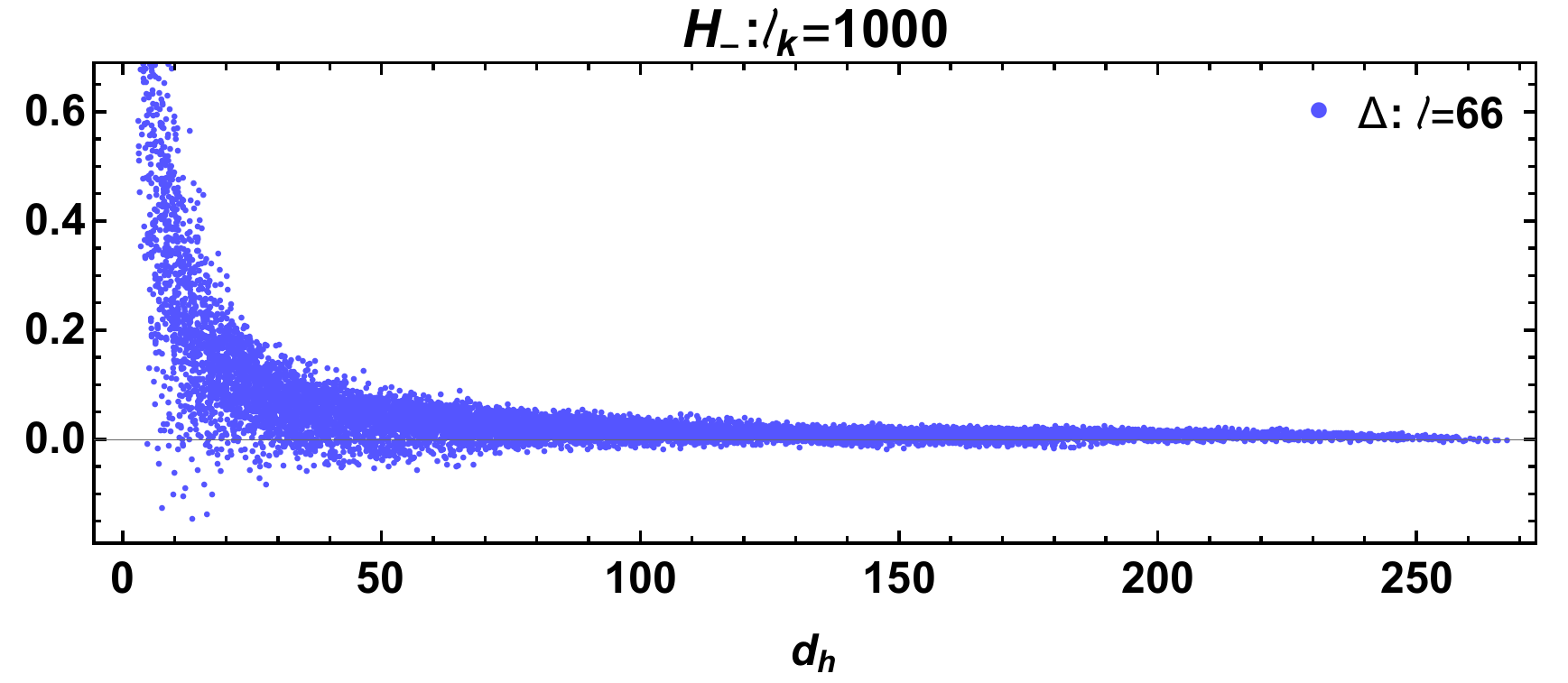}\quad\quad \includegraphics[width=0.5\linewidth, keepaspectratio]{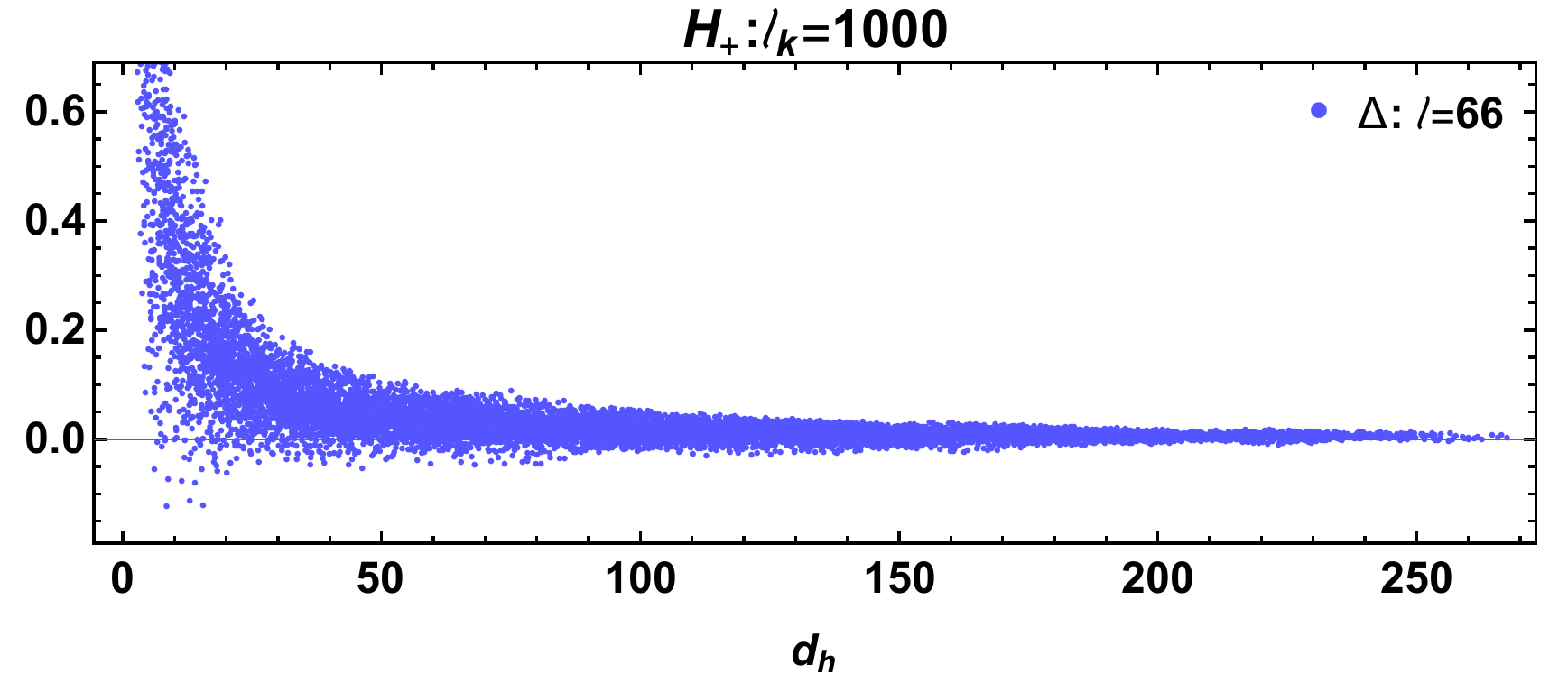}
\caption{The error in the discrete spatial distance is plotted as a function of the continuum induced distance
    on $\Sigma$ for causal
    sets in $\mink^2$ for  $\Sigma$ of both constant negative and constant positive  extrinsic curvature. The discrete spatial distance always overestimates the continuum distance around the discreteness
    scale giving rise to ``discrete asymptotic silence''. For larger distances, when there is a good separation of
    scales, the discrete distance gives a good approximation to the continuum induced distance.}
\label{distance.fig}
\end{figure}

\subsection{The d'Alembertian  for a scalar field } 
\label{ssec:dalem}

One of the very first questions that comes to mind in the continuum approximation of CST is whether a tangent space  can
be defined naturally on a causal set. To answer this (unfortunately in the negative), we need to examine  the non-local nature of a
manifold-like causal set in more detail. 
The nearest neighbours of an element $e$ are
those that it is linked to, both in its future and its past. In a causal set approximated by Minkowski spacetime for
example, and  as discussed in Sect.~\ref{sec:cst}, every element has an infinite number of nearest neighbours (see Fig.~\ref{valency.fig}).  Similarly, the 
``next nearest'' neighbours  to $e$ are those for which the interval $|\cAlex[e,e']| =1$ or $| \cAlex[e',e]|=1$.\footnote{Note that this is the \emph{exclusive} interval and hence there exists exactly
  one element $e''$ such that $e\prec e'' \prec e'$.} 
Thus, in keeping with the covariance of the causal set, we say
that if $e\prec e' $ and $|\cAlex[e,e']|=k$ (or $e'\prec e$ and $|\cAlex[e',e]|=k$),  then    
$e'$ is the $k$-nearest neighbour of
$e$.  Examples of past $k$-nearest
neighbours of an element in a Minkowski-like causal set are shown in Fig.~\ref{nn.fig}.
\begin{figure}[!htb]
  \centering \resizebox{3.0in}{!}  {\includegraphics[width=\textwidth]{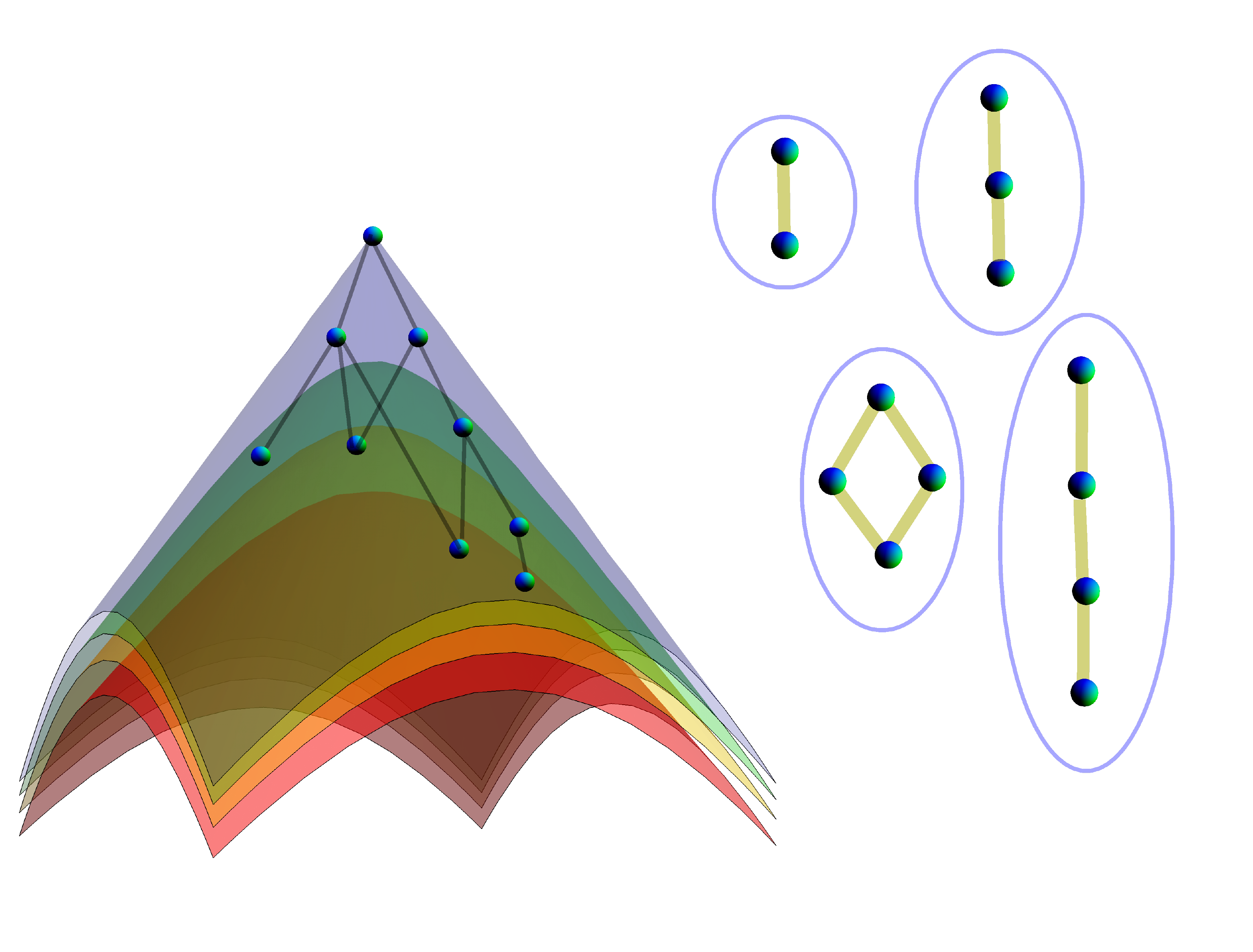}}
\caption{The layered structure of neighbourhoods. The nearest neighbours are the links or zero intervals, the
    next to nearest neighbours are the 1-element intervals, the etc. Here we depict the types of $0,1,2$ element
    intervals. In the figure two examples of  $3$ element intervals are also shown.}
\label{nn.fig}
\end{figure}

It is already clear from the picture that emerges in $\mink^d$ that,  unlike a regular lattice,  a simple construction of  a locally defined tangent space
from the set of links or next to nearest neighbours to $e$ is not possible, since the valency of
the graph is infinite. This means in particular that derivative
operators cannot also be simply defined. 
How then can we look for the effect of discreteness on  the propagation of fields?  We will discuss this in more detail in Sect.~\ref{sec:matter} but for now we
notice that the best way forward is to look for scalar quantities, rather than more general tensorial ones,  in making the
discrete-continuum correspondence.  

A scalar field d'Alembertian is a good first step. In \cite{sorkinnonlocal,joereviewtwo}, a proposal was given for a discrete
d'Alembertian of a free
scalar field on a causal set approximated by $\mink^2$, and extended in \cite{bd,dionthesis,dg} to higher
dimensions. For a real scalar field on  a causal set $\phi: C \rightarrow \re$ define the  $d=4$ dimensionless
discrete operator
\begin{equation} 
B\phi(e) \equiv \frac{4}{\sqrt{6}}\biggl[-\phi(e)  + \biggl( \sum_{e'\in L_0(e)}  - \, \,  9\!\!\!\!\sum_{e' \in L_1(e)}  + \, \, 16
\!\!\!\!\sum_{e'\in L_2(e)}   - \, \, 8 \!\!\!\!\sum_{e' \in L_3(e)} \biggr) \phi(e')\biggr],
\label{eq:disbk} 
\end{equation} 
where $L_k(e)$ denotes the set of $k$-nearest neighbours to the past  of $e \in C$. This is a highly non-local
operator since it depends on the number of all the (possibly infinite) nearest $k=0,1,2,3$ neighbours. Notice  
the alternating sum whose precise coefficients turn out to be very important to the continuum limit.  The expectation value of the  random variable $\bB\phi(x)$  associated with $\cC(\mink^4,\rc)$ 
at  $x\in \mink^4$ is  
\begin{eqnarray} 
  \!\!\!\!\!\frac{1}{\sqrt{\rc }}\av{\bB\phi(x)}\! &=& \!  \frac{4 \sqrt{\rc}}{\sqrt{6}}\biggl[-\phi(x)  + \nn \\
  && \!\! \rc \!\! \int_{y \in J^-(x)}\!\!\!\!\! \!\!\!\!\! \!\!\!\!\! d^4 y \, \phi(y) \, e^{-\rc
                                         v} \biggl(1-9\rc v + 8 (\rc v)^2 -\frac{4}{3} (\rc v)^3\biggr) \biggr],   
\end{eqnarray} 
where  $v\equiv \vol(\Alex(y,x))$ and we have used the probability $P_n(v) $ for $v$ to contain $n$ elements,
Eq.~(\ref{eq:poisson}). We have also made the expression dimensionful, in order to be able to make a direct comparison with the continuum. 
 Let us consider
the past of $x$ in $\mink^2$ and choose a frame $\cF_\phi$  such that $\phi(y)$ varies slowly in  the immediate past of  $x$
with respect to $\cF_\phi$. 
As   was
shown in $\mink^2$ by \cite{sorkinnonlocal} and in $\mink^4$ by \cite{bd} (see also \citealt{dionthesis}),
for   $\phi$ of  compact
support  there are miraculous cancellations that make the contributions far down the light cone negligible, thus
making the operator effectively local. 

In order to evaluate this integral, we first note that since $\phi$ is of compact support, the region of integration
is compact.  In $\cF_\phi$, a
small $|y-x|$ expansion of  $\phi(y)$ around $\phi(x)$ can be done.  Following \cite{sorkinnonlocal,bd,dionthesis},  the non-compact region
of integration $J^-(x)$ can be  split into 3 non-overlapping regions,
$\cW_1,\cW_2,\cW_3$ in $\cF_\phi$:  $\cW_1$ is  a neighbourhood of $x$,  $\cW_2$ is a neighbourhood of $\partial J^-(x)$ but
bounded away from the origin and $\cW_3$ is bounded away from $\partial J^-(x)$.  The integral over  $\cW_3$ was shown
in \cite{dionthesis} to be bounded from above by an integral that  tends to zero faster than any power of $\rc^{-1}$, while
the integral over $\cW_2$ was shown to go to zero faster than $\rc^{-3/2}$.  The local contribution from $\cW_1$  
dominates so that 
\begin{equation} 
\lim_{\rc \rightarrow \infty} \frac{1}{\sqrt{\rc }}\av{\bB\phi(x)} = \Box \phi(x). 
\end{equation}  
Thus, $B(\phi)$ is  ``effectively local'' since its dominant contribution comes from $\cW_1$ which is a local
neighbourhood of $x$ defined by the frame $\cF_\phi$.  In \emph{this}  frame, the contribution to $\bB\phi(x)$ is
dominated by the restrictions of $L_k$ to $\Alex(p,q)\cap J^-(x)$. Thus,   while $\bB\phi(x)$ is not determined just by
the value of $\phi$ at $x$, it depends on $\phi$ only in  an appropriately defined compact neighbourhood of $x$,
rather than all of $J^-(x)$.  This ``restoration of locality'' is an important subtlety in CST kinematics.

How does a scalar field on a causal set evolve under this non-local d'Alembertian?  There are indications that while the
evolution in $d=2$ is stable, it is  unstable in $d=4$ as suggested by \cite{gendalem}. Hence it is  desirable to look for
generalisations  of the $B_\kappa$ operator.  An infinite family of  non-local d'Alembertians has been constructed by 
\cite{gendalem} and shown to give the right continuum limit. It is still an open question whether there is a subfamily of
these operators that lead to a  stable evolution.

An interesting direction that has been explored by \cite{yasamanspectral} is to use the spectral information of the
d'Alembertian operator to obtain all the information about the causal set. This was explored for $\Alex_2[p,q] \subset
\mink^2$ and it  was shown that the spectrum of the d'Alembertian (or Feynman propagator) gives the link matrix (see
Eq.~(\ref{eq:linkm}) below), i.e., the matrix of all linked pairs using which the  entire causal set can be reconstructed
via transitivity. Extending these results to higher dimensions is an interesting open question.   

\subsection{The Ricci scalar and the Benincasa--Dowker action} 
\label{ssec:bdaction}

Next we describe a very important development in CST : the construction of the discrete Einstein--Hilbert action or the
\emph{Benincasa--Dowker (BD) action}  for a causal set \citep{bd,dg}. The approach of \cite{bd} was to generalise
  $\bB\phi(x)$ to an RNN in curved spacetime in $d=2,d=4$. Again, the region of integration can be split into three
  parts as was done for  flat spacetime. The
contribution from $\cW_3$ i.e., away from a neighbourhood of $\partial J^-(x)$ can again be shown to be 
bounded from above by an integral that  tends to zero faster than any power of $\rc^{-1}$.   In the limit, the contribution from the
near region $\cW_1 $ contained in an RNN  is such that 
\begin{equation} 
\lim_{\rc \rightarrow \infty} \frac{1}{\sqrt{\rc }}\av{\bB\phi(x)} \lvert_{\cW_1}  = \Box \phi(x) -\frac{1}{2} \cR(x)\phi(x).
\end{equation} 
where $\cR(x)$ is the Ricci scalar \citep{bd,dionthesis}. However, the calculation in region $\cW_2$ which is in the neighbourhood of
$\partial J^-(x)$ but bounded away from the origin,  is non-trivial, and needs a further set of assumptions to show
that it does not contribute in the $\rc \rightarrow \infty$ limit.  A painstaking calculation in \cite{bbd} using Fermi Normal Coordinates shows that
this is indeed the case  in an approximately flat region of a four dimensional spacetime.  Generalising this calculation
to arbitrary spacetimes is
highly non-trivial but is an important open question in CST.

What is of course exciting about this form for the d'Alembertian Eq.~\ref{eq:disbk} is that it can be used to find the discrete Ricci
curvature and hence the action. Assuming that 
\begin{equation} 
\lim_{\rc \rightarrow \infty} \frac{1}{\sqrt{\rc }}\av{\bB\phi(x)}|_{\cW_2}  = 0 
\label{eq:assumptionbd} 
\end{equation} 
holds in all spacetimes, and  putting\footnote{By doing so, we violate the condition that $\phi$ is of
  compact support. However, given that the regions $\cW_3$ and by assumption $\cW_2$  contribute  negligibly, we can
  always ensure this by only requiring constancy of $\phi$ in a neighbourhood of   $\cW_1$.}   $\phi(x)=1$    
\begin{equation} 
\lim_{\rc \rightarrow \infty} \frac{1}{\sqrt{\rc }}\av{\bB\phi(x)} = -\frac{1}{2} \cR(x).
\end{equation} 
Thus we can write the dimensionless discrete Ricci curvature at an element $e \in  C$ \citep{dionthesis} as   
\begin{equation} 
R(e) =  \frac{4}{\sqrt{6}}\biggl[ 1 - N_0(e) + 9N_1(e) -16 N_2(e) +8N_3(e)\biggr], 
\end{equation}  
where $N_k(e)\equiv|L_k(e)|$.  Summing over the $n$ elements of a finite element causal set gives the dimensionless discrete action  
\begin{equation} 
\label{BD} 
S^{(4)} (C)= \sum_{e\in C} R(e)=\frac{4}{\sqrt{6}} \biggl[ n-  N_0 + 9N_1-16N_2+8N_3\biggl],      
\end{equation}  
where $N_k$ is the total number of $k$-element order intervals in $C$.  

\cite{bd} (see also \citealt{dionthesis}) showed  that (under the assumption Eq.~(\ref{eq:assumptionbd})) the random
variable $\bS^{(4)}$ associated with $\cC(M,g)$  gives the Einstein--Hilbert action in the continuum limit  
\begin{equation} 
\lim_{\rc \rightarrow \infty} \hbar \frac{\lc^2}{\lp^2} \av{\bS^{(4)}(C)} = \cS_{EH}(g),   
\label{eq:bdehlimit} 
\end{equation} 
up to (as yet unknown)  boundary  terms.  

Equation~(\ref{eq:bdehlimit})  is exactly true in an approximately flat  region of a four dimensional spacetime  as
shown in  \cite{bbd}.  Proving Eq.~(\ref{eq:assumptionbd})  in general is however non-trivial since there are caustics
in a generic spacetime which complicate the calculation. On the other hand, numerical
simulations suggest that again, up to boundary terms,  the Benincasa--Dowker action $S$ \emph{is} the Einstein--Hilbert
action \citep{dionthesis,willthesis}. We will discuss these boundary terms below. 

Before doing so, we note that crucial to the validity of the causal set action are its fluctuations in a given causal
set. These were shown in \cite{sorkinnonlocal} to be large for the operator $B$ in $\mink^2$ . This can be traced to the
fact that the elements in $L_k(e)$ for $k=0,1,2,3$ are very close to the discreteness scale and hence the d'Alembertian
is susceptible to  large Poisson fluctuations at small volumes. In order to ``shield'' the continuum from these
fluctuations, a new \emph{mesoscale} $\llk > \lc$ and its associated density $\rk$ 
was introduced in \cite{sorkinnonlocal}. 
Thus instead of a single discrete operator $B$, we have a one parameter family of
operators:
\begin{equation} 
B_\kappa\phi(e) \equiv \frac{4}{\sqrt{6}}\biggl[-\phi(e)  + \epsilon \sum_{e'\prec e} f(n(e',e), \epsilon) \phi(e')\biggr],  
\label{eq:bkoperator}
\end{equation} 
where $\epsilon \equiv {\rk}/{\rc}$  is a \emph{non-locality parameter},\footnote{$\epsilon$ is a new free parameter in the theory, whose 
value should  ultimately be decided by the fundamental dynamics.} $n(e,e')=|I(e,e')| $ and 
\begin{equation} 
f(n,\epsilon) = (1-\epsilon)^n\biggl[1-\frac{9\epsilon n}{1-\epsilon} +  \frac{8\epsilon^2 n(n-1)}{(1-\epsilon)^2} -
\frac{4\epsilon^3 n(n-1)(n-2)}{3(1-\epsilon)^3} \biggr]. 
\end{equation}   
This function ``smears out'' the contributions of the $N_k$ into four ``layers'' which appear with alternating
sign, as shown in  Fig.~\ref{fne4d.fig}. Each layer is thus thickened from a single value of $k$ to a range of $k$
values depending on $\epsilon$. When this mesoscale matches the discreteness scale, i.e., $\epsilon=1$, each layer
collapses to a single value of $k$.  
\begin{figure}[ht]
  \centering \resizebox{3in}{!}{\includegraphics[width=\textwidth]{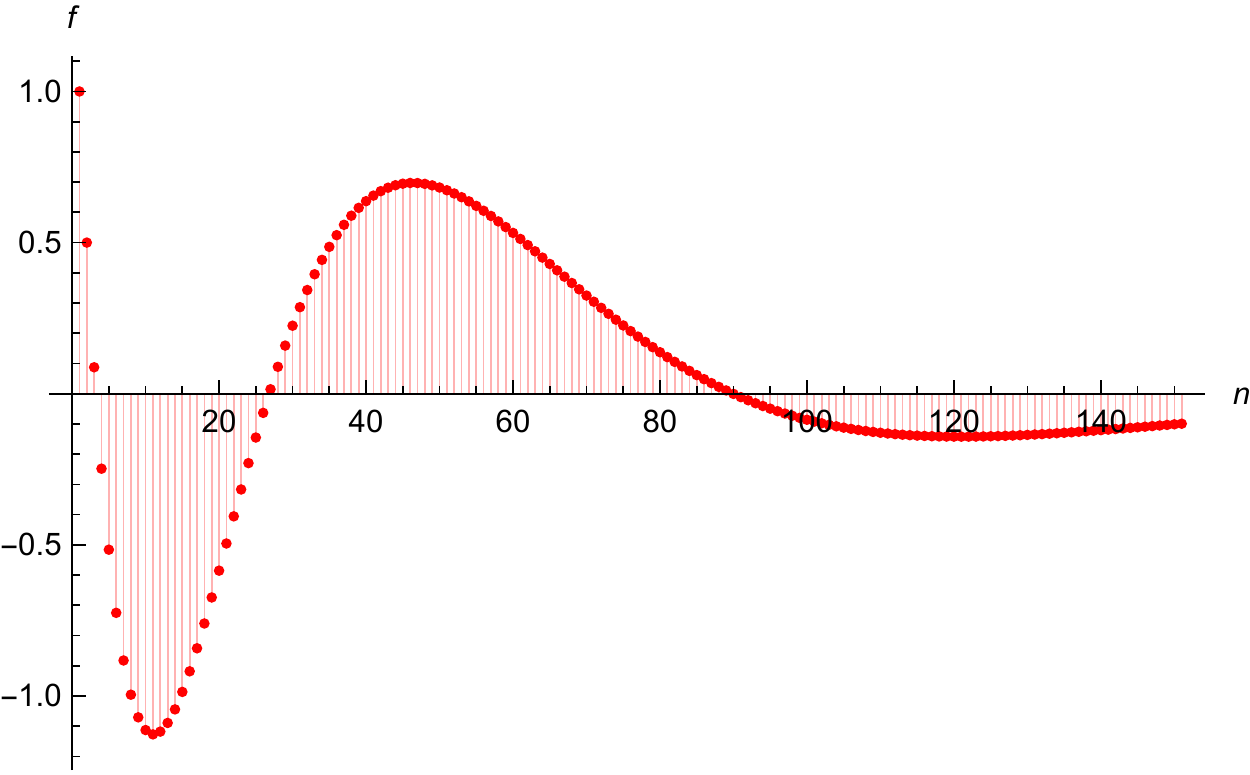}} \hskip 0.1cm 
  \caption{The function $f(n,0.05)$. There are 4 regions of alternating sign corresponding to 4
  ``smeared out'' layers.}
\label{fne4d.fig}
\end{figure}
This then gives us a one-parameter family of actions $S_\kappa(C,\epsilon)$, where  $\epsilon$ can be viewed as a tunable
coupling constant.  As we will see in  Sect.~\ref{sec:dynamics}, this gives rise to an interesting phase structure in 2d CST.

The  result  for $d=2,4$  are due to   \cite{bd,dionthesis} and  were generalised to arbitrary dimensions by \cite{dg,glaser},
using   a dimension dependent smearing function $f_d(n,\epsilon)$.  
 
There have been other attempts to obtain the action of a causal set. In \cite{bomsver}, the curvature at the centre of an
Alexandrov interval $\Alex_d[p,q]$ in a RNN was obtained using the leading order corrections to the volume of a small
causal diamond \citep{gs} 
\begin{equation} 
V=V_0\left(1-\frac{d}{24(d+1)(d+2)} R(0) T^2 + \frac{d}{24(d+1)} R_{00}T^2\right), 
\end{equation} 
where $T$ is the proper time from $p$ to $q$ and $V_0$ is the flat spacetime volume. 
The expression obtained is in terms of the discrete volume and the length of the longest chain from $p$ to $q$.  Since
$R$ is approximately a constant in $\Alex_d[p,q]$,  this also gives the
approximate action.  Extending it to an action on the full spacetime is however quite tricky since it is unclear
how to localise the calculation. 

The calculation for  the abundance of $k$-chains $C_k$ in an RNN  in \cite{rss} also gives an expression for the curvature   
\begin{equation} 
\label{R}
R(0)= 
-{2(n+2)(2n+2)(3n+2)} 2^{\frac{3n+2}{3n}} n^{\frac{4}{3n}-1} \frac{(K_1-2K_2+K_3)}{\,\,\,\,\,(J_1 - 2J_2 +
  J_3)^{\frac{3n+2}{3n}}}.
\end{equation} 
 where 
\begin{eqnarray} 
\label{moreidentitiestwo}
J_k &\equiv& (kn+2) K_k \nno \\ 
K_k &\equiv& ((k+1)n+2)Q_k, 
\end{eqnarray} and 
\begin{equation} 
Q_k  \equiv  \biggl(\frac{\av{\bC_k}}{\rho^k\zeta_k }\biggr)^{3/k}
=\frac{1}{\wzeta^3}\biggl(\frac{\av{\bC_k}}{\rho^k\chi_k }\biggr)^{3/k}. 
\end{equation} 
While this expression is compact, it is not  defined on a single causal set, but rather, over the ensemble.  Whether this
can be expressed as a function on a single  causal set or not is an interesting open  question and under current
investigation. As in the previous case, having obtained $R(0)$, however, it is non-trivial to construct the action,
without a localisation requirement as was done for the BD action.

\subsection{Boundary terms for the causal set action} 
\label{ssec:ghy}

Although the BD action gives the bulk Einstein--Hilbert action in the
continuum approximation, the role of boundary terms is less clear. As shown by 
\cite{gaussbonnet} the expectation value for the BD action does not vanish for  $\cC(\Alex_2,\rc)$, where $\Alex_2[p,q]
\subset \mink^2$ as one might expect,
but instead converges to a constant
as $\rc \rightarrow \infty$ and is independent of $\vol(\Alex_2)$.   
\cite{bdjs} showed more generally that  for $\cC(\Alex_d,\rc)$ with   $d \geq 2$ that 
\begin{equation} 
\lim_{N \rightarrow \infty} \frac{1}{\hbar}\left\langle \tbS_{BDG}^d \right\rangle=  \frac{1}{l_p^{d-2}} \vol
(\mathcal{J}^{(d-2)}) \,,  
\label{result} 
\end{equation} 
where $\mathcal{J}^{(d-2)} \equiv \partial J^+(p) \cap \partial J^-(q)$ is the co-dimension 2 ``joint'' of the causal
diamond $\Alex_d$, which is a round sphere $S^{d-2}$.  In  $d=2$ this is the volume of a  zero sphere $S^0$ which is the
constant found in 
\citep{gaussbonnet}.  This in turn corresponds to the Gibbons--Hawking--York (GHY) null   boundary term of \citep{jsss,lmps} for a
particular choice of the null affine parameter.\footnote{It is an interesting question whether the choice of affine
  parameter along ``almost''  null directions can be obtained from the causal set.} Extending this calculation to curved spacetime  is 
challenging but would provide additional evidence that the BD action contains the null GHY term ({\it Dhingra, Glaser and
S. Surya, work in progress}).      

Simulations of causal sets corresponding to different regions of $\mink^2$  moreover suggest that while the  BD action contains timelike boundary terms, it does not  contain spacelike boundary terms.  Recent  efforts by \cite{tlbdry}  have been made to obtain 
time like boundaries  in a causal set using numerical methods for $d=2$, but it is an open question whether they admit a simple
characterisation in arbitrary dimensions. 
 
Unlike timelike boundaries, spacelike boundaries are naturally defined in a finite element causal set: a future/past spatial
boundary is the future-most/past-most inextendible antichain in the causal set, which we denote as $\cF_0,\cP_0$
respectively.  GHY terms for spacelike boundaries play an important role in the additivity of the action in the
continuum path integral (though  such an additivity is far from guaranteed in a causal set because of non-locality).

The spatial causal set GHY  terms were found by \cite{bdjs}, and we will describe that construction here briefly. Let
$(M,g)$ be a spacetime with initial and final spatial boundaries $(\Sigma^\pm,h^\pm)$ .  The GHY term on $(\Sigma^\pm,h^\pm)$ can be re-expressed as
\begin{equation} 
\int_{\Sigma^\pm} d^{d-1}x\: \sqrt{h^\pm}\, K^\pm= \frac{\partial}{\partial n}\int_{\Sigma^\pm} d^{d-1}x\: \sqrt{h^\pm}
=\frac{\partial}{\partial n} A_{\Sigma^\pm},
\label{eq:GHYterm}
\end{equation} 
where $\frac{\partial}{\partial n}$ is the normal derivative, and $A_{\Sigma^\pm}$ is the co-dimension $1$ volume of $\Sigma^\pm$. 
Using  the $n\sim \rc v$ correspondence, this suggests that $A_{\Sigma^\pm}$ should be  given by the cardinality 
$\F{0}\equiv |\cF_0|$ or $\P{0}\equiv |\cP_0|$ with the normal gradient represented by the  change in the
cardinality.  But of course this is subtle, 
since  apart from the future most $\cF_0$ or pastmost $\cP_0$ antichains, one needs  another ``close by'' antichain.
Let us focus on $(\Sigma^+,h^+)$ without loss of generality. There are two  ways of finding
this nearby antichain. To begin with if $(M,g) \subset (M',g')$ such that $(\Sigma^+,h^+)$ is not
a boundary in $(M',g')$, then we can use this embedding  to define the two antichains, in any $C \in \cC(N,\rc)$: one to
its immediate past $\cF_0(\Sigma^+)$ and one
to its immediate future $\cP_0(\Sigma^+)$. Thus the GHY term should be proportional to the difference in the cardinality of these two
antichains.  

However, this partitioning is not intrinsic to the causal set. Instead consider a partition  $C = C^-\cup C^+$, such that $C^+\cap C^-=\emptyset$, and $\fut(C^-)=C^+$, $\past(C^+)=C^-$. 
Let $\cF_0^-$ and $\cP_0^+$, be the
future-most and past-most antichains of $C^-$ and $C^+$ respectively.  We can then define the  
dimensionless causal set ``boundary term'' \citep{bdjs}
\begin{equation} 
S_{\rm CBT}^d[C,C^-,C^+] \equiv \frac{a_d}{2 \Gamma\left(\frac{2}{d}\right)}\biggl( \F{0}[C^-]-\P{0}[C^+]\biggr),  
\label{eq:cbt} 
\end{equation} 
where
\begin{equation} 
\label{eq:Cn}
a_{d}=\frac{d (d+1)}{(d+2)}\left(\frac{\vs_{d-2}}{d(d-1)}\right)^{\frac{2}{d}},
\end{equation} 
and $\vs_d=(d+1)\pi^{\frac{d+1}{2}}/\Gamma\left (\frac{d+1}{2}+1\right)$ is the volume of the unit $d$-sphere.

To make
contact with the continuum, let $(M,g)$ be a spacetime with compact Cauchy hypersurfaces. For a given Cauchy
hypersurface $(\Sigma,h)$ let $M^\pm=J^\pm(\Sigma)$ and let $C^\pm \in \cC(M^\pm,\rc)$.  It was shown by \cite{bdjs} that in
the limit $\rc \rightarrow \infty$ 
\begin{equation} 
\negs\negs \lim_{\rc\rightarrow \infty}\left (\frac{\lc}{\lp}\right)^{\! \! d-2}\negs \left\langle\textbf{S}^{
    (d)}_{\rm CBT}[M, \Sigma,\rc]\right\rangle= \frac{1}{l_p^{d-2}}\int_{\Sigma} d^{d-1}x\: \sqrt{h}\: K = {S}_{GHY}(\Sigma,
M^-), \label{eq:ghyresult} 
\end{equation}
where $\textbf{S}^{(d)}_{\rm CBT}$ is the associated random variable in $(M,g)$.  To obtain this expression, the volume  of a
half
cone $J^+(p) \cap J^-(\Sigma)$ was calculated using a combination of RNN coordinates 
and GNN coordinates\footnote{This calculation has later been extended by \cite{jubb} to higher orders to
  obtain more information about the spatial geometry.}
\begin{equation} 
V_\blacktriangle (T,\mathbf x)
=\frac{S_{d-2}}{d(d-1)}T^d\left (1+\frac{d}{2 (d+1)}K (0,\mathbf{x})T\right)
+O (T^{d+2}),   \label{eq:TopVolumeWithK} 
\end{equation}
for $p \in J^+(\Sigma)$  sufficiently close to $\Sigma$, where $T$ is the proper time from $p$ to $\Sigma$. 
As might be expected from dimensional considerations, the leading order correction to the flat spacetime volume of the
half cone comes  from the trace of the extrinsic curvature of $\Sigma$ from which the GHY contribution can be obtained.

If on the other hand, $(\Sigma,h)$ is a future boundary of $(M,g)$, then we require a second
antichain in $\past(\cF_0)$  for $C \in \cC(M,\rc)$,. Define the antichain $\cF_1$ in $C^-$ to
be the set of elements in $C^-$ such that $\forall e \in \cF_1$, $|\fut(e) \cap C^-|=1$ (where $\fut(e)$ excludes the
element $e$).\footnote{Note that while  $\cF_1\cap \cF_0=\emptyset$,  $\cF_1$ is not necessarily an inextendible
  antichain.}  The boundary  term can then be expressed as 
\begin{equation} 
S_{\rm CBT}^d[C,C^-,C^+] \equiv \frac{a_d}{\Gamma\left(\frac{2}{d}\right)}\biggl( d\F{1}[C^-]-\F{0}[C^+]\biggr),  
\label{eq:cbttwo} 
\end{equation} 
which again yields the GHY term Eq.~(\ref{eq:ghyresult})  in the limit. Indeed, a whole  family of  of boundary terms was
obtained using the antichains $\cF_k[C^-] = \{e \in C^-|  |\fut(e)|=k\}$, $\cP_k[C^+]= \{e \in C^+ |
|\past(e)|=k\}$ each of which gives the GHY term in the limit Eq.~(\ref{eq:ghyresult}).\footnote{The expression in \cite{bdjs} holds for \emph{any} two subsets of $C$ not just those we
  consider here.}

A by-product of the analysis of \cite{bdjs} is that for the partitioned causal set $C=C^-\cup C^+$ described above, the quantities 
\begin{equation} 
A_+^d[C^-]\equiv \frac{b_d}{\Gamma(\frac{1}{d})}\F{0}[C^-], \quad A_-^d[C^+]\equiv \frac{b_d}{\Gamma(\frac{1}{d})}\P{0}[C^+]
\end{equation} 
for $a_d=\frac{d+1}{d(d+2)} b_d^2$ limit to the spatial volume of $\Sigma$
\begin{equation} 
\lim_{\rc \rightarrow \infty} \left(\frac{\lp}{\lc}\right)^{d-2} \av{A_\pm^d[C^\mp]} = \frac{1}{\lp^{d-1}}\int_\Sigma
  d^{d-1} x \sqrt{h} = A_\Sigma. 
\end{equation} 
Again, as for the boundary terms, one can construct a  whole family of functions $A^d[C]$ each of 
which limit to the spatial volume of $\Sigma$ as $\rc \rightarrow \infty$.

\subsection{Localisation in a causal set} 
\label{ssec:localisation} 

In these calculations generalisations are made to curved
spacetime using an RNN which represent a local region of a spacetime.  How are we to find such local regions in a
causal set using a purely order theoretic quantities?  For a causal set  a natural  definition of  a local region
is given by  the size of an interval, but for a manifold-like causal set, this will not necessarily correspond  to regions in
which the curvature is small. On the other hand, many of the order invariants we have obtained so far correspond to
geometric invariants only in such RNN-type regions.

A characterisation of intrinsic localisation was obtained by \cite{intervals}  using the abundance $N_m^d$ of $m$ element order
intervals for $C\in\cC(\Alex_d,\rc)$.  They  found the following closed form expression for the associated
expectation value 
\begin{align}
	\av{\bN_m^d(\rho,V)}=& \frac{(\rho V)^{m+2}}{(m+2)!} \frac{\G{d}^{2}}{\Poch{\frac{d}{2} (m + 1) + 1}{d - 1}} \frac{1}{\Poch{\frac{d}{2} m + 1}{d - 1}} \nonumber \\
	& \mFm{d}{1+m,\frac{2 }
	{d}+m,\frac{4}{d}+m, \dots ,\frac{2 (d-1)}{d}+m)}
	{3+m,\frac{2 }{d}+m+2,\frac{4}{d}+m+2, \dots ,\frac{2 (d-1)}{d}+m+2}
	{- \rho V} \;,
\label{eq:flatclosedform} 
\end{align} 
The distribution of $\av{\bN_m^d}$ with $m$ therefore has a characteristic form which depends on dimension, and as a by-product,
can be used as a  dimension estimator. However, it can also be used look for intervals in a manifold-like causal set
which are approximately flat by comparing the interval abundances $N_m^d$ to the above expression for $\av{\bN_m^d}$.  While one might expect  the fluctuations for a given causal
set $C$ to be  large,  numerical simulations show that there is typically a ``self averaging'' which results in 
 relatively small fluctuations even for a given realisation. This makes it an ideal
diagnostic tool for checking whether a neighbourhood in a manifold-like causal set is approximately flat or not. Once such local
neighbourhoods have  been found, a local check of geometric estimators can be made.

In \cite{intervals}, the analytic curves
were compared  against simulations for a range of different causal sets including those that are not manifold-like . While
curvature affects the abundance of the intervals, the distribution retains its characteristic form. Hence
the dependence of the abundance of intervals with size  also becomes a test for manifold-likeness.
\begin{figure}[ht]
  \centering \resizebox{4in}{!} {\includegraphics[width=\textwidth]{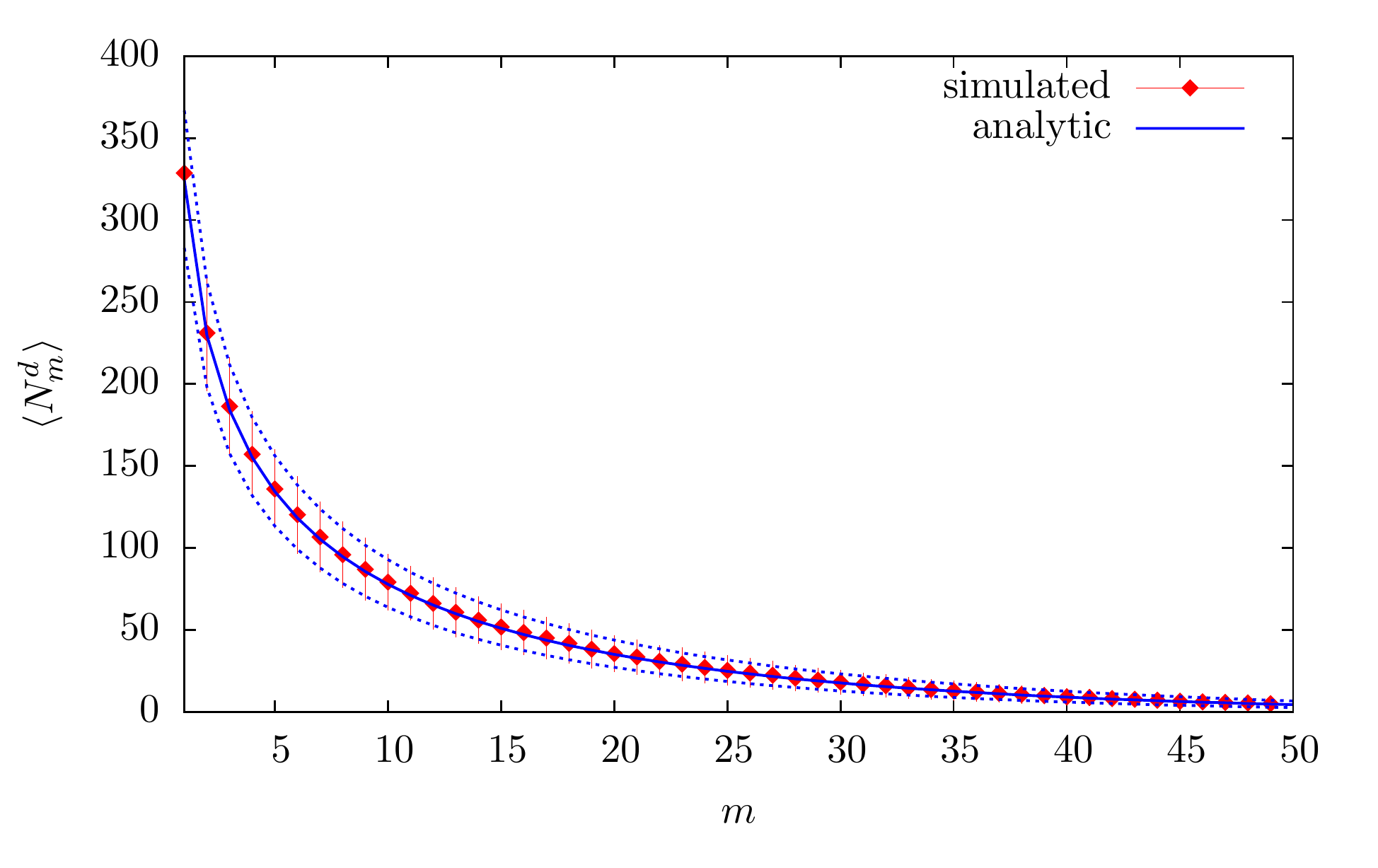}} 
  \caption{{The expectation value of interval abundances in a 100 element causal set $\sim \mink^2$ as a function of
      interval size $m$. The red dots depict the average value obtained from simulations with  1000 realisations,  along with error
  bars. The solid blue line depicts the analytic expectation value for $n=100$ and the blue dotted lines for $n\pm
  \sqrt{n}$. } 
  }
\label{intabund.fig}
\end{figure}

There are other ways of testing for manifold-likeness.  In a similar approach, the distribution of the longest  chains or linked paths of length $k$ in a finite element causal set $C$ has been
studied in $\mink^d$, $d=2,3,4$ and shown to have a dimension-dependent peak \citep{bomemad}. 
 In \cite{nodedegree},  a novel way to test for manifold-likeness was given, using the order invariant obtained from counting
 the number of elements with a fixed valency  in a finite
element causal set. In   \cite{joeinterval},  an  algorithm for  determining  the  embeddability of a causal set in
$\mink^2$ was given,  which
again gives an intrinsic characterisation of manifold-likeness in $d=2$. Extending and expanding on these studies using
causal sets obtained from sprinklings into different types of spacetimes would be a straightforward but useful exercise.      

\subsection{Kinematical entropy} 
\label{ssec:kinent}

Since the classical continuum geometry itself is fundamentally statistical in CST, it is interesting to ask if a
kinematic 
entropy can be assigned even classically to the continuum. In \cite{dousorkin},  a kinematic entropy was associated
with a horizon $H$ and a  spatial or null hypersurface $\Sigma$ in a  dimensionally reduced $d=2$ black hole spacetime by counting links between 
elements in $J^-(\Sigma) \cap J^-(H)$ and those in $J^+(\Sigma) \cap J^+(H)$, with the additional requirement that the
former is future-most and the latter past-most in their respective regions. A dimensionally reduced calculation showed that the number of
links is proportional to the horizon area. Importantly, the  calculation yields the same
constant for a dimensionally reduced dynamical spacetime where  a collapsing shell of null matter eventually forms a
black hole. However, extending this calculation  to higher dimensions proves to be  tricky.  In \cite{sarahthesis}, an entropy
formula was proposed for higher dimensions by replacing links with other sub-causal sets. While these ideas  hold
promise, they  have not as yet been fully explored. 

In analogy with Susskind's entropy bound,  the
maximum causal set entropy  associated with a finite
spherically symmetric spatial hypersurface $\Sigma$  was defined by \cite{rz} as the number of maximal or future most elements
in its future domain of dependence $D^+(\Sigma)$. It was shown  that for several such examples this bound
limits to the Susskind entropy bound in the continuum approximation. Again, extending this discussion to more general
spacetimes is an interesting open question. 

In \cite{dionthesis},  the  mutual information between  different regions of a causal set was defined  using the
BD action. The source of this entropy is non-locality which implies that $S_{BD}$ is not in general additive.  Dividing
a causal set $C$ into two (set-wise) disjoint regions $C_1$ and $C_2$, so that $C=C_1 \sqcup C_2$,  we see that in
general $S_{BD}(C) \neq S_{BD}(C_1) + S_{BD}(C_2)$.  This is because 
there can be order intervals between elements in $C_1$
and in $C_2$ that are not counted by either $S_{BD}(C_1)$ or $S_{BD}(C_2)$. The  mutual information is thus 
defined as
\begin{equation}
  MI[C_,C_2]\equiv S_{BD}(C_1) +S_{BD}(C_2)-S_{BD}(C). 
\end{equation}
In  \citep{dionthesis}  a spacetime region with a horizon $\cH$ and a  spacelike or
null hypersurface $\Sigma$ was considered. Defining $X=J^+(\cH) \cap J^-(\Sigma)$ and $Y=J^-(\cH) \cap J^-(\Sigma)$ the mutual
information between $X$ and $Y$ was calculated from  a causal set obtained from sprinkling  into $X\cup Y$.  Under
certain assumptions, this equal to the area of $\cH\cap \Sigma$. These results are suggestive, but currently incomplete.

As we will see in the next section, the {Sorkin spacetime  entanglement entropy (SSEE)} for a free scalar field provides a
different avenue for exploring entropy.


\subsection{Remarks} 

To conclude this section  we note that several order invariants have been constructed on manifold-like causal sets  whose expectation
values limit to manifold  invariants  as $\rc \rightarrow \infty$. At finite $\rc$ there are  
fluctuations that serve to distinguish the fundamental discreteness of causal sets from the continuum, and these have potential
phenomenological consequences.  Numerical simulations are often important in 
assessing the relative importance of these fluctuations. 

For each of these invariants, one has therefore proved an  \emph{$\cO$-Hauptvermutung}.  While this collection of
order invariants is not sufficient to prove   the full Hauptvermutung, they lend it strong support. These order
invariants are moreover important observables for the full theory.   In addition to these
manifold-like order invariants, there are several other order invariants that can be constructed, 
some of which may be important to the deep quantum regime but by themselves hold no direct continuum interpretation.

\section{Matter on a continuum-like causal set}
\label{sec:matter}

Before passing on to the dynamics of CST, we look at a phenomenologically important question, namely how quantum fields
behave on a fixed manifold-like causal set.   The simplest matter field  is the free scalar field on a
causal set in $\mink^d$.  As we noted in the previous Section, this is the only class of matter fields that
we know how to study, since at present no well defined representation of non-trivial tensorial fields on causal sets is
known. However, as we will see, even this very simple class of matter fields  brings with it both exciting new insights and
interesting conundrums.  

\subsection{Causal set Green functions for a free scalar field } 
\label{ssec:Green} 
Consider the real scalar field $\phi: \mink^d \rightarrow \re$ and its CST counterpart,    $\phi:C \rightarrow
\re$  where $C \in \cC(\mink^d,\rc)$. The Klein Gordon  operator of the continuum is replaced on the causal set by the $B_\kappa$ operator
of Sect.~\ref{sec:kinematics}, Eq.~(\ref{eq:bkoperator}).   In the continuum $\Box^{-1}$ gives the Green function, and we can do
the same with $B_\kappa$ to obtain the discrete Green function $B_\kappa^{-1}$.

However, there are more direct ways of obtaining the Green function
as was shown in \cite{daughton,salgado,johnston,dsx}.  The \emph{causal matrix} 
\begin{equation}
C_0(e,e') \equiv
\left\{
        \begin{array}{ll}
                1  & \mbox{if } e' \prec e \\
                0 & \mbox{} \mathrm{otherwise}
        \end{array}
\right.
\end{equation}
on a causal set $C$.    For $C \in \cC(\mink^d,\rc)$,  $C_0(e,.)$ is therefore zero everywhere except within  the past light cone of $e$ at
which it is $1$.  In $d=2$, this is just half the massless retarded Green's function $G^{(2)}_0(x,x') =\frac{1}{2}
\theta(t-t') \theta(\tau^2(x,x'))$.  Hence, we find the almost trivial relation  
\begin{equation} 
\bC_0(x,x') = 2 G^{(2)}_0(x,x'), 
\end{equation} 
without having to take an expectation value, so that  the  dimensionless  massless causal set retarded Green
function is  \citep{daughton}  
\begin{equation}
\label{massless2d}
K^{(2)}_0(x,x')\equiv  \frac{1}{2}C_0(x,x').
\end{equation}

To obtain the $d=4$ massless causal set Green function we use the \emph{link matrix}
\begin{equation}
L_0(x,x'):=
\left\{
        \begin{array}{ll}
                1  & \mbox{if } x' \prec x {\mathrm \ is \ a \ link}\\
                0 & \mbox{} \mathrm{otherwise}
        \end{array}
      \right. 
      \label{eq:linkm}
    \end{equation}
  For $C \in \cC(\mink^4,\rc)$ the expectation value of the associated random variable is 
\begin{equation}
\label{linkexp}
\av{\bL_0(x,x')}=\theta(x_0-x'_0) \theta( \tau^2(x,x'))\exp(-\rc V(x,x')),
\end{equation}
where $V(x,x') = \vol(J^-(x) \cap J^+(x')) = \frac{\pi}{24}\tau^4(x,x')$. Since the exponential in the above expression
is a Gaussian which, in the $\rc \rightarrow \infty $ limit  is proportional to $\delta(\tau^2)$, we see that it resembles the
massless retarded Green function in $\mink^4$, 
\begin{equation} 
\label{linklim}
\lim_{\rc \rightarrow \infty} {\sqrt{\frac{\rc}{6}}\av{\bL_0(x,x')}} =  \theta(x_0-x_0') \delta(\tau^2) =2 \pi
G^{(4)}_0(x,x'). 
\end{equation} 
Hence we can write the  dimensionless massless causal set scalar retarded Green function as \citep{johnston,johnstonthesis} 
\begin{equation}
\label{eq:massless4d}
K^{(4)}_0(x,x')= \frac{1}{2 \pi} \sqrt{\frac{1}{6}} L_0(x,x')\,. 
\end{equation} 

In the continuum the massive Green function can be
obtained from the massless Green function in $\mink^d$ via the formal expression \citep{dsx} 
\begin{equation} 
\label{eq:conv} 
G_m=G_0 - m^2\,G_0*G_0 + m^4 \,G_0*G_0*G_0+ \ldots = \sum_{k=0}^\infty(-m^2)^k \underbrace{G_0 * G_0* \ldots G_0}_{k+1}
\end{equation}
where 
\begin{equation}
(A\ast B)(x,x')\equiv \int d^dx_1 \sqrt{-g(x_1)} A(x,x_1)
B(x_1,x')\,.
\label{eq:conv}
\end{equation}  
Using this as a template, with the discrete convolution operation given by  matrix multiplication,  
\begin{equation}
(A \ast B)(e,e')\equiv  \sum_{e''} A(e,e'') B(e'',e)\, , 
\end{equation}  
a candidate for the  $d=2$ dimensionless massive causal set Green function is 
\begin{equation}
\label{eq:2dGf} 
K^{(2)}_M(x,x') =  \frac{1}{2} \sum\limits_{k=0}^\infty (-1)^k \, \frac{M^{2k} }{2^k}  C_k(x,x').   
\end{equation} 
Here $M$ is dimensionless and  we have used the relation $C_k(x,x')=C_0^k(x,x')$, where the product is defined by the
convolution operation Eq.~\ref{eq:conv} and, $C_k(x,x')$ counts the number of
$k$-element chains from $x$ to $x'$.  For $C \in \cC(\mink^2, \rc)$  it can  be shown
that \citep{johnston,johnstonthesis}  
\begin{equation} 
\av{\bK^{(2)}_M(x,x')} = G^{(2)}_m(x,x') \label{proof3}\,, 
\end{equation} 
when $M^2=\frac{m^2}{\rc}$.  
Similarly, a candidate for the $d=4$ massive causal set Green function is 
\begin{equation}
\label{eq:4dGF}
K^{(4)}_M(x,x') =  \frac{1}{2\pi\sqrt{6}} \sum\limits_{k=0}^\infty
(-1)^k \, \left(\frac{M^2}{2\pi\sqrt{6}}\right)^{k} L_k(x,x')\, ,
\end{equation} 
where we have used the fact that the number of $k$-element linked paths $L_k(x,x')=L_0^k(x,x')$. For  $C \in \cC(\mink^4,\rc)$, 
\begin{equation} 
\lim_{\rc \rightarrow \infty} \sqrt{\rc} \av{\bK^{(4)}_M(x,x')} = G^{(4)}_m(x,x')\,, 
\end{equation} 
when  $M^2=\frac{m^2}{\sqrt{\rc}}$. 

These  massive causal set Green function were  first obtained by \cite{johnston,johnstonthesis} using an evocative  analogy between Feynman paths
and the $k$-chains or  $k$-linked paths (see Fig.~\ref{hopstop.fig}). ``Amplitudes'' $a$ and $b$ are  assigned to a  ``hop'' between two
elements in the Feynman path, and to a ``stop'' at an 
intervening element, respectively. This gives a total ``amplitude'' $a^{k+1}b^k$ for each
chain or linked path, so that the massive Green functions can be expressed as  
\begin{equation} 
K_m^{(2)}(e,e') \equiv \sum_{k=0} a_2^{k+1} b_2^{k} C_k(e,e'), \quad K_m^{(4)}(e,e') \equiv \sum_{k=0} a_4^{k+1} b_4^{k}
L_k(e,e'), 
\end{equation}   
where the coefficients $a_d,b_d$ are set by comparing with the continuum.
\begin{figure}[ht]
  \centering \resizebox{0.75in}{!} {\includegraphics[width=\textwidth]{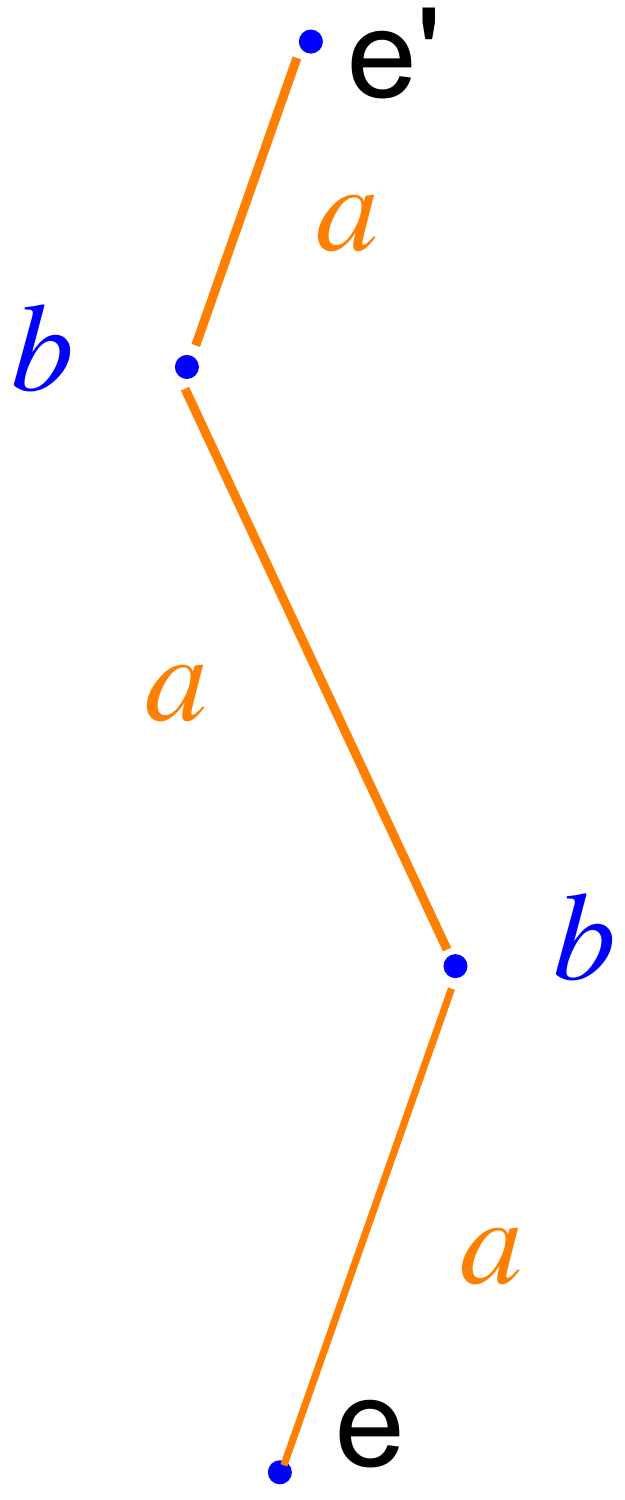}} 
  \caption{The hop and stop amplitudes $a$ and $b$ on a 2-element chain from $e$ to $e'$ for a massive scalar field on a causal set.}
\label{hopstop.fig} 
\end{figure}

Finding causal set Green functions for other spacetimes is more challenging, but there have been some recent results \citep{dsx}
which show that the flat spacetime form  of \cite{johnston, johnstonthesis} can be used in a wider context. These
include (a) a causal diamond in an RNN of a $d=2$ spacetime with $M^2={\rc}^{-1}({m^2+\xi R(0)})$, where $R(0)$  is the
Ricci scalar at the centre of the diamond and $\xi$ is the non-minimal
coupling,  (b)  a causal diamond in an RNN of a $d=4$ spacetime with $R_{ab}(0)
\propto g_{ab}(0)$ and  $M^2={\rc}^{-1}({m^2+\xi R(0)})$  when (c) $d=4$ de Sitter and anti de Sitter spacetimes with $M^2={\rc}^{-1}({m^2+\xi })$. 

The de Sitter causal set Green function in particular  allows us to  explore cosmological consequences  of  discreteness, one of which
we will describe below.  It would be useful to extend this construction to other conformally flat spacetimes of cosmological relevance like the flat FRW
spacetimes.  Candidates for causal set Green functions in $\mink^3$ have also  been obtained using both the volume of the causal interval and the
length of the longest chain \citep{johnstonthesis,dsx}, but the comparisons with the continuum need
further study.

As the attentive reader would have noticed, in $d=4$ the causal set Green function matches the continuum only for $\rc
\rightarrow \infty$, unlike in $d=2$. At finite $\rc$,  there  can be potentially observable
differences with the continuum. Comparisons with observation can therefore  put constraints on CST. \cite{dhstwo}  examined
a model for the propagation of a classical massless scalar field from a source to a
detector on a background causal set. In $\mink^d$, an oscillating point source 
with scalar charge $q$, frequency $\omega$ and amplitude $a$, and a ``head-on'' rectangular shaped detector  
was considered, so that the  field produced by the source is 
\begin{equation} 
\phi(y)=\int_P G(y,x(s)) q ds
\end{equation} 
where $\mathcal P$ is the world line of the source and $s$ the proper time along this world line. If $\mathcal{D}$ represents the
spacetime volume swept out  by the detector during its detection time $T$ then the output of the detector is 
\begin{equation} 
F= \int_{\mathcal{D}} \phi(y) d^4y= q\int_{\mathcal P} ds \int_{\mathcal{D}} d^4y  G(y,x(s)) \approx \sqrt{\frac{1+\nu}{1-\nu}}
\frac{q}{4 \pi R} v_{\mathcal{D}}
\label{eq:detres}
\end{equation}  
where $R$ is the distance between the source and detector,  $\nu$ is the component of the velocity along the
displacement vector between the source and detector and $v_{\mathcal{D}}$ is the spacetime volume of the detector region
$\mathcal D$. Here, $R>>a$ and  $R >> \omega^{-1}$ which in turn is much larger than  the spatial and temporal extent of the detector region
$\mathcal D$. The causal set detector output can then be defined as 
\begin{equation} 
\tF = q\frac{1}{2\pi \sqrt{6}}\sum_{e \in \tilde{\mathcal P} }\sum_{e' \in \tilde{\mathcal D}}L_0(e',e)
\end{equation}   
where $\tilde{\mathcal D} $ and $\tilde{\mathcal P}$ correspond to the detector and source subregions in the causal set
and the causal set function $L(e,e')$ is equal to some normalisation constant $\kappa$ when $e$ and $e' $ are linked and
is zero otherwise. For $C \in
\cC(\mink^4,\rc) $ it was shown that,  with the above constraints on $R, \omega, a$ and the dimensions of the detector,
that $\av{\btF}$  approximates to  same continuum expression
Eq.~(\ref{eq:detres}) when $R>> \rc^{-\frac{1}{4}}$. A detailed calculation gives an  upper bound  on the
fluctuations, which,  for a particular AGN model is one  part in $10^{12}$ for $\rc=\rp$. Hence the discreteness does not seem to mess
with the coherence of waves from distant sources. As we will see in Sect.~\ref{sec:phen}  there are other potential signatures of the
discreteness that may have phenomenological consequences \citep{swerves,lambdaone,lambdatwo,lambdathree}.

\subsection{The Sorkin--Johnston (SJ) vacuum} 
\label{ssec:SJ} 
 
Having obtained the classical Green function and the d'Alembertian operator in $\mink^2$ and $\mink^4$, the obvious next step
is to build a full quantum scalar field theory on the causal set. As we have mentioned earlier, the
canonical route to quantisation is not an option for causal sets nor for fields on causal sets and hence there is a need to look at more covariant
quantisation procedures. 

\cite{johnstontwo,johnstonthesis}  used the the covariantly defined {Peierls'}   bracket 
\begin{equation} 
[\hP(x),\hP(y)] = i\Delta(x,y) 
\end{equation} 
as the starting
point for quantisation, where 
\begin{equation} 
\Delta(x,y) \equiv G^R(x,x')-G^A(x,x') 
\end{equation}
is the Pauli Jordan function, and $G^{R,A}(x,x')$ are the retarded and advanced Green's
functions, respectively.  As we have seen,  these Green functions  can be defined on certain manifold-like causal sets and
hence provide a 
natural starting point for quantisation. 

However, even  here, the standard route to quantisation involves the
mode decomposition of the space of solutions of the Klein Gordan operator,  $\ker(\Box-m^2)$.  In $\mink^d$ the space of
solutions has a unique  split  into positive
and negative frequency classes of modes  with respect to which a vacuum can be defined.  
In his quest for a Feynman propagator, \cite{johnstontwo} made a bold proposal, which as we
will describe below,  has led  to a very interesting new direction in quantum field theory  even in the
continuum. This is  the  \emph{Sorkin--Johnston or SJ vacuum} for a free quantum scalar field theory. 

Noticing that the  Pauli--Jordan function on a {finite}  causal set $C$ is a Hermitian operator, and that
$\Delta(e,e')$ itself is antisymmetric,  Johnston used the fact that the eigenspectrum of $i\Delta$ 
\begin{equation}
i \hD \circ \vkpj(e) \equiv \sum_{e'\in C} \,i\Delta(e,e')\vkpj(e')=\lk \vkpj(e)
\end{equation} 
splits into  pairs $(\lk, -\lk)$, with  eigenfunctions $(\vkpjp, \vkpjm)$, $\vkpjm={\vkpjp}^\ast$.  
This provides a natural  split  into a positive part and a negative part, without explicit reference to $\ker(\Box
-m^2)$.\footnote{The identification of $\ker(\Box -m^2)$ with $Im(i\Delta)$ is in fact well known \citep{waldqft} when the latter is
restricted to functions of compact support.} A spectral decomposition of $i\hD$ then gives 
\begin{equation} 
i \Delta(e,e')=\lk\sum_{\mathbf{k}}\vkpjp(e){\vkpjp}^*(e')-\vkpjp(e)^*{\vkpjp}(e').
\label{modeexp} 
\end{equation}
This decomposition is used to define the SJ Wightmann function as the positive part of $i \Delta$ 
\begin{equation} 
W_{SJ}(e,e') \equiv \lk\sum_{\mathbf{k}}\vkpjp(e){\vkpjp}^*(e'). 
\end{equation} 
Importantly, for a non-interacting theory with a  Gaussian state, the Wightmann function is sufficient to describe
the full theory and thus the vacuum state. Simulations in $\mink^d$ for $d=2,4$  give a  good agreement with the
continuum \citep{johnstontwo,johnstonthesis}.     

\cite{sorkinsj}  noticed  that the construction on the causal set, which was born out of necessity, provides a
new way of thinking of the quantum field theory vacuum.  
A well known feature of quantum field theory in a general curved spacetime is that the vacuum obtained from mode decomposition
in  $\ker(\hB-m^2)$ is observer dependent and hence not unique.  Since the  SJ vacuum is intrinsically defined, at least in finite spacetime
regions, one has a uniquely defined vacuum.   As a result, the SJ state has generated some interest in the broader algebraic field theory community
\citep{fv12,bf14,fewsterart}.  For example, while not in itself Hadamard in general, the SJ vacuum  can be used to generate a new class of Hadamard
states \citep{bf14}.

In the  continuum,  the SJ  vacuum was constructed for the  massless scalar field in the $d=2$ causal
diamond \citep{sj2ddiamond} and recently extended to the small mass case \citep{mathursurya}.  It has also been obtained  for the trousers topology and shown to produce a divergent
energy along both the future and the past light cones associated with the Morse point singularity \citep{sjtrousers}.
Numerical simulations of the SJ vacuum on causal sets are are approximated by 
de Sitter spacetime suggest that the
causal set SJ state differs significantly from the Mottola--Allen $\alpha$ vacuua \citep{sxy}.  This has potentially far
reaching observational consequences which need further investigation.  

\subsection{Entanglement entropy} 
\label{ssec:SSEE}
 
 Using
the Pauli Jordan operator $i\hD$ and the associated Wightman  $\hW$, \cite{sorkinEE} defined a spacetime entanglement entropy, \emph{Sorkins' Spacetime Entanglement Entropy (SSEE)} 
\begin{equation} 
S = \sum_{i} \lambda_i \ln|\lambda_i|
\end{equation} 
where $\lambda_i$ are the  generalised eigenvalues satisfying 
\begin{equation} 
\hW\circ v_i = i \lambda_i \hD \circ  v_i. 
\label{eq:genev} 
\end{equation} 
It was shown by \cite{yasamaneecont} that for a  causal diamond  sitting at  the centre of a larger
one  in $\mink^2$, $S$ has the expected
behaviour in the limit that the size of the smaller diamond $l$ is much smaller than that of the larger diamond,    
\begin{equation} 
S=b \ln \left(\frac{l}{l_{uv}}\right) +c, 
\end{equation} 
where $l_{uv}$ is the UV cut-off and $b,c$ are constants that can be determined.

One of the promises that discretisation holds is of curing the UV divergences of quantum field theory and in particular
those coming from  the calculation of the entanglement entropy of \cite{bklsEE}.  As shown by \cite{causetee} the
causal set version of the above calculation is proportional to the volume rather than
the above ``area'', thus differing from the continuum. This can be traced to the fact that the 
continuum spectrum of eigenvalues (Eq.~\ref{eq:genev})  agrees with the discrete eigenvalues only up to a 
``knee'', beyond  which the effects of discreteness become important, as shown in Fig.~\ref{spectrum.fig}.
Using a double truncation of the spectrum -- once in the larger diamond and once in the smaller one, \cite{causetee}
obtained the requisite area law. This raises
very interesting and as yet unanswered puzzles about the nature of SSEE in the causal set. It is for example possible that in a
fundamentally non-local theory like CST an area law is less natural than a volume law. Such a radical understanding could
force us to rethink continuum inspired ideas about Black Hole entropy. 

\begin{figure}[ht]
  \centering \resizebox{3in}{!}{\includegraphics{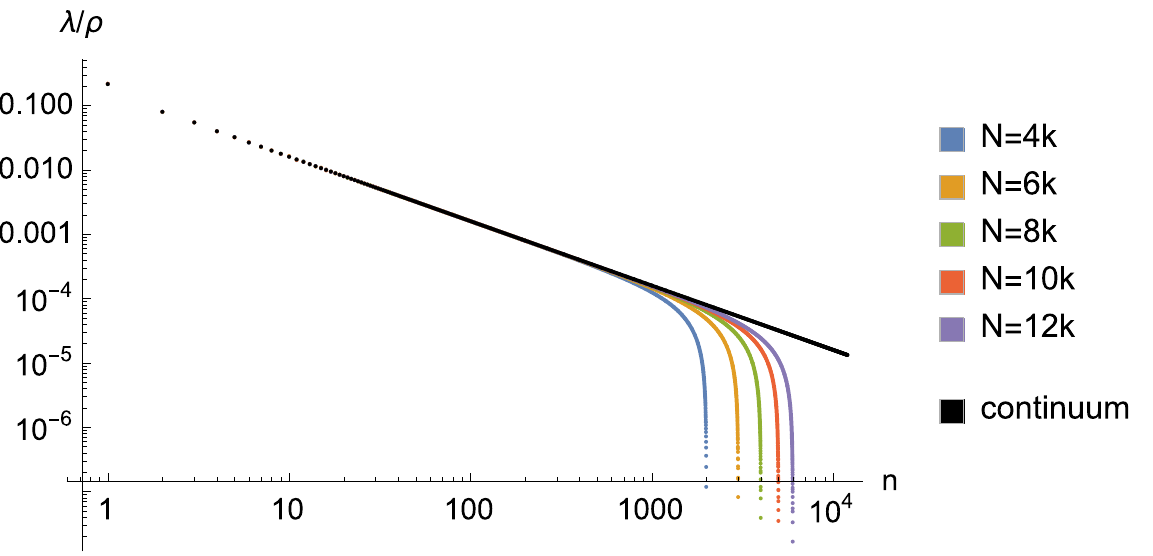}} 
  \caption{A log-log plot depicting the SJ spectra  for causal sets in a causal diamond in $\mink^2$. A comparison with the
      continuum (the straight black line) shows that the causal set SJ spectrum matches the continuum in the IR but has
      a characteristic ``knee'' in the UV after which it deviates significantly from the continuum. As the density of the
      causal set increases, this knee shifts to the UV.}
\label{spectrum.fig} 
\end{figure}

Extending the above calculation to actual black hole spacetimes is an important open problem. Ongoing simulations for
causal sets obtained from sprinklings into 4d de
Sitter spacetime show that this double truncation  procedure gives the right de Sitter horizon entropy ({\it Dowker, Surya,
Sumati, X  and Yazdi, work in progress}),
but one first needs to make an ansatz for 
locating  the knee in the causal set $i \Delta$ spectrum.

\subsection{Spectral dimensions}
\label{ssec:specdim}

An interesting direction in causal set theory has been to calculate the spectral dimension of the causal set
\citep{em,diondr,carlipdr}. \cite{carlipdr} has argued that $d=2$ is special in the UV limit, and that several theories of quantum
gravity lead to such a dimensional reduction. In light of how we have presented CST, it seems that
this continuum inspired description must be limited. It is nevertheless interesting to ask if causal sets that
are manifold-like might exhibit such a behaviour around the discreteness scales at which the continuum approximation is
known to break down. As we have seen earlier (Sect.~\ref{ssec:distance}), one such behaviour is discrete asymptotic silence \citep{ems}.

\cite{em} calculated the spectral dimension on a causal set using  a random walk on a finite element causal
set. It was found that in contrast, the dimension at small scales goes up rather than down.
On the other hand, \cite{diondr} showed  that  causal set inspired  non-local d'Alembertians do give a spectral
dimension of $2$ in all dimensions. As we noted in  Sect.~\ref{sec:kinematics},  \cite{carlipdrone}  showed that  dimensional reduction
of causal sets occurs for the Myrheim--Meyer dimension as one goes to smaller scales.  Recently \citep{esvsd}, the spectral dimension
was calculated on a maximal antichain for a causal set obtained from sprinklings into $\mink^d$, $d=2,3$ using the
induced distance function of \cite{esv}. It was seen to decrease at small scales, thus bringing the results closer to
those from other approaches.

\section{Dynamics} 
\label{sec:dynamics}        

Until now our focus has been on manifold-like causal sets, since the aim was to find useful manifold-like covariant
observables as well as to make contact with phenomenology.  However, as discussed in Sect.~\ref{sec:cst}, the arena for CST is a sample space
$\Omega$ of locally finite posets which replaces the space of $4$-geometries, and contains non-manifold-like causal
sets.  A CST  dynamics is given by the measure 
triple $(\Omega, \fA, \mu)$ where $\fA$ is an event algebra and $\mu$ is either a classical or a quantum  measure. 
 We will define these quantities  later  in this section.

To begin with, $\Omega$ itself can be chosen depending on the particular physical situation in mind.  In the
context of initial conditions for cosmology, for example,  it is appropriate to  restrict to the sample space of past finite
countable causal sets $\Omega_{g}$, while for a  unimodular type dynamics using the Einstein--Hilbert action,   the
natural restriction is to $\Omega_n$  the sample space  of causal sets of fixed cardinality $n$.  We will see that
dimensional  restrictions on the  sample space are also of interest  and
can lead to a closer comparison with other approaches to quantum gravity.

As discussed in Sect.~\ref{sec:cst} and \ref{sec:kinematics},  in the asymptotic $n \rightarrow \infty$ limit  the sample space $\Omega_n$ is
dominated  by the non-manifold-like  KR  causal sets depicted in Fig.~\ref{kr.fig}. This is the ``entropy problem'' of
CST. 
These posets have approximately just three ``moments'' of time and hence should not  play a role in the classical or
continuum approximation of the theory.

For a  quantum dynamics of CST  we would like to start  
with a few  basic axioms, including discrete general covariance and dynamical causality. A very important step in this
direction was made by the \emph{classical sequential growth}  models (CSG)\citep{csg} , which are Markovian
growth models. We will describe these in Sect.~\ref{ssec:csg} and \ref{ssec:beable}. 

One of the main challenges in CST is to build a viable \emph{quantum sequential growth}(QSG) dynamics. The  appropriate
framework for the dynamics is as a quantum measure space which  is a natural quantum
generalisation of classical stochastic dynamics \citep{qmeasureone,qmeasuretwo,sorkinqmeasure}. This means replacing the
classical probability measure $P$ in the
measure space triple $(\Omega, \cA, \mu_c)$ with a \emph{quantum measure} $\mu$.  The quantum measure is defined via a
decoherence functional  and can also be defined as a \emph{vector measure}  in a  corresponding \emph{histories Hilbert
  space}. We will discuss this in Sect.~\ref{ssec:qsg}. 

It is also of interest to construct  an effective   continuum-inspired dynamics, where the discrete
Einstein--Hilbert or BD action is used to give the measure for the discrete path integral or path sum. The quantum
partition function can either be evaluated directly or converted into a statistical partition function over causal sets using an analytic
continuation. This makes it amenable to  Markov Chain Monte Carlo (MCMC) simulations as we  will see below in Sect.~\ref{ssec:partn}.

\subsection{Classical sequential growth models}
\label{ssec:csg}

The \cite{csg} classical sequential growth or CSG models are a class of stochastic dynamics in which causal sets are grown
element by element, with the dynamics satisfying a few basic principles \citep{csg,csgtwo,csgrg,davidthesis,rv}. The
stochastic dynamics  finds a natural expression in measure theory and allows for an explicit definition of 
covariant classical observables \citep{observables,observablesds}. This measure theoretic
structure provides an important template for the quantum theory, and hence we will first flesh it out in some detail
before discussing quantum dynamics.      

Let us start with a naive picture. Imagine living on a \emph{classical} causal set
universe, with our universe represented by a single causal set.  Since causal sets are locally finite,  the ``passage of time'' occurs with the addition of a new
element. If we are to
respect causality, this new element cannot be added so as to disturb the past. Instead it can be added to the future of some of the existing events or it can be  unrelated to all of them. Every such
``atomic change'' in spacetime corresponds to the causal set changing cardinality or ``growing''  by one. Starting with a causal set 
$\tc_n$ of cardinality $n$, the passage of time means transitioning from  $\tc_n\rightarrow \tc_{n+1}$ where the new
element in $\tc_{n+1}$ is to the future of some of the elements of $\tc_n$, but never in their past.  In the infinite ``time'' limit,  $n\rightarrow
\infty$, the dynamics, either deterministic, probabilistic or quantum, will take you from $\tc_n$ to a countable causal
set. 

Working  \emph{backwards}, on the other hand, leads us  to a ``beginning'', with $n=0$. This gives the most natural
initial condition\footnote{Of course, we could insist that there is no beginning,  in which case $n$ is never finite.}  for
causal sets: begin with the empty set $\emptyset$.  The only way to go forward from here, is to
make $n=1$, i.e., we have  a single element. For $n=2$, the new element could either be to the future of the existing
element or unrelated to it, as in Fig.~\ref{growthtwo.fig}. 

\begin{figure}[ht]
  \centering \resizebox{2in}{!} {\includegraphics[width=\textwidth]{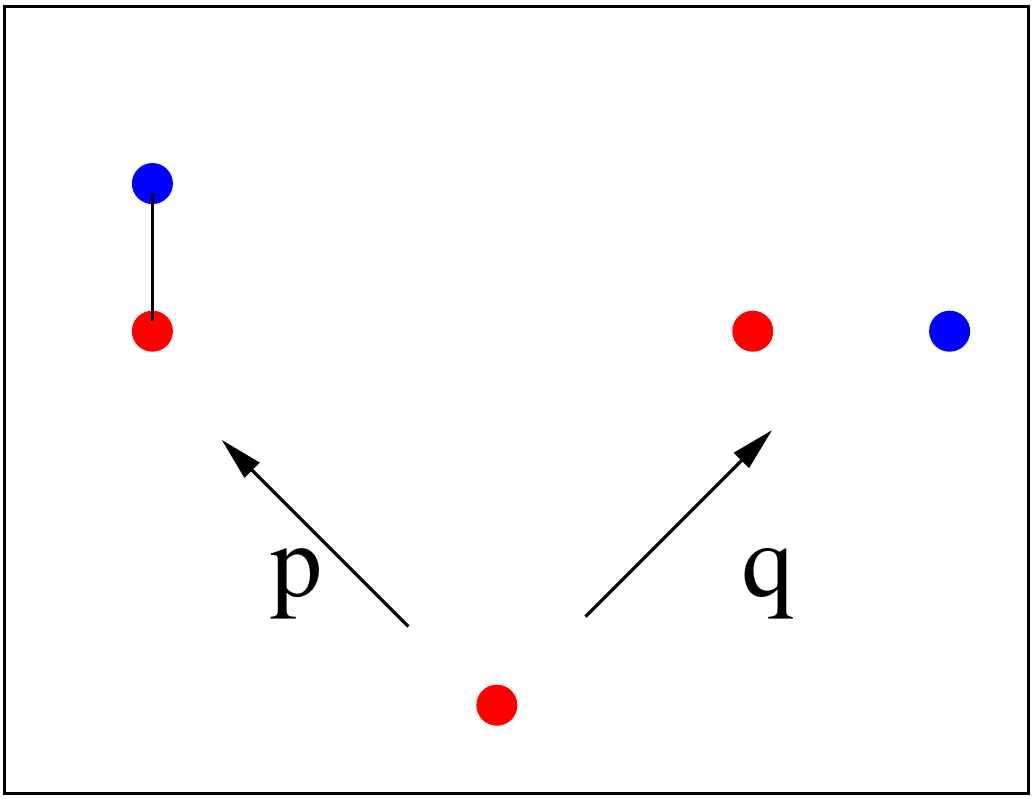}} 
  \caption{The first two stages of a classical sequential growth(CSG)  dynamics. The probability for a single element
      (red) 
    to appear at coordinate time $n=1$ is 1. Subsequently, the new element (blue) at $n=2$ is added either to the future of the
  existing element with probability $p$ or is unrelated to it with probability $1-p$.}
\label{growthtwo.fig}
\end{figure}

Thus, one can build up the tree $\csgtree$ of causal sets as $n\rightarrow \infty $ as shown in
Fig.~\ref{csgtree.fig}. As $n$ increases, the number of possibilities grows superexponentially as expected from the KR
theorem \citep{kr},  and there is no easy enumeration of this space. 
The growth process generates a sample space $\tOg$ 
of countable causal sets which are are all \emph{past finite} and labelled by the ``time'' at which each element is
added. A causal set $\tc$ in $\tOg$ is said to be \emph{naturally  labelled}, i.e., there exists an injective map   
$L: \tc \rightarrow \mathbb N$ (the natural numbers) which 
preserves the order relation in $\tc$, i.e., $e \prec e' \Rightarrow L(e) < L(e')$. In the growth process, this label is the
coordinate time. 

\begin{figure}[ht]
  \centering \resizebox{4in}{!} {\includegraphics[width=\textwidth]{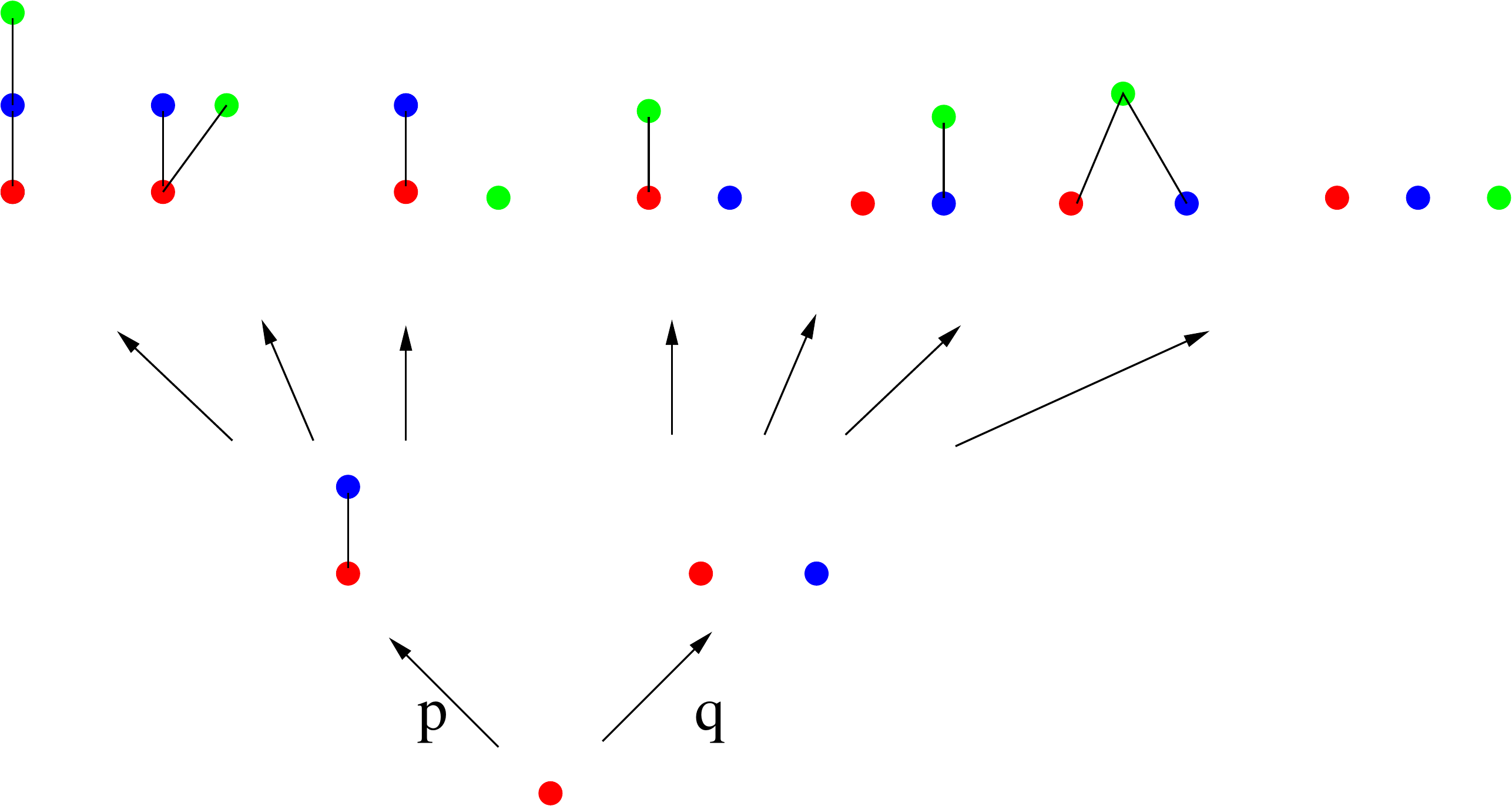}} 
  \caption{The CSG tree $\csgtree$. There are three ways to get the  3-element unlabelled causal set
      whose natural labellings are given by the 3rd, 4th and 5th 3-element labelled causal sets in the figure. One path
      is via the 2-element chain and the other two are from the 2-element antichain. Covariance
      demands that the probability along each path is the same.}
    \label{csgtree.fig}
\end{figure}

In the spirit of covariance,  however, we cannot take  the  time label to be  fundamental; the dynamics and the
observables cannot depend on the order in which the elements are born. Thus, the probability to get a labelled causal set $\tc_n$
and any of its relabellings,  $\tc_n'$ must be  the same.  Identifying relabelled causal sets as the same object in the CST tree
$\csgtree$ gives us a non-trivial poset of causal sets or the ``postcau''  $\postcau$ of \cite{csg}. On $\postcau$, 
a  covariant dynamics is thus path-independent: if there is more than one path from an unlabelled initial
causal set $c_{n_{i}}$ to an unlabelled final causal set $c_{n_{f}}$ in $\postcau$, then in order to satisfy covariance,
the measure on both
paths should be  the same. 

Apart from covariance, this  dynamics also satisfies an internal causality condition, dubbed  \emph{Bell
  causality}.   Consider the transition $\tc_n \rightarrow \tc_{n+1}$ with probability $\alpha_n$ where  the new element
$e_{n+1}$ is added to the future of a
``precursor''   set $p_n \subset \tc_n$, and is unrelated to a ``spectator set'' $s_n \subset \tc_n$.  Causality suggests
that the 
probability for the transition should not depend on the spectator set $s_n$.  For non-empty $s_n$ with
$|s_n|<n$,  consider
the causal sets $\tc_m=\tc_n\backslash s_n$
and $\tc_{m+1}=\tc_{n+1} \backslash s_n$,  
where $\backslash$ denotes set difference  and $m+|s_n| =n$.  The  transition probability $\alpha_m$ for  $\tc_m\rightarrow
\tc_{m+1}$   should then be proportional to $\alpha_n$. If $\tc_n \rightarrow \tc_{n+1}'$ is another transition from $\tc_n$, then
defining $p'_n, s_n'$, $ \alpha_n'$, and $\tc_{m+1}'=\tc_{n+1}'\backslash s_n'$, analogously, the condition of Bell
causality is 
\begin{equation}
\frac{\alpha_n(\tc_n \rightarrow \tc_{n+1})}{\alpha_n'(\tc_n \rightarrow \tc_{n+1}')}= \frac{\alpha_m(\tc_m \rightarrow
  \tc_{m+1})}{\alpha_m'(\tc_m \rightarrow \tc_{m+1}')} 
\end{equation}
Though relatively easy to implement classically, a quantum version of Bell causality has been hard to find 
\citep{joequantumbell}.    

The triple requirements of (a) covariance, (b) Bell causality and (c) Markovian evolution define the classical
sequential growth dynamics of \cite{csgone}.  Starting from the empty set, a causal set is thus grown element by
element, assigning probabilities to each transition $\tc_n$ to a $\tc_{n+1}$, consistent with these requirements. Because of
it being a Markovian evolution, the probability associated with any finite $c_n$ is given by the product of the
transition probabilities along a path in $\postcau$.   

The dynamics was shown in \cite{csg} to be  fully determined by the infinite set of coupling constants, $t_n$, one for
each stage of the growth. If  $q_k$ denotes  the transition probability from  the $k$-element antichain to the $k+1$-element
antichain, these  coupling constants can be expressed as 
\begin{equation}
t_n\equiv  \sum_{k=0}^n (-1)^{n-k} \binom{n}{k}\frac{1}{q_k}.
  \end{equation}  
In general, the $t_n$  can be independent of each other. Including relations between the different $t_n$ thus simplifies  the dynamics. The simplest example is that of \emph{transitive percolation}
determined by the probability  $(1-q) \geq 0$ of adding an element to the immediate future of an existing element,\footnote{By this we mean that the new
  element is ``linked'' to an existing one, not just related to it.} and $q$ of
being unrelated to it.  Thus, the
probability of adding a new element to the immediate future of $m$ elements of   $c_n$ and of being  unrelated to  $m'$ others
is $(1-q)^mq^{m'}$.  In terms of the general coupling constants,  $t_n=t^n\equiv \left(\frac{1-q}{q}\right)^n$.

In \cite{rv} and \cite{observablesds},
a generalisation of the dynamics was explored,  where some of the transition probabilities were allowed to vanish, consistent with
(a) (b) and (c). This requires  a  generalisation of the Bell causality condition. The resulting dynamics exhibits a
certain ``forgetfulness'' when these transition probabilities vanish, but are otherwise very similar to the CSG
models.

Since the generic dynamics  consistent with  (a), (b) and (c) does not by itself lead to constraints on the 
 $t_n$,  this is an embarrassment of riches. Does nature pick out one set over another? 
In \cite{csgrg}, an evolutionary mechanism for doing so was suggested using cosmological bounces  which
give rise to new epochs which ``renormalise'' the coupling constants towards fixed points.  
A cosmological bounce in a causal set is naturally described by the appearance of a \emph{post} 
which is an inextendible antichain of cardinality 1.

\begin{figure}[ht]
  \centering \resizebox{3in}{!} {\includegraphics[width=\textwidth]{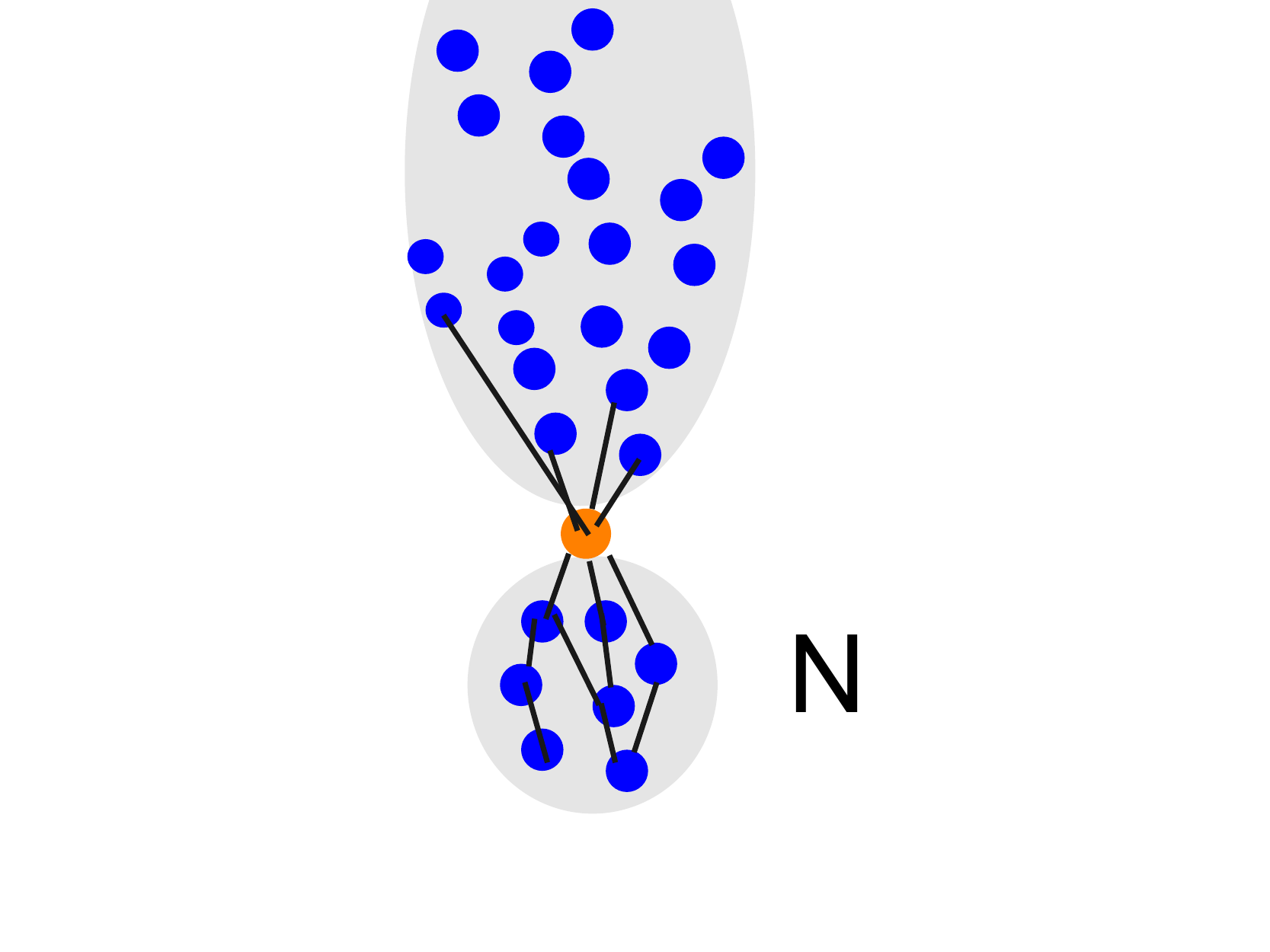}} 
  \caption{A post is an analogue of a bounce in causal set cosmology.}
\label{post.fig}
\end{figure}

Thus, every element in $c$ either lies to its past or to its
future. Moreover, because it is a single element maximal antichain, there are no ``missing links'' (see
Fig.~\ref{Missinglinks.fig}), and the post is indeed a
summary of its past. The post is  the causal set equivalent  to a ``bounce'' but is non-singular in the  causal set.  We
define the causal set between two posts as an ``epoch'',  with the
last epoch being the one after the last post. Let  $e$ be a post in $c$ and let  $r=|\past(e)|$. Then a 
set of ``effective''  coupling constants in the epoch after $e$ can be defined as \citep{csgrg} 
\begin{equation}
\tilde t_n^{(r)}=\sum_{k=0}^r \binom{r}{k} t_{n+k}.
  \end{equation} 
Thus, the memory of the past of the post, which is common to all the elements to the future of the post is 
``washed'' out, but not without ``dressing'' up the new effective coupling constants. Denoting the set of effective couplings by $\cT^{(i)}\equiv \{t_0^{(i)}, t_1^{(i)}, \ldots  \} $ with $i=0$ being the original set of couplings,  this    corresponds to applying
$r$ copies of  the transform $M: \cT^{(i)} \rightarrow \cT^{(i+1)}$ where $t_n^{(i+1)}=t_n^{(i)} + t_{n+1}^{(i)}$, $i=0,
\ldots r-1$. In \cite{csgrg}, it was shown that the fixed points of the map $M$  give $t_n=t^n$ (transitive percolation) for some $t \geq 0$ and
moreover  $M$ does not have any other cycles.  Starting from any set  $\cT^{(0)}$ for which
$\lim_{n\rightarrow \infty} (t_n^{(0)})^{{1}/{n}}$ is finite, $M^r: \cT^{(0)} \rightarrow \cT^{(r)}$, is such that
$\cT^{(r)}$ converges pointwise to $t^{(r)}_n=t^n$ for $t=\lim_{n\rightarrow \infty} (t_n^{(0)})^{{1}/{n}}$. While
this result does not guarantee that every $\cT^{(0)}$ will converge to transitive percolation,
\cite{csgrg} examined several cases, and conjectured that the deviation from percolation-like values are ``rare'' and
that typically,  $\cT^{(r)}$  will be nearly like transitive percolation.

Such an evolutionary renormalisation thus brings  the infinite dimensional coupling constant space to a one dimensional
space, which is remarkable. Assuming that this is indeed the case in general,  a sufficiently late epoch will likely
have a transitive
percolation dynamics.

What can one say about the causal sets generated from this dynamics?  A very important result from
transitive percolation is that the typical causal sets obtained are not 
KR like posets and hence the dynamics beats their entropic dominance. The question of whether there is a
continuum-like limit for transitive percolation dynamics was explored in \cite{csgtwo}, using a comparison
criterion. The abundance of fixed small subcausal sets was examined as a function of the coupling, by fixing the
density relations. Comparisons with Poisson sprinklings in flat spacetime showed a convergence, suggestive of  a continuum
limit. In \cite{davidmaqbool}, it was shown that the dynamics typically yields an exponentially expanding
universe. Moreover, for $(1-q) \ll 1 $ and $n \gg \frac{1}{1-q}$, after a post the universe enters a tree like phase and then
a de Sitter-like phase, in which the cardinality of large causal diamonds are de Sitter like functions of the discrete
proper time.  In \cite{intervals}, it was shown that despite this, the abundance of causal intervals is not de~Sitter
like, and thus, this is not strictly a manifold-like phase. In \cite{grnick} and \cite{grahammalwina}, moreover, it was shown
explicitly that in the asymptotic limit $n \rightarrow \infty$ the causal sets limit to ``semi-orders'' which,  though temporally
ordered,  have no spatial structure at all,  and are hence non-manifold-like. Nevertheless, the dominance of measure
over entropy is important and the hope is that it will be reflected in the right quantum version of the dynamics.

Recently, \cite{faystav} proposed a method for  dealing with black hole singularities in CSG models. As in the case of cosmological
bounces a  new epoch is  created beyond the singularity. Using ``breaks'' which are multi-element versions of a post,
they demonstrated that a renormalisation of the coupling constants occurs in the new epoch. 

\subsection{Observables as beables}
\label{ssec:beable}

As mentioned in the introduction to this section, a dynamics for CST is given by the triple $(\Omega, \fA, \mu)$.  In
CSG this is  a probability measure space, where the sample space $\tOg$ is the set of all past finite naturally labelled
causal sets.

The event algebra $\fA$ can be constructed from the sequential growth process as follows.  
 We define  a \emph{cylinder set}  $\cyl(\tc_n) \subset \tOg$ as
the set of all labelled causal sets in $\tOg$ whose first $n$ elements are the causal set
$\tc_n$. Figure~\ref{cylinder.fig} depicts an example of a cylinder set.\footnote{A useful example to keep in mind is the 1-d random walk. Let $\gamma^T$ be a finite element path in the $t-x$ plane from $t=0$
to $t=T$.  A cylinder set $\cyl(\gamma^T)$ is then the set of all infinite time paths, which coincide with $\gamma^T$
from $t=0$ to $t=T$.}   For every finite
element causal set $\tc_n$,  $\cyl(\tc_n) \subseteq  \tOg$, and in the trivial $n=1$ case, $\cyl(\tc_1)=\tOg$.   The cylinder
sets in CSG satisfy a nesting property. Namely, if $n'>n$ and $\cyl(\tc_{n'})\cap \cyl(\tc_n)
\neq \emptyset$, then $\cyl(\tc_{n'})\subset  \cyl(\tc_n)$. Thus, a non-trivial intersection of two different cylinder sets is 
possible only if one is strictly a subset of the other.

\begin{figure}[ht]
  \centering \resizebox{4in}{!} {\includegraphics[width=\textwidth]{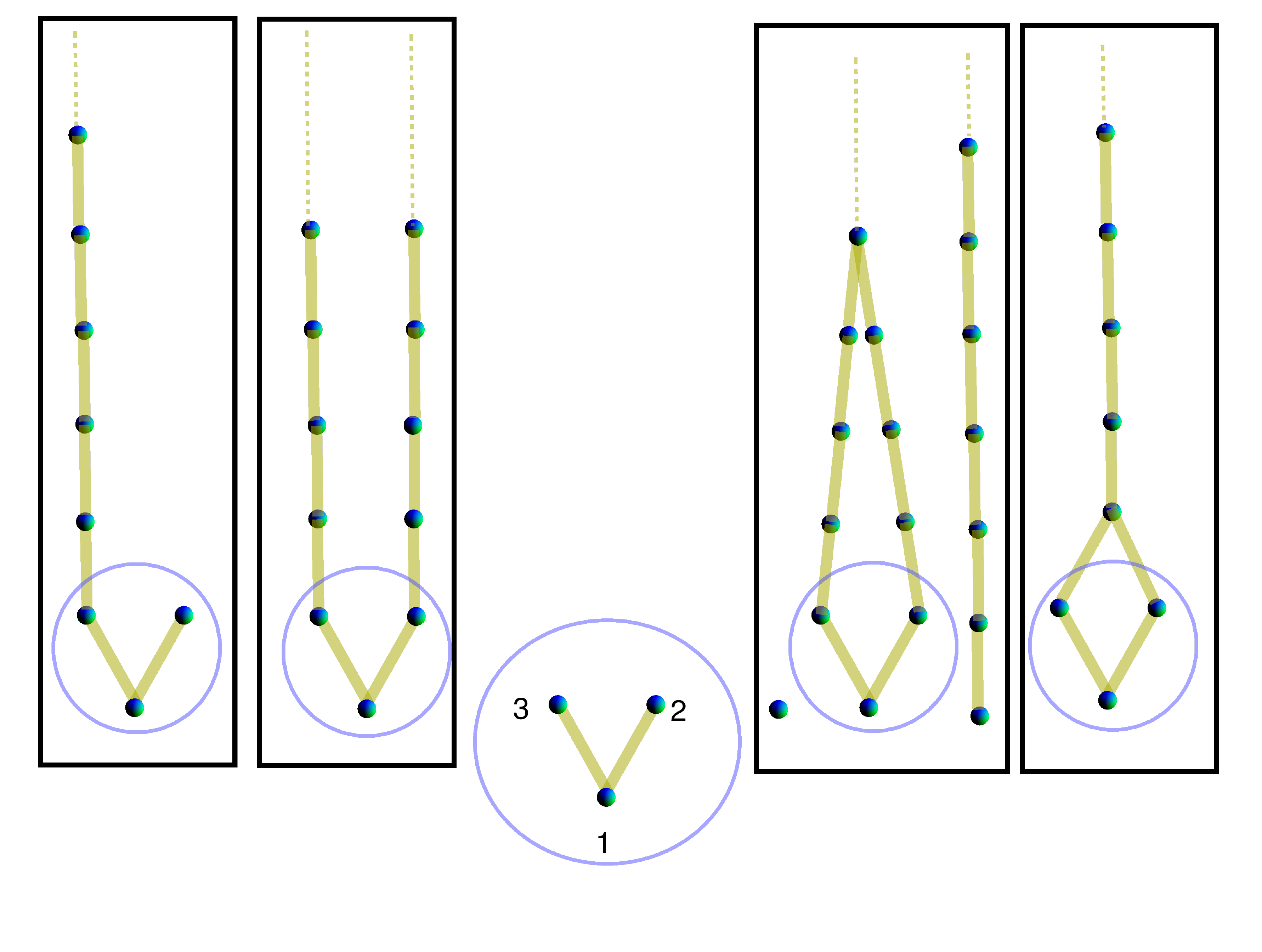}} 
  \caption{The cylinder set for the ``{\bf V}'' poset consists of all countable causal sets
      in $\tOg$ whose first three elements are the labelled ``{\bf V}'' poset. Examples of causal sets that lie in $\cyl(\mathbf{V})$
      are depicted in the boxes.}
\label{cylinder.fig}
\end{figure}

The \emph{event algebra $\tfA$} is generated from the cylinder  sets via
finite unions,  intersections and set differences. It is closed under finite set operations and
contains the null set $\emptyset$ as well as $\tOg$. In the growth process we assign a probability $\mu(\tc_n)$ to every finite
labelled causal set $\tc_n$. By identifying $\tc_n$ with its cylinder set $\cyl(\tc_n)$, we define the measure 
$\mu(\cyl(\tc_n))\equiv \mu(\tc_n)$ and hence on all elements of $\tfA$, since $\mu$ is finitely additive.   This makes 
$(\tOg, \tfA, \tmu')$ a ``pre-measure''  space.  

An \emph{event}  $\alpha$ is  an element of $\fA$ deemed to be  \emph{covariant}  as  a measurable subset $\alpha
\subset \tOg$ if for every $\tc \in \alpha$,  its relabelling $\tc'$ also
belongs to $\alpha$. Since a relabelling can happen arbitrarily far into the future,  no event in $\fA$
is covariant, since  $\fA$ is closed only  under finite set operations. Take for example the covariant \emph{post event}
which is the set of all  causal sets which  have a post. This is a covariant event, and is the equivalent of the return event in the random walk. In both cases, the event cannot be defined using
only countable set operations, and hence the post event does not belong to $\fA$.

One route to obtaining  covariant events is to pass
to the full sigma algebra $\tfS$  generated  by $\tfA$, which  is closed under countable set operations.
For classical measure spaces, the  Kolmogorov--Caratheodory--Hahn extension theorem allows us to extend  $\tmu'$ to $\tfS$
and hence pass  with ease to a full measure space $(\tOg, \tfS, \tmu)$, where $\tmu|_{\tfA} = \tmu'$.  Not every event
in $\fS$ is covariant, but we  can restrict our attention to covariant events, i.e., sets that are invariant under
relabellings. If  $\sim$ denotes the equivalence up to relabellings one can define the quotient algebra $\fS=\tfS/\sim$
of covariant events. An element of $\fS$ is measurable covariant set, or a \emph{covariant observable (or beable)}. Our
example of the  post event belongs to $\fS$.  Another example of a covariant event is the set of \emph{originary} causal
sets, i.e.,  causal sets with a
single initial element  to the past of all other elements.  Constructing  more physically interesting covariant
observables in $\fS$  is important, since it tells us what covariant questions we can ask of causal set quantum gravity. 

A more covariant way to proceed is to generate the event algebra not via the cylinder sets in
$\tOg$ but by using covariantly defined sets in $\Og$, the sample space of unlabelled causal sets. Because causal sets are past finite we can use the analogue of past sets
$J^-(X)$ to characterise causal sets in a covariant way.   A finite unlabelled
sub-causal set $c_n$ of $ c \in \Og $ is said to be a \emph{partial stem}  if it
contains its own past. A
\emph{stem set}  $\stem(c_n)$ is then a subset of $\Og$ such that every $c \in \stem(c_n)$ contains the partial stem
$\tilde{c_n}$. Let $\mathcal S$ be the sigma algebra generated by the stem sets. Although $\mathcal S $ is a strictly
smaller subalgebra of $\fS$, it differs on sets of measure zero for the CSG and extended CSG models as shown by 
\cite{observables} and \cite{observablesds}.  Thus, one can characterise all the observables of CSG in terms of stem sets. This is
a non-trivial result and the hope is that some version of it will carry over to the quantum case.

\subsection{A route to quantisation: The quantum measure} 
\label{ssec:qsg} 

The generalisation of CSG to QSG  is,  at least formally, very  straightforward. One ``quantises'' 
the  classical covariant probability space $(\Og, \fS, \mu_c)$, by simply replacing the classical probability $\mu_c$
with a \emph{quantum measure}  $\mu: \fS \rightarrow \re^+$, where $\mu$ satisfies the quantum sum rule \citep{qmeasureone,qmeasuretwo,salgadoqm,sorkinqmeasure}\footnote{We will not discuss the very rich and interesting literature on the co-event interpretation of  the quantum measure, which though incomplete, contains essential features that one would seek for a theory of quantum
  gravity \citep{coevents}.}  
\begin{equation}
\mu(\alpha \cup \beta \cup \gamma) = \mu(\alpha \cup \beta) + \mu(\alpha \cup \gamma) + \mu(\beta \cup \gamma) - \mu(\alpha) -\mu(\beta) -\mu(\gamma),   
\end{equation}
for the mutually disjoint sets $\alpha,\beta,\gamma \in \fS$. $\mu(.)$ is not in general a probability measure since it
does not satisfy additivity $ \mu(\alpha \cup \beta) \neq \mu(\alpha) + \mu(\beta)$ for  $\alpha \cap \beta =\emptyset$.
As in the classical case, observables in this theory are simply the quantum measurable sets in
$\fS$.  
The quantum measure $\mu(.)$  can be obtained from  a  decoherence  functional $D: \fS \times \fS \rightarrow \complex$ of quantum theory
with 
\begin{equation}
\mu(\alpha) = D(\alpha,\alpha), 
\end{equation} 
where $D$ satisfies 
\begin{itemize}
\item Hermiticity:  $D(\alpha,\beta)=D^*(\beta,\alpha)$ 
\item  Countable  biadditivity: $D(\alpha, \sqcup_i \beta_i)=\sum_i D(\alpha, \beta_i)$ and $D(\sqcup_i\alpha_i,
  \beta)=\sum_i D(\alpha_i, \beta)$  
\item Normalisation: $D(\Omega,\Omega)=1$ 
\item Strong
  positivity: $M_{ij}\equiv D(\alpha_i, \alpha_j)$ for any finite collection $\{\alpha_i\} $ is positive semi-definite
\end{itemize}
In a QSG model the transition probabilities of CSG are replaced by the decoherence functional
  $D$ or quantum measure.  Leaving aside Bell causality, the other principle of the growth dynamics are easy to
  implement.  In \cite{djs}, a simple complex percolation dynamics was studied, given by a product decoherence function
  $\tD(\alpha, \beta) =A^*(\alpha) A(\beta)$ on $\fA\times \fA$, where $A(\alpha)$ is obtained from the transition
  amplitudes $q \in \mathbb C$, similar to transitive percolation.  Thus, as in the case of CSG models, one starts with
  the labelled event algebra $\fA$ generated by the cylinder sets, and a quantum pre-measure $\tD'$.  Again, in order to
  obtain covariant observables one has to pass to the full sigma algebra $\fS$ associated with $\fA$.  However, unlike a
  classical measure $\tD$ need not extend to a full sigma algebra.  In \cite{djs}, the quantum pre-measure was shown to
  be a \emph{vector pre-measure} $\hmu'$ in the associated histories Hilbert space \citep{hhh}. Extension of $\hmu'$ to $\fS$ is then
  possible provided certain convergence conditions are satisfied.\footnote{In general, these are given by  the conditions
    in the 
    Kolmogorov--Caratheodory--Hahn--Kluvanek theorem \citep{du}.} 

Although the vector measure is 1-dimensional in   complex percolation dynamics, it  was shown in \cite{djs}  not to
satisfy this  convergence condition and hence one cannot pass to $\fS$ to construct  covariant observables.
However a
smaller algebra may be sufficient for answering physically interesting questions, which require far weaker 
convergence condition as suggested by \cite{ec}. This relaxation of conditions means that some simple measurable covariant
observables can be constructed in complex percolation, including for the originary event ({\it Sorkin and Surya, work in progress}).  Whether these
results on extension are shared by  all QSG models  or not is of course an interesting question. Another possibility is
that an extension of the
measure in QSG could,   for example,  be a criterion for limiting the parameter space of QSG. Very recently a class of
QSG 
dynamics that does admit an extension  has been found ({\it Surya and Zalel, work in progress}). 

 The space of QSG models is largely unexplored. It is however critical to study it extensively in order to find the
 right CST  quantum dynamics based on first principles.

\subsection{A continuum-inspired dynamics} 
\label{ssec:partn}

As we have seen, at  a fundamental level the quantum dynamics of causal sets looks very different
from that of a continuum theory of quantum gravity, even if the latter is formulated as a path integral.  However, as
one approaches the continuum approximation of the theory, it is possible that the effective  quantum dynamics begins to resemble  the
continuum path integral.  In CST,  the quantum partition function is 
\begin{equation} 
  Z_\Omega \equiv  \sum_{c\in \Omega} e^{\frac{i S(c)}{\hbar} }
  \label{eq:pathsum} 
\end{equation}
where  $S(c)$ is an action for causal sets, and the choice of sample space $\Omega$ is determined by the
problem at hand. One might also consider more generally a  decoherence functional $D(c_1,c_2) $ on causal sets, inspired
by the continuum, where $D(c_1,c_2) =e^{-i \frac{1}{\hbar}
  (S(c_1)-S(c_2))}f(c_1,c_2)$ with $f(c_1,c_2)$  a causal set analog of the delta
  function associated with 
  unitarity quantum theories. This is currently an unexplored direction and we will not discuss it further in this
  work.

The natural  choice for  $S(c)$ is the $d$ dimensional BD action $S_{BD}^{(d)}(c)$ which limits to the Einstein--Hilbert
action in the continuum. As discussed in Sect.~\ref{sec:cst}, the sample space $\Omega_n$ of causal sets  of
cardinality $n$ is dominated by KR type causal sets.  An important question is whether the  action $S_{BD}^{(d)}(c)$ can
overcome the KR entropy  in the large $n$ limit.

Indeed, there is a hierarchy of sub-dominant causal sets which are  non manifold-like \citep{dharone,dhartwo,kr,pst}, with the
set of bilayer posets $\cB$ being the next  subdominant class.
A recent  calculation  by  \cite{carliploomis} shows that $\cB$  is suppressed by the
BD action when the mesoscale and dimension satisfy certain conditions. The only relations  in a bilayer poset are
links. Given that the maximum number of relations is $\binom{n}{2}$ the causal sets in $\cB$ can be   classified  by the
linking fraction $p$ given by the ratio of the total number of  links $N_0$ to the maximal possible number of links
$\binom{n}{2}$. Moreover, the action itself reduces to a simple sum over $n$ and $N_0$.  
In the limit of large $n$, \cite{carliploomis} consider $p$ to be a continuous variable using which the partition
function $Z_{\cB}$ can be expressed as an integral over $p$ 
\begin{equation} 
Z_{\cB}= \int d p|\cB_{p,n}|e^{iS(p)/\hbar}  = e^{i\mu n} \int dp|\cB_{p,n}| e^{\frac{1}{2}i \mu \lambda_0 p n^2 +o(n^2)}  
\end{equation} 
where  $\cB_{p,n}$ denotes the  class of $n$-element  causal sets in $\cB$ with linking fraction $p$ and $\mu,
\lambda_0$ are related to the mesoscale $\epsilon$ and  function $f_d(n,\epsilon)$ that appears in $S_{BD}^{(d)}(c)$. The
challenge is then shifted to  calculating  $|\cB_{p,n}|$. Using another parameter $q$ which gives the cardinality of the upper layer as a further subclassification of $\cB_{p,n}$, the leading order contribution to $|\cB_{p,n}|$ was found. The resulting partition function was then shown to be strictly suppressed when $\mu \lambda_0$  satisfy the condition
\begin{equation}
\tan(-\mu\lambda_0/2) > \left(\frac{27}{4} e^{-\frac{1}{2}} -1 \right).  
\end{equation} 
This is an important analytic calculation and paves the way for a more rigorous understanding of the CST partition
function. 

More than the partition function, however,  it is the expectation value of observables or order invariants   
\begin{equation}
\av{\cO} = \frac{1}{Z_c} \sum_{c \in \Omega} \cO[c] e^{i \frac{1}{h} S[c]} 
\end{equation}
that is of physical significance.\footnote{We leave out interpretational questions!}  Evaluating this for larger values of $n$  is however a big challenge and we turn to numerical simulations to help us.

One route could be to simply ``perform''  the sum above. However, given that $|\Omega_n|$ grows superexponentially (to  leading order it is  $\sim 2^{\frac{n^2}{4}}$), this is computationally challenging even for relatively small values of $n$. On the other hand, Markov Chain Monte Carlo (MCMC)  methods for sampling the space $\Omega$ can be used if we can convert $Z_{\Omega}$ into a statistical partition function.

In CST, there is no analogue of a Wick rotation: since the order relation derives from the causal structure, it cannot
be ``Euclideanised''. On the other hand, there are other ways to analytically  continue $Z_{\Omega}$ (see
\citealt{sorkinlouko} for a continuum example). One option, first explored in \cite{2dqg} is to introduce a new parameter
$\beta$ such that  
\begin{equation} 
Z_{\Omega,\beta} \equiv  \sum_{c \in \Omega} e^{i  \frac{\beta}{\hbar} S(c)}. 
 \end{equation}  
This allows us to analytically continue $Z_{\Omega,\beta}$ from real to imaginary values of $\beta$, thus rendering the
quantum partition function into a statistical partition function. We can then  use  standard tools in statistical
physics, including MCMC methods,  to find the expectation values of suitable observables \citep{2dqg,2dhh,fss,ising}.

In \cite{onset}, MCMC methods were used to examine the sample space of {naturally labelled} posets $\tilde{\Omega}_n$
to determine the onset of the KR regime, using the uniform measure ($\beta=0$).
The Markov Chain was  generated via a set of moves that sample  $\Omega_n$. A  mixture of two
moves,  the \emph{link move} and the \emph{relation move},  was used to obtain  the quickest thermalisation.

To illustrate the complexity of these  moves we describe in detail the 
\emph{link move}.  A  pair of elements $e,e'$ are picked randomly and independently from the causal set $c$, and retained
if  $L(e) < L(e')$, where $L$ is the natural labelling defined in Sect.~\ref{ssec:csg}. If $e \prec
e'$ and moreover the relation is a  link, then the move is to ``unlink'' them. Those relations implied by this link via
transitivity also need to be removed. These are  relations between elements in $\ipast(e)$  and those in $\ifut(e')$
which are ``mediated'' solely either by $e$ or $e'$. On the other hand if $e$ and $e'$ are not related, then one adds in a
link between $e$ and $e'$, \emph{provided} that there are no existing  links between elements in $ \ipast(e)$ and $\ifut(e')$,
after which the transitive closure is taken.  In the relation move, although the existence or non-existence  of a link from $e$ to
$e'$ is also required, the move doesn't care about the sanctity of links, but is in other ways more restrictive.  Thus,
for both moves, picking of a pair of elements at random in $c$ does not always lead to a possible move, let alone a probable
one, and hence this  MCMC model is slow to thermalise. Trying to find a more efficient move is however non-trivial
precisely because of transitivity. 

The simulations of \cite{onset} suggest that the onset of the asymptotic  KR regime occurs for $n$ as small as $n
\approx 90$.  $\Omega_n$ is very large even for $n=90$ ($\sim 2^{90^2}$ !) and hence thermalisation  becomes  a problem
very quickly. 
Recently, steps have been taken to incorporate the action ($\beta \neq 0$)  into the measure, but again, because of
thermalisation issues, the size of the posets are fairly small. 

Instead of taking the full sample space, one can restrict $\Omega_n$ to causal sets that capture some gross features of
a class of spacetimes. As discussed above, for large enough $n$,  $\Omega_n$ contains causal sets that are approximated
by spacetimes of arbitrary dimensions. It is thus of interest to restrict the sample space so that those causal sets
that are  manifold-like in the sample space are approximated only by spacetime regions of a given dimension. Such a
restriction is typically hard to find, since it requires ``tailoring'' $\Omega$ using non-trivial order theoretic
constraints determined by dimension estimators of the kind we have encountered in Sect.~\ref{sec:kinematics}.  

Somewhat fortuitously, this restriction is very natural in $d=2$. Here, the sample space of  ``{2-orders}''
$\Omega_{\twod}$  is one in which the continuum dimension and a particular  order
theoretic dimension coincide \citep{2dorders,es,winkler}. The latter is defined only for a certain class of posets,
namely those obtained by the ``intersection''  of  $d$ totally ordered sets. For example,  an  $n$ element   \emph{2-order} is the
intersection of two linear orders $U=(u_1, u_2, \ldots u_n)$ and $V= (v_1, v_2, \ldots v_n)$ where each $u_i$ and $ v_i$
are valued on a set $S_n$ of $n$ non-overlapping points in $\re$. 
$U$ and $V$ are  therefore ``totally  ordered''  by the relation $<$ in $\re$. Their  intersection is the poset 
\begin{equation}
  U\cap V \equiv \{(u_i,v_i) \in U \times V |  (u_i,v_i)\prec (u_j,v_j)  \Leftrightarrow u_i < u_j \, \, \& \, \, v_i<v_j\}.
\end{equation} 
Similarly, one can define a \emph{$d$-order} as the intersection of $d$ linear orders. This is the  \emph{order theoretic
  dimension} referred to above.

For $d=2$, the total orders $U$, $V$ can be thought of as the set of light-cone coordinates of
a causal set obtained from an embedding (not necessarily faithful)  into a causal diamond in $\mink^2$.  Of special
interest is the 2-order obtained from a Poisson sprinkling,   an example of which is
shown in Fig.~\ref{2drandom.fig}.   As shown in \cite{2dorders} this is equivalent to choosing the entries of  $U$ and $V$ from a fixed $S_n$ at
random and independently. Importantly, this \emph{random order}  dominates $\Omega_{\twod}$ in the large $n$ limit as shown in
\cite{es,winkler}, and grows as {$|\Omega_{\twod}| \sim n!/2 $}. Thus, unlike $\Omega_n$,  the sample space is dominated
by manifold-like causal sets, though it also contains causal sets that are distinctly non-manifold-like. This makes
it an ideal starting point to study the non-perturbative  quantum dynamics of causal sets. Moreover, as shown in
\cite{2dorders}, 2-orders also have trivial spatial homology in the sense of \cite{homology}  (see Sect.~\ref{sec:kinematics})  
and hence $\Omega_{\twod}$ is the sample space of topologically trivial $\twod$ causal set quantum gravity. 

The continuum-inspired partition function for 2-orders or topologically trivial \emph{2d CST}  is 
\begin{equation}
Z_{\twod}(\beta, n)=\sum_{c \in \Omega_{\twod}} \exp^{\frac{i}{\hbar} S_{\twod}(c,\epsilon)} \,,
  \end{equation} 
where $S_{\twod}(c,\epsilon)$ is the BD action for $d=2$ with the non-locality parameter $\epsilon=l_p^2/l_c^2 \in (0,1]$
(see Eq.~(\ref{eq:bkoperator})).   Taking  $\beta
\rightarrow i \beta$ allows one to obtain the expectation values of order invariants using  MCMC techniques as was done
by \cite{2dqg}.  The  MCMC move  in $\Omega_{\twod}$ is very straightforward,  unlike that in $\Omega_n$: a pair of elements is picked independently and at random in either $U$
or $V$, and swapped.  For example, if $u_i \leftrightarrow u_j$, then the elements 
$(u_i,v_i)$ and $(u_j,v_j)$ in $U\cap V$ are replaced by $(u_i'=u_j,v_i'=v_i)$  and $(u_j'=u_i,v_j'=v_j)$, hence
changing the poset. \emph{Every} move is possible, and hence one saves considerably on efficiency and thermalisation times.

Importantly, the  MCMC simulations of \cite{2dqg}  give rise to a phase transition  from a continuum phase at low $\beta$ to a non-manifold-like phase at high
$\beta$.  This is very similar to the disordered to ordered phase transition in an Ising model.  The 
$\beta^2$ versus $\epsilon$    phase diagram moreover indicates that the continuum phase survives the analytic
continuation for any value of $\epsilon$.   

It was  recently  demonstrated by \cite{fss}
using finite size scaling  arguments that that this is a first order phase
transition.  The analysis moreover suggests that the continuum phase
corresponds to a spacetime with negative cosmological constant. This is an explicit example of a non-perturbative theory
of quantum gravity in which the cosmological constant is generated via
the dynamics. 

This simple system also allows us to examine other physically relevant questions. Of particular interest is the
Hartle--Hawking wave function using the no-boundary proposal. In $2d$ CST, this was constructed by \cite{2dhh} using a 
natural no-boundary condition for causal sets, namely requiring the existence of an ``initial'' element $e_0$ to the past of all the other
elements. $\psi_{\rm HH} (\cA_f)$ is the wave function for a final antichain of cardinality $|\cA_f|$, where one is summing
over all causal sets that have an initial element $e_0$ and final boundary $\cA_f$.

The MCMC simulations give the expectation value of the action $S_{\twod}$ from which the partition
function can be calculated by numerical integration, up to normalisation. The normalisation itself was determined in \cite{2dhh} 
using a combination of analytic and numerical calculations.  The results of the extensive analysis was that the
Hartle--Hawking wave function $\psi_{\rm HH}(\cA_f)$ peaks at low $\beta $
on antichains of small cardinality,  with the  peak jumping  at higher $\beta$ to  antichains with cardinality $\sim n/2$.
Interestingly, in 
the latter, high $\beta$ (low temperature) phase,  the dominant causal sets satisfy some of the rudimentary features of early universe cosmology: 
(a) the growth from a single element to a large antichain takes place rapidly and (b) each element in
$\cA_f$ is causally related to all the elements in its immediate past which makes $\cA_f$ ``homogeneous''. However, this is a non manifold-like phase,
and it is an open question how one exits this phase into a manifold-like phase.  If there is a dynamical mechanism that
makes $\beta$ small, then this would be a promising new mechanism for generating cosmologically relevant initial
conditions for the universe. 

Will this analysis survive higher dimensions? One of the issues at hand is that even for 2-orders the cardinality of $\Omega_{2d}$ grows rapidly
with $n$ and hence thermalisation can become a major stumbling block. However, the  finite sized scaling analysis of \cite{fss}
and the techniques used therein, tell us that it suffices to be in the asymptotic regime. For 2-orders, this is
already true around $n \sim 80 $ and hence the results of \cite{2dqg} and \cite{2dhh} are at least qualitatively
robust. Nevertheless, to get to the asymptotic regime in $d=4$  will require far more extensive computational
power. Recently, using new sophisticated computational techniques \citep{willthesis},  the algorithms of
\cite{2dqg} have been updated, so that $n \sim 300 $ simulations can be done in a reasonable time. 

An important question, however is how to obtain a dimensionally restricted $\Omega_n$ more generally. While  2-orders
are a good representation of $\twod$ (topologically trivial) causal set quantum gravity,  this is not true for
 higher order theoretic dimension.   For $d>2$ a   $d$-order is an embedding into a space with ``light-cubes'' rather
 than lightcones. Though potentially interesting, this does not serve our more narrowly defined goal of obtaining a
 continuum-inspired dimensionally reduced sample space.

 Recently,  a   lattice inspired method  has been
 investigated to generate sample spaces which are both dimensionally and topologically restricted. These are obtained as
 embeddings (not necessarily faithful) into a fixed spacetime, and thus include manifold-like causal sets. In
$d=2$, the simplest example comes from causal sets obtained from sprinkling into the flat cylinder spacetime
$ds^2= - dt^2+d\theta^2$, $\theta \in [0,2\pi]$.  Recent  simulations ({\it Cunningham and Surya, work in progress}) suggest that the results of
the topologically trivial case are largely unchanged. The next step is to include a wider class of embeddings as well as
topology change into the model, and hence bring it closer to a full 2d theory of quantum gravity.  

Of course, 2d causal set quantum gravity without matter does not have a continuum counterpart, since 2d continuum quantum gravity 
is coupled to a
scalar field (for example,   Liouville gravity).  Studying 2d CST with matter is therefore an open interesting question.  In \cite{ising}, Ising
spins were coupled to the causal set by placing a spin $s_i=\pm 1 $ at every element $e_i$ and coupling spins along the links, i.e.,
\begin{equation}
S_{I}(j)\equiv j \sum_{i k} s_i s_k L_{ik} \,,
  \end{equation}  
where $L_{ik}$ is the link matrix and $j$ the spin coupling constant.  The phase structure of this model coupled to the
BD action is substantially richer. In particular, the hope is that some of the resulting phase transitions are of higher
order and hence comparisons with conformal field theories might be possible. Further analysis of this model would
definitely be useful and interesting. 

In the MCMC simulations discussed above,  labelled posets are used for practical reasons, since this is how they are stored on the computer. A single  unlabelled poset admits many relabellings  or ``automorphisms'', but the 
number of relabellings varies from poset to poset even  for the same cardinality.
{For example, in the list of coloured or labelled 3-element causal sets in } 
Fig.~\ref{csgtree.fig}, we see that there is only one  3-element causal set with
three distinct natural labellings, while 
all {the} others admit only one {natural labelling}. Enumerating the
number of automorphisms {for a given causal set} quickly becomes {very difficult}  as $n$ increases.

{In the
continuum path integral, the ``correct'' measure in a gauge theory {involves} the volume of the gauge orbits. In this discrete setting, as we have
discussed above, the analogous gauge  orbits corresponding to 
to the automorphisms, are not of the same cardinality for each $c\in \tilde{\Omega}_n$. }

{Indeed, the choice of measure is not obvious in CST since it is not merely a discretisation of the continuum
  theory, with the  path sum Eq.~(\ref{eq:pathsum})  including  causal sets that are non-manifold-like. There  is no underlying order theoretic
  reason to pick the specific BD action; we have done so, ``inspired'' by the continuum.  For continuum like causal
  sets of a fixed dimension the number of relabellings is approximately  the same, so that they appear
  roughly with the same weight in the path integral. However, it is the relative weight compared the non-continuum-like
  causal sets that depends on the relabellings. In the classical sequential growth model described above, the
  labelling is related to temporality and hence   
  the choice of a uniform measure on the set of labelled causal sets $\tOg$ is a natural one.
In the MCMC simulations,
therefore we pick  a measure }  that is uniform on $\tilde{\Omega}_n$, rather than on the unlabelled sample space $\Omega_n$. Causal
sets that admit more relabellings come with a higher natural weight than those that admit fewer relabellings.  However,
{discrete} 
covariance {or label invariance} is not compromised since the observables themselves are label independent. 

While these numerical simulations have uncovered a wealth of information about the statistical thermodynamics of causal
sets, one must pause to ask how this is related to the quantum dynamics, as $\beta \rightarrow -i \beta$.  
There is for example no analogue of the Osterwalder--Schrader theorems to protect the results we have obtained in the
MCMC simulations.  Pursuing
these questions further is important, though finding definitive and rigorous answers is perhaps beyond the scope of our
present understanding of CST.

\section{Phenomenology} 
\label{sec:phen}

While the deep realm of quantum gravity is extremely well shielded from experimental probes in the foreseeable future, it is possible that certain
{properties}  of quantum gravity  can ``leak'' into observationally accessible regimes. This is the reason for the push, in the last
couple of decades, for exploring quantum gravity phenomenology. Without a full theory of quantum gravity, of course
there is little hope that any phenomenology is entirely believable, since it requires assumptions about an
incomplete theory. Nevertheless, quantum gravity phenomenology can be  useful  in setting realistic bounds on
these leaked out properties, and hence constrain theories of quantum gravity, albeit weakly.  Models of quantum gravity
phenomenlogy moreover use distilled properties of the underlying theory to build reasonable
models that can be tested. Some of these properties are unique to a given approach.

In CST spacetime discreteness takes
a special form and brings with it a special type of  non-locality that can  affect observable physics. We have already
encountered the possibility of voids in Sect.~\ref{sec:cst} as well as the propagation of scalar fields from distance
sources in Sect.~\ref{sec:matter}. 
The continuum approximation of CST  is Lorentz invariant and consistent with stringent observational bounds as
summarised in \cite{li}.
In
addition, as suggested by \cite{swerves}, there is the possibility of generating very high energies particles through long time diffusion in
momentum space. This arises  from the randomness of CST  discreteness, which cause particles to ``swerve'',  or suddenly
change their momentum, 
as they traverse the causal set underlying our universe \citep{swervestwo,faypolarisation}.
This spacetime
Brownian motion was calculated in $\mink^d$ and  can be constrained by observations \citep{swerveconstraint}, but an
open question is how to extend the calculation  to our FRW universe. 

There have been some very interesting recent ideas by \cite{nonlocalqftfour} for testing CST type  non-locality  via its
effect on propagation in the continuum using the d'Alembertian operator. \cite{nonlocalfive}  have looked at the
associated quantum field theory which contain critical instabilities. These can be removed by modifying the
d'Alembertian, but the relationship to CST is unclear.  \cite{darkmatter}  have proposed a candidate for dark
matter as off-shell modes of the non-local CST  d'Alembertian. This is an exciting proposal and should be investigated
in more detail. 

We will not review these very interesting ideas on CST phenomenology here, except one, namely the prediction of
$\Lambda$.

\subsection{The 1987 prediction for $\Lambda$} 
\label{ssec:lambda}

One of the most outstanding questions in theoretical physics is understanding the origin of ``dark energy''  which
observationally has been seen to make up $\sim 70\%$ of the total energy of the universe. The current observational value
is  $\sim 2.888 \times 10^{-122}$ in Planck units. Quantum field theory predictions for dark energy interpreted as the
energy of vacuum fluctuations of quantum fields on the other hand gives a huge value, perhaps as large as  $\sim 1$ in Planck units.  The gross
conflict with observation obviously implies that this cannot be the source of $\Lambda$.\footnote{On the other hand, it
  would be interesting to understand \emph{why} the back of the envelope quantum field theory calculation is not
  observationally relevant. Interesting insights into this question could come
  from a better understanding of the SJ vacuum in de~Sitter spacetime.}

In light of this conundrum, the CST prediction  for $\Lambda$  due to \cite{lambdaone} is startling in its simplicity and
accuracy, especially since it was made several years before the  {1998} observation.
One begins with the framework of unimodular gravity \citep{lambdatwo,unruhwald} in which the spacetime
volume element is fixed. 
$\Lambda$ then appears as a Lagrange multiplier in the action, with $\Lambda \int dV= \Lambda V ={\mathrm{constant}}$, for any finite spacetime region of volume $V$. In a canonical formulation of the
theory, therefore $\Lambda$ and $V$ are conjugate to each other, so that on quantisation there is an uncertainty relation 
\begin{equation}
\Delta V \Delta \Lambda \sim 1.
  \end{equation} 
Using the fact that $\Delta V $ is generated from Poisson fluctuations of the underlying causal set ensemble 
\begin{equation} 
  \Delta V \sim \sqrt{V}. 
\end{equation}
{ Assuming}  $\av{\Lambda}=0$, moreover, we see that 
\begin{equation}
  \Lambda \sim \frac{1}{\sqrt{V}}\sim H^2 =\frac{1}{3}\rho_{\mathrm{critical}}
\end{equation}
where $H$ is the Hubble constant. If  $V$ is taken to be the  volume  of the visible universe, 
\begin{equation}
  \Lambda=\Delta \Lambda \sim 10^{-120}, 
\end{equation}
in Planck units. This is very close to the subsequently observed value of $\Lambda$! Importantly, the prediction also states that
that under these assumptions, $\Lambda$ always tracks the critical density and is hence ``everpresent''. 

This argument is general and requires three 
important ingredients: (i)  the assumption of unimodularity and hence the  conjugacy between $\Lambda$ and $V$, (ii) the
number to volume correspondence $V \sim n$ and (iii)  that there \emph{are}  fluctuations in $V$ which are  Poisson, with
$\delta V = \sqrt{V} \sim \sqrt{n}$. While (i) can be
motivated by a wide range of theories of quantum gravity, (ii) and (iii)  are both  \emph{distinctive} to causal set theory.  No other discrete approach to quantum gravity makes
the $n \sim V$ correspondence at a fundamental level and also  incorporates  Poisson
fluctuations kinematically in the continuum approximation. Quoting from \cite{lambdaone}, {``Fluctuations in $\Lambda$  arise as residual nonlocal quantum effects of
spacetime discreteness''}.   Interestingly, as shown by \cite{sorkinledl},  if spacetime admits large extra directions, then the contribution to $V$ is
very different and gives the wrong answer for $\Delta \Lambda$.   

Of course, an important question that arises in this quick calculation is  why we should assume that $\av{\Lambda}=0$.\footnote{In
  \cite{samsupurna}, a very striking analogy was made between   a fluctuating $\Lambda$ and the surface tension $T$ of  a
  fluid
  membrane. In addition, using the atomicity of the model, the mean value of $T$ was shown to be zero, with a suggestion
of how this might extend to  CST.} The answer to
this may well lie in the  full and as yet unknown quantum dynamics. Nevertheless, phenomenologically  this assumption leads to further predictions that can
already be tested. The first conclusion is that a fluctuating $\Lambda$ must violate conservation of the stress energy
tensor, and hence the Einstein field equations. 

In \cite{lambdathree}, a dynamical model for generating fluctuations of $\Lambda$ was constructed, starting with the flat
$k=0$ FRW spacetime. In order to accommodate a fluctuating $\Lambda$, one of the two Friedmann equations   
must be dropped.  In \cite{lambdathree}, the Friedmann equation
\begin{equation} 
3\biggl(\frac{\dot{a}}{a}\biggr)^2= \rho + \rho_\Lambda 
\end{equation}
was retained,\footnote{Subsequently, more general ``mixed equation'' models were examined in \cite{eptwo}, which
  indicate that the results of \cite{lambdathree} are robust to these modifications.}  with  
\begin{equation}
  \rho_\Lambda =\Lambda, \qquad p_\Lambda= -\Lambda -\dot{\Lambda}/3H, 
 \end{equation} 
and  $\Lambda$ modelled as a stochastic function of $V$, such that
 \begin{equation}
   \Delta \Lambda \sim \frac{1}{\sqrt{V}}.
   \end{equation} 

More generally, $\Lambda$ can be thought of as the action $S$  per unit volume, which for causal sets  means that $\Lambda \sim S/V$.  A  very
   simple  stochastic dynamics is then generated by assuming that every element contributes $\pm \hbar$ to $S$, so that 
   \begin{equation}
     S=\sum_{\mathrm{elements}} \pm \hbar \Rightarrow S/\hbar \sim \pm \sqrt{N} \sim \pm \sqrt{V/l_p^4} \Rightarrow \Lambda\sim
     \pm \frac{\hbar /l_p^2}{\sqrt{V}} \,,  
\end{equation} 
 where we have equated the discreteness scale $l_c$ with the Planck length $l_p$. One then gets the integro-differential equations  
\begin{eqnarray} 
\frac{da}{a} &=&\sqrt{\frac{\rho+\Lambda}{3}}d\tau \nonumber \\ 
V d\Lambda &=& Vd(S/V)=dS-\Lambda \dot{V} d\tau \,, \nonumber 
\end{eqnarray} 
where
\begin{equation}
  V(\tau) = \frac{4\pi}{3} \int_0^t dt' a(t')^3 \biggl( \int_0^{t'} dt" \frac{1}{a(t")}\biggr)^3
  \end{equation} 
is the volume of the entire causal past of an event in the FRW spacetime.  
The stochastic equation is then generated as follows. At the $i^{\tth}$ step 
one has the variables $a_i$ (scale factor), $N_i,V_i,S_i $ and $ \Lambda_i$. The scale factor is updated using the
discrete  Friedmann equation  
$a_{i+1}=a_i+a_i\sqrt{\frac{\rho+\Lambda}{3}}(\tau_{i+1}-\tau_i)$, from which $V_i=V(\tau)$ can be calculated and thence 
$N_{i+1}=V_{i+1}/\ell^4$. The  action is then updated  via $S_{i+1}=S_i+ \alpha \, \, \xi\sqrt{N_{i+1}-N_i}$, where
$\quad\xi$ is a Gaussian random variable, with $\quad\Delta \xi = 1$, and $\alpha$ is a tunable free parameter which
controls the magnitude of the fluctuations.   Finally, $\Lambda_{i+1}=S_{i+1}/V_{i+1}$, with   $S_0=0$.  It was shown in
\cite{lambdathree} that in order to be consistent with astrophysical observations,  $0.01 < \alpha < 0.02$. The results of
simulations moreover suggest  that $\Lambda$ is \emph{``everpresent''} and tracks the energy density of the universe.

This model assumes spatial homogeneity and it is important to check how inhomogeneities affect these results.  In
\cite{barrow} and \cite{zuntz},  inhomogeneities were modelled by taking $\Lambda(x^\mu)$, such that $\Delta \Lambda(x)$ is
dependent only on $\Lambda(y)$ for $y \in J^-(x)$.  This would mean that well separated patches in the CMB sky would
contain uncorrelated fluctuations in $\Omega_\Lambda$, which in turn are strongly constrained to $< 10^{-6}$ by 
observations and hence insufficient to account for $\Lambda$.  In \cite{lambdathree} and \cite{recentlambda}, it was suggested that  quantum Bell
correlations may be a possible way to induce correlations in the CMB sky. However, incorporating inhomogeneities into the
dynamics in a systematic way remains  an
important open question. 

In \cite{recentlambda}, a phenomenological model was adopted which uses the homogeneous temporal fluctuations in $\Lambda$ to model a  quintessence type spatially inhomogeneous scalar field with a potential term
that varies   from realisation to realisation. Using MCMC methods to sample  the cosmological parameter space, and
generate different stochastic realisations, it was shown  that  these CST inspired models agrees
with the observations as well as $\Lambda CDM$ models and in fact does better for the   Baryonic Acoustic Oscillations (BAO)
measurements.  The very extensive and detailed analysis of \cite{recentlambda} sets the stage for direct comparisons
with future observations and heralds an exciting phase of  quantum gravity phenomenology.

\section{Outlook} 
\label{sec:outlook}

CST has come a long way in the last three decades, despite the fact that there are only a few  practitioners who have been able
to dedicate their time to it. Over the last
decade, in particular, there has been a growth of interest with inputs from the wider quantum gravity community. 
This is  heartening, since an  extensive exploration of the theory is required in order to make significant progress. It
is our hope  that this review will spark the interest of the larger quantum gravity  community, and continue what has
been a productive dialogue. 

We have in this review touched upon several open questions, many  of which are challenging but some of which are
straightforward to carry out. We will not summarise these but just pick two that are of utmost importance. One is the
the pursuit of CST-inspired inhomogeneous models of fluctuating $\Lambda$ which can be tested against the most recent observations. The second, on the other side of the quantum gravity spectrum, is the construction from first
principles of a viable quantum dynamics for causal sets. Between these two ends lie myriad interesting
questions. We invite you to join us.

\begin{acknowledgements}
I am indebted to Rafael Sorkin for his deep insights and  vast
knowledge,  that have directly and indirectly shaped  this review. 
I  am also deeply indebted  to Fay Dowker for our  interactions  and
collaborations over the past 25 years, which have helped enrich my understanding of quantum gravity. I am 
grateful  to my  other
collaborators, 
including David Rideout, Joe Henson, Graham Brightwell, 
Petros Wallden, Lisa Glaser, Denjoe O'Connor,  Ian Jubb, Yasaman Yazdi and my students Nomaan X
and Abhishek Mathur, for their active and continuous engagement with
the questions in CST, which have led to fruitful discussions,
arguments, disagreements and  debates over the years. Finally, I would like to thank Yasaman Yazdi and  Stav Zalel  for
a careful reading through the first draft of the manuscript and giving me useful feedback. This research was supported
in part by the Emmy Noether Fellowship (2017\,--\,2018) and also by a  Visiting Fellowship (2019\,--\,2022) at the Perimeter
Institute of Theoretical Physics. 
\end{acknowledgements}

\appendix
\normalsize

\section{Notation and terminology} 
\label{sec:aone}

We list some of the more widely used definitions as well as the abbreviations used in the paper.

\subsection*{Definitions}

  \begin{description}
     \item[\textit{Relation:}] $e,e' \in C$ are said to be \emph{related} if $e \prec e'$ or   $e \prec e'$. 
     \item[\textit{Link:}]  $e \prec e' \in C$ is said to be  a \emph{link} if  $\nexists \, \, e''\in C$ such
      that $e''\neq e,e'$ and $e\prec e''\prec e'$. 
     \item[\textit{Hasse diagram:}] In a \emph{Hasse diagram}, only the nearest neighbour relations or {links} are depicted with the remaining relations following from
 transitivity (see Fig.~\ref{2drandom.fig}). 
     \item[\textit{Valency:}] The \emph{valency} $v(e)$ of an element $e$ in a causal set $C$ is the set of elements in $C$
   that are linked to $e$. 
     \item[\textit{Order Interval:}]  The \emph{order interval} between the pair $e_i, e_j \in C$  is the set $\cAlex[e_i,e_j]
   \equiv \fut(e_i) \cap \past(e_j)$ where $\fut(x), \past(x)$ are the exclusive future and past of $x$.  
     \item[\textit{Labelling:}] A \emph{labelling} of the causal set $C$ of cardinality $n$  is an injective  map $L:C   \rightarrow  \bbN$,
    where $\bbN$ is the set of natural numbers.  
     \item[\textit{Natural Labelling:}] A labelling  $L:C   \rightarrow  \bbN $ is called \emph{natural}
      if $e_i \prec e_j \Rightarrow L(e_i) < L(e_j)$.
     \item[\textit{Total Order:}] A poset $C$ is \emph{totally ordered}  if for each pair $e_i, e_j \in C$, either $e_i \prec
   e_j$ or $e_j \prec e_i$.     
     \item[\textit{Chain:}] A $k$-element set  $C$ is called a \emph{chain} (or \emph{$k$-chain}) if it is a totally ordered
       set, i.e., for every $e_i,e_j \in C$ either $e_i \prec e_j $ or $e_j \prec e_i$.    
     \item[\textit{Length of a chain:}] The \emph{length of a $k$-chain} is  $k-2$.
       
     \item[\textit{Antichain:}] A causal set $C$ is an  \emph{antichain} if no two elements are related to
       each other.
       
     \item[\textit{Inextendible Antichain:}] A subset  $\cA \subseteq  C$ is  an  \emph{inextendible antichain} in $C$
       if it is an antichain and for every 
        element $e \in C \backslash\cA$  (where $\backslash$ is set difference)  either $e \in \past(\cA)$ or $e\in \fut(\cA)$ (see Eq.~(\ref{eq:futpast})).
     \item[\textit{Order Invariant:}] $\cO: \rightarrow \re$ is an \emph{order invariant}  if it is independent of the
        labelling of the causal set $C$. It is possible to generalise from $\re $ to a  more general field, but since this has not been
        explicitly used here, the above definition is sufficient.
        
     \item[\textit{Manifold-like:}] A causal set $C$ is said to be manifold-like if $C$ has a continuum approximation.
     \item[\textit{Alexandrov interval:}] This is the generalised causal diamond in $(M,g)$,   $\Alex[p,q]\equiv
          I^+(p)\cap I^-(q)$, $p,q \in M$.
    
     \item[\textit{Sample Space $\Omega$:}] This is a collection or  space of causal sets.
     \item[\textit{non-locality parameter:}] $\epsilon \equiv {\rk}/{\rc}$  appears in the BD action. 
  \end{description}

\subsection*{Abbreviations in alphabetical order}

    \begin{description}
       \item[BD action:] Benincasa--Dowker action (see Sect.~\ref{ssec:bdaction}). 
       \item[BLMS:] Bombelli, Lee, Meyer and Sorkin's CST proposal \citep{blms}.
       \item[CSG:] Classical Sequential Growth Dynamics  (see Sect.~\ref{ssec:csg}). 
       \item[CST:] Causal Set Theory. 
       \item[GHY:] Gibbons--Hawking--York  (see Sect.~\ref{ssec:ghy}). 
       \item[GNN:] Gaussian Normal Neighbourhood.  
       \item[HKMM theorem:] Hawking--King--McCarthy--Malament theorem (see Sect.~\ref{sec:history}).  
       \item[KR posets:] Kleitman--Rothschild posets (see Sect.~\ref{ssec:haupt}). 
       \item[MCMC:] Markov Chain Monte Carlo (see Sect.~\ref{ssec:partn}). 
       \item[QSG:] Quantum Sequential Growth Dynamics (see Sect.~\ref{ssec:qsg}). 
       \item[RNN:] Riemann Normal Neighbourhood. 
       \item[SJ vacuum:] Sorkin-Johnston vacuum (see Sect.~\ref{ssec:SJ}). 
       \item[SSEE:] Sorkin Spacetime Entanglement Entropy  (see Sect.~\ref{ssec:SSEE}).
    \end{description} 

\bibliographystyle{spbasic}      
\bibliography{Masterref}   

\begin{thebibliography}{165}
\providecommand{\natexlab}[1]{#1}
\providecommand{\url}[1]{{#1}}
\providecommand{\urlprefix}{URL }
\expandafter\ifx\csname urlstyle\endcsname\relax
  \providecommand{\doi}[1]{DOI~\discretionary{}{}{}#1}\else
  \providecommand{\doi}{DOI~\discretionary{}{}{}\begingroup
  \urlstyle{rm}\Url}\fi
\providecommand{\eprint}[2][]{\url{#2}}

\bibitem[{Abajian and Carlip(2018)}]{carlipdrone}
Abajian J, Carlip S (2018) {Dimensional reduction in manifoldlike causal sets}.
  Phys Rev D 97:066007, \doi{10.1103/PhysRevD.97.066007}, \eprint{1710.00938}

\bibitem[{Afshordi et~al(2012)Afshordi, Buck, Dowker, Rideout, Sorkin, and
  Yazdi}]{sj2ddiamond}
Afshordi N, Buck M, Dowker F, Rideout D, Sorkin RD, Yazdi YK (2012) {A Ground
  State for the Causal Diamond in 2 Dimensions}. JHEP 10:088,
  \doi{10.1007/JHEP10(2012)088}, \eprint{1207.7101}

\bibitem[{Aghili et~al(2018)Aghili, Bombelli, and Pilgrim}]{bomemad}
Aghili M, Bombelli L, Pilgrim BB (2018) {Statistical Lorentzian geometry and
  the dimensionality of Minkowski space}. arXiv e-prints \eprint{1807.08701}

\bibitem[{Ahmed and Rideout(2010)}]{davidmaqbool}
Ahmed M, Rideout D (2010) {Indications of de Sitter Spacetime from Classical
  Sequential Growth Dynamics of Causal Sets}. Phys Rev D 81:083528,
  \doi{10.1103/PhysRevD.81.083528}, \eprint{0909.4771}

\bibitem[{Ahmed and Sorkin(2013)}]{eptwo}
Ahmed M, Sorkin R (2013) {Everpresent $\Lambda$. {II}. {S}tructural stability}.
  Phys Rev D 87:063515, \doi{10.1103/PhysRevD.87.063515}, \eprint{1210.2589}

\bibitem[{Ahmed et~al(2004)Ahmed, Dodelson, Greene, and Sorkin}]{lambdathree}
Ahmed M, Dodelson S, Greene PB, Sorkin R (2004) Everpresent $\Lambda$. Phys Rev
  D 69:103523, \doi{10.1103/PhysRevD.69.103523}

\bibitem[{Ashtekar and Pullin(2017)}]{ashtekar}
Ashtekar A, Pullin J (2017) {Applications}. In: Ashtekar A, Pullin J (eds) Loop
  Quantum Gravity: The First 30 Years, World Scientific, p 181,
  \doi{10.1142/9789813220003_others03}

\bibitem[{Aslanbeigi et~al(2014)Aslanbeigi, Saravani, and Sorkin}]{gendalem}
Aslanbeigi S, Saravani M, Sorkin RD (2014) {Generalized causal set
  d`Alembertians}. JHEP 06:024, \doi{10.1007/JHEP06(2014)024},
  \eprint{1403.1622}

\bibitem[{Bachmat(2007)}]{bachmat}
Bachmat E (2007) {Discrete spacetime and its applications}. arXiv e-prints
  \eprint{gr-qc/0702140}

\bibitem[{Barrow(2007)}]{barrow}
Barrow JD (2007) {A Strong Constraint on Ever-Present Lambda}. Phys Rev D
  75:067301, \doi{10.1103/PhysRevD.75.067301}, \eprint{gr-qc/0612128}

\bibitem[{Beem et~al(1996)Beem, Ehrlich, and Easley}]{BE}
Beem J, Ehrlich P, Easley K (1996) Global {L}orentzian Geometry. Marcel Dekker,
  New York

\bibitem[{Belenchia et~al(2015)Belenchia, Benincasa, and
  Liberati}]{nonlocalfive}
Belenchia A, Benincasa DMT, Liberati S (2015) {Nonlocal Scalar Quantum Field
  Theory from Causal Sets}. JHEP 03:036, \doi{10.1007/JHEP03(2015)036},
  \eprint{1411.6513}

\bibitem[{Belenchia et~al(2016{\natexlab{a}})Belenchia, Benincasa, and
  Dowker}]{bbd}
Belenchia A, Benincasa DMT, Dowker F (2016{\natexlab{a}}) {The continuum limit
  of a 4-dimensional causal set scalar d'Alembertian}. Class Quantum Grav
  33:245018, \doi{10.1088/0264-9381/33/24/245018}, \eprint{1510.04656}

\bibitem[{Belenchia et~al(2016{\natexlab{b}})Belenchia, Benincasa, Liberati,
  Marin, Marino, and Ortolan}]{nonlocalqftfour}
Belenchia A, Benincasa DMT, Liberati S, Marin F, Marino F, Ortolan A
  (2016{\natexlab{b}}) {Testing Quantum Gravity Induced Nonlocality via
  Optomechanical Quantum Oscillators}. Phys Rev Lett 116:161303,
  \doi{10.1103/PhysRevLett.116.161303}, \eprint{1512.02083}

\bibitem[{Belenchia et~al(2016{\natexlab{c}})Belenchia, Benincasa, Marciano,
  and Modesto}]{diondr}
Belenchia A, Benincasa DMT, Marciano A, Modesto L (2016{\natexlab{c}})
  {Spectral Dimension from Nonlocal Dynamics on Causal Sets}. Phys Rev D
  93:044017, \doi{10.1103/PhysRevD.93.044017}, \eprint{1507.00330}

\bibitem[{Bell and Kort\'e(2016)}]{weyl}
Bell JL, Kort\'e H (2016) Hermann Weyl. In: Zalta EN (ed) The Stanford
  Encyclopedia of Philosophy, winter 2016 edn, Metaphysics Research Lab,
  Stanford University,
  \urlprefix\url{https://plato.stanford.edu/archives/win2016/entries/weyl/}

\bibitem[{Benincasa and Dowker(2010)}]{bd}
Benincasa DM, Dowker F (2010) The Scalar Curvature of a Causal Set. PhysRevLett
  104:181301, \doi{10.1103/PhysRevLett.104.181301}

\bibitem[{Benincasa et~al(2011)Benincasa, Dowker, and Schmitzer}]{gaussbonnet}
Benincasa DM, Dowker F, Schmitzer B (2011) The Random Discrete Action for
  2-Dimensional Spacetime. Class Quantum Grav 28:105018,
  \doi{10.1088/0264-9381/28/10/105018}

\bibitem[{Benincasa(2013)}]{dionthesis}
Benincasa DMT (2013) The action of a casual set. PhD thesis, Imperial College
  London

\bibitem[{Bolognesi and Lamb(2016)}]{nodedegree}
Bolognesi T, Lamb A (2016) {Simple indicators for Lorentzian causets}. Class
  Quantum Grav 33:185004, \doi{10.1088/0264-9381/33/18/185004},
  \eprint{1407.1649}

\bibitem[{Bombelli(1987)}]{bomthesis}
Bombelli L (1987) Space-time as a Causal Set. PhD thesis, Syracuse University

\bibitem[{Bombelli(2000)}]{bomclose}
Bombelli L (2000) {Statistical Lorentzian geometry and the closeness of
  Lorentzian manifolds}. J Math Phys 41:6944--6958, \doi{10.1063/1.1288494},
  \eprint{gr-qc/0002053}

\bibitem[{Bombelli and Meyer(1989)}]{bommeyer}
Bombelli L, Meyer DA (1989) {The Origin of Lorentzian Geometry}. Phys Lett A
  141:226--228, \doi{10.1016/0375-9601(89)90474-X}

\bibitem[{Bombelli and Noldus(2004)}]{bomnoldus}
Bombelli L, Noldus J (2004) {The Moduli space of isometry classes of globally
  hyperbolic space-times}. Class Quantum Grav 21:4429--4454,
  \doi{10.1088/0264-9381/21/18/010}, \eprint{gr-qc/0402049}

\bibitem[{Bombelli et~al(1986)Bombelli, Koul, Lee, and Sorkin}]{bklsEE}
Bombelli L, Koul RK, Lee J, Sorkin RD (1986) {A Quantum Source of Entropy for
  Black Holes}. Phys Rev D 34:373--383, \doi{10.1103/PhysRevD.34.373}

\bibitem[{Bombelli et~al(1987)Bombelli, Lee, Meyer, and Sorkin}]{blms}
Bombelli L, Lee J, Meyer D, Sorkin R (1987) {Space-Time as a Causal Set}. Phys
  Rev Lett 59:521--524, \doi{10.1103/PhysRevLett.59.521}

\bibitem[{Bombelli et~al(2009)Bombelli, Henson, and Sorkin}]{bomhensor}
Bombelli L, Henson J, Sorkin RD (2009) Discreteness without symmetry breaking:
  {A} {Theorem}. ModPhysLett A24:2579--2587, \doi{10.1142/S0217732309031958}

\bibitem[{Bombelli et~al(2012)Bombelli, Noldus, and Tafoya}]{bomnoldustwo}
Bombelli L, Noldus J, Tafoya J (2012) {Lorentzian Manifolds and Causal Sets as
  Partially Ordered Measure Spaces}. arXiv e-prints \eprint{1212.0601}

\bibitem[{Brightwell and Georgiou(2010)}]{grnick}
Brightwell G, Georgiou N (2010) Continuum limits for classical sequential
  growth models. Rand Struct Alg 36:218--250

\bibitem[{Brightwell and Gregory(1991)}]{bg}
Brightwell G, Gregory R (1991) {The Structure of random discrete space-time}.
  Phys Rev Lett 66:260--263, \doi{10.1103/PhysRevLett.66.260}

\bibitem[{Brightwell and Luczak(2011)}]{gmtwo}
Brightwell G, Luczak M (2011) Order-invariant measures on causal sets. Ann Appl
  Probab

\bibitem[{Brightwell and Luczak(2012)}]{gmone}
Brightwell G, Luczak M (2012) Order-invariant measures on fixed causal sets.
  Comb Prob Comput

\bibitem[{Brightwell and Luczak(2015)}]{grahammalwina}
Brightwell G, Luczak M (2015) {The mathematics of causal sets}. arXiv e-prints
  \eprint{1510.05612}

\bibitem[{Brightwell et~al(2003)Brightwell, Dowker, Garcia, Henson, and
  Sorkin}]{observables}
Brightwell G, Dowker HF, Garcia RS, Henson J, Sorkin RD (2003) Observables in
  causal set cosmology. PhysRev D67:084031, \doi{10.1103/PhysRevD.67.084031}

\bibitem[{Brightwell et~al(2008)Brightwell, Henson, and Surya}]{2dorders}
Brightwell G, Henson J, Surya S (2008) A 2D model of causal set quantum
  gravity: the emergence of the continuum. Class Quantum Grav 25:105025,
  \doi{10.1088/0264-9381/25/10/105025}

\bibitem[{Brum and Fredenhagen(2014)}]{bf14}
Brum M, Fredenhagen K (2014) {`Vacuum-like' Hadamard states for quantum fields
  on curved spacetimes}. Class Quantum Grav 31:025024,
  \doi{10.1088/0264-9381/31/2/025024}, \eprint{1307.0482}

\bibitem[{Buck et~al(2015)Buck, Dowker, Jubb, and Surya}]{bdjs}
Buck M, Dowker F, Jubb I, Surya S (2015) {Boundary Terms for Causal Sets}.
  Class Quantum Grav 32:205004, \doi{10.1088/0264-9381/32/20/205004},
  \eprint{1502.05388}

\bibitem[{Buck et~al(2017)Buck, Dowker, Jubb, and Sorkin}]{sjtrousers}
Buck M, Dowker F, Jubb I, Sorkin R (2017) {The Sorkin--Johnston state in a
  patch of the trousers spacetime}. Class Quantum Grav 34:055002,
  \doi{10.1088/1361-6382/aa589c}, \eprint{1609.03573}

\bibitem[{Carlip(2017)}]{carlipdr}
Carlip S (2017) {Dimension and Dimensional Reduction in Quantum Gravity}. Class
  Quantum Grav 34:193001, \doi{10.1088/1361-6382/aa8535}, \eprint{1705.05417}

\bibitem[{Christ et~al(1982)Christ, Friedberg, and Lee}]{tdlee}
Christ NH, Friedberg R, Lee TD (1982) {Random Lattice Field Theory: General
  Formulation}. Nucl Phys B 202:89, \doi{10.1016/0550-3213(82)90222-X}

\bibitem[{Contaldi et~al(2010)Contaldi, Dowker, and Philpott}]{faypolarisation}
Contaldi CR, Dowker F, Philpott L (2010) {Polarization Diffusion from Spacetime
  Uncertainty}. Class Quantum Grav 27:172001,
  \doi{10.1088/0264-9381/27/17/172001}, \eprint{1001.4545}

\bibitem[{Cort{\^e}s and Smolin(2014)}]{lee}
Cort{\^e}s M, Smolin L (2014) {Quantum energetic causal sets}. Phys Rev D
  90:044035, \doi{10.1103/PhysRevD.90.044035}, \eprint{1308.2206}

\bibitem[{Cunningham(2018{\natexlab{a}})}]{tlbdry}
Cunningham W (2018{\natexlab{a}}) {Inference of Boundaries in Causal Sets}.
  Class Quantum Grav 35:094002, \doi{10.1088/1361-6382/aaadc4},
  \eprint{1710.09705}

\bibitem[{Cunningham(2018{\natexlab{b}})}]{willthesis}
Cunningham WJ (2018{\natexlab{b}}) {High Performance Algorithms for Quantum
  Gravity and Cosmology}. PhD thesis, Northeastern U., \eprint{1805.04463}

\bibitem[{Daughton(1993)}]{daughton}
Daughton AR (1993) The Recovery of Locality for Causal Sets and Related Topics.
  PhD thesis, Syracuse University

\bibitem[{Dhar(1978)}]{dharone}
Dhar D (1978) Entropy and phase transitions in partially ordered sets. J Math
  Phys 19(8)

\bibitem[{Dhar(1980)}]{dhartwo}
Dhar D (1980) Asymptotic enumeration of partially ordered sets. Pacific J Math
  90(2)

\bibitem[{Diestel and Uhl(1977)}]{du}
Diestel J, Uhl J (1977) Vector Measures. American Mathematical Society

\bibitem[{Dou and Sorkin(2003)}]{dousorkin}
Dou D, Sorkin RD (2003) {Black hole entropy as causal links}. Found Phys
  33:279--296, \doi{10.1023/A:1023781022519}, \eprint{gr-qc/0302009}

\bibitem[{Dowker(2005)}]{fayreview}
Dowker F (2005) {Causal sets and the deep structure of spacetime}. In: Ashtekar
  A (ed) 100 Years Of Relativity: space-time structure: Einstein and beyond,
  World Scientific, pp 445--464, \doi{10.1142/9789812700988_0016},
  \eprint{gr-qc/0508109}

\bibitem[{Dowker and Ghazi-Tabatabai(2008)}]{kochenspecker}
Dowker F, Ghazi-Tabatabai Y (2008) {The Kochen-Specker Theorem Revisited in
  Quantum Measure Theory}. J Phys A 41:105301,
  \doi{10.1088/1751-8113/41/10/105301}, \eprint{0711.0894}

\bibitem[{Dowker and Glaser(2013)}]{dg}
Dowker F, Glaser L (2013) Causal set d'Alembertians for various dimensions.
  Class Quantum Grav 30:195016

\bibitem[{Dowker and Surya(2006)}]{observablesds}
Dowker F, Surya S (2006) {Observables in extendcarliped percolation models of
  causal set cosmology}. Class Quantum Grav 23:1381--1390,
  \doi{10.1088/0264-9381/23/4/018}, \eprint{gr-qc/0504069}

\bibitem[{Dowker and Zalel(2017)}]{faystav}
Dowker F, Zalel S (2017) {Evolution of Universes in Causal Set Cosmology}.
  Comptes Rendus Physique 18:246--253, \doi{10.1016/j.crhy.2017.03.002},
  \eprint{1703.07556}

\bibitem[{Dowker et~al(2004)Dowker, Henson, and Sorkin}]{swerves}
Dowker F, Henson J, Sorkin RD (2004) {Quantum gravity phenomenology, Lorentz
  invariance and discreteness}. Mod Phys Lett A 19:1829--1840,
  \doi{10.1142/S0217732304015026}, \eprint{gr-qc/0311055}

\bibitem[{Dowker et~al(2010{\natexlab{a}})Dowker, Henson, and Sorkin}]{dhstwo}
Dowker F, Henson J, Sorkin R (2010{\natexlab{a}}) {Discreteness and the
  transmission of light from distant sources}. Phys Rev D 82:104048,
  \doi{10.1103/PhysRevD.82.104048}, \eprint{1009.3058}

\bibitem[{Dowker et~al(2010{\natexlab{b}})Dowker, Johnston, and Sorkin}]{hhh}
Dowker F, Johnston S, Sorkin RD (2010{\natexlab{b}}) {Hilbert Spaces from Path
  Integrals}. J Phys A 43:275302, \doi{10.1088/1751-8113/43/27/275302},
  \eprint{1002.0589}

\bibitem[{Dowker et~al(2010{\natexlab{c}})Dowker, Johnston, and Surya}]{djs}
Dowker F, Johnston S, Surya S (2010{\natexlab{c}}) {On extending the Quantum
  Measure}. J Phys A 43:505305, \doi{10.1088/1751-8113/43/50/505305},
  \eprint{1007.2725}

\bibitem[{Dowker et~al(2017)Dowker, Surya, and X}]{dsx}
Dowker F, Surya S, X N (2017) {Scalar Field Green Functions on Causal Sets}.
  Class Quantum Grav 34:124002, \doi{10.1088/1361-6382/aa6bc7},
  \eprint{1701.07212}

\bibitem[{Eichhorn(2018)}]{astridcg}
Eichhorn A (2018) {Towards coarse graining of discrete Lorentzian quantum
  gravity}. Class Quantum Grav 35:044001, \doi{10.1088/1361-6382/aaa0a3},
  \eprint{1709.10419}

\bibitem[{Eichhorn and Mizera(2014)}]{em}
Eichhorn A, Mizera S (2014) {Spectral dimension in causal set quantum gravity}.
  Class Quantum Grav 31:125007, \doi{10.1088/0264-9381/31/12/125007},
  \eprint{1311.2530}

\bibitem[{Eichhorn et~al(2017)Eichhorn, Mizera, and Surya}]{ems}
Eichhorn A, Mizera S, Surya S (2017) {Echoes of Asymptotic Silence in Causal
  Set Quantum Gravity}. Class Quantum Grav 34(16):16LT01,
  \doi{10.1088/1361-6382/aa7d1b}, \eprint{1703.08454}

\bibitem[{Eichhorn et~al(2018)Eichhorn, Surya, and Versteegen}]{esv}
Eichhorn A, Surya S, Versteegen F (2018) {Induced Spatial Geometry from Causal
  Structure}. arXiv e-prints \eprint{1809.06192}

\bibitem[{Eichhorn et~al(2019)Eichhorn, Surya, and Versteegen}]{esvsd}
Eichhorn A, Surya S, Versteegen F (2019) {Spectral dimension on spatial
  hypersurfaces in causal set quantum gravity}, arXiv:1905.13498

\bibitem[{El-Zahar and Sauer(1988)}]{es}
El-Zahar MH, Sauer NW (1988) Asymptotic Enumeration of Two-dimensional Posets.
  Order 5:239

\bibitem[{Fewster(2018)}]{fewsterart}
Fewster CJ (2018) {The art of the state}. Int J Mod Phys D 27:1843007,
  \doi{10.1142/S0218271818430071}, \eprint{1803.06836}

\bibitem[{Fewster and Verch(2012)}]{fv12}
Fewster CJ, Verch R (2012) {On a Recent Construction of 'Vacuum-like' Quantum
  Field States in Curved Spacetime}. Class Quantum Grav 29:205017,
  \doi{10.1088/0264-9381/29/20/205017}, \eprint{1206.1562}

\bibitem[{Feynman(1944)}]{Feynman}
Feynman R (1944) The Character of Physical Law. Modern Library

\bibitem[{Finkelstein(1969)}]{finkelstein}
Finkelstein D (1969) Space-Time Code. Phys Rev 184:1261--1271,
  \doi{10.1103/PhysRev.184.1261}

\bibitem[{Gibbons and Solodukhin(2007)}]{gs}
Gibbons GW, Solodukhin SN (2007) {The Geometry of small causal diamonds}. Phys
  Lett B649:317--324, \doi{10.1016/j.physletb.2007.03.068},
  \eprint{hep-th/0703098}

\bibitem[{Glaser(2014)}]{glaser}
Glaser L (2014) A closed form expression for the causal set d’Alembertian.
  Class Quantum Grav 31:095007

\bibitem[{Glaser(2018)}]{ising}
Glaser L (2018) {The {Ising} model coupled to 2d orders}. Class Quantum Grav
  35:084001, \doi{10.1088/1361-6382/aab139}, \eprint{1802.02519}

\bibitem[{Glaser and Surya(2013)}]{intervals}
Glaser L, Surya S (2013) {Towards a Definition of Locality in a Manifoldlike
  Causal Set}. Phys Rev D 88:124026, \doi{10.1103/PhysRevD.88.124026},
  \eprint{1309.3403}

\bibitem[{Glaser and Surya(2016)}]{2dhh}
Glaser L, Surya S (2016) {The Hartle--Hawking wave function in 2D causal set
  quantum gravity}. Class Quantum Grav 33:065003,
  \doi{10.1088/0264-9381/33/6/065003}, \eprint{1410.8775}

\bibitem[{Glaser et~al(2018)Glaser, O'Connor, and Surya}]{fss}
Glaser L, O'Connor D, Surya S (2018) {Finite Size Scaling in 2d Causal Set
  Quantum Gravity}. Class Quantum Grav 35:045006,
  \doi{10.1088/1361-6382/aa9540}, \eprint{1706.06432}

\bibitem[{Greene and Plesser(1991)}]{string}
Greene BR, Plesser MR (1991) {Mirror manifolds: A Brief review and progress
  report}. In: {2nd International Symposium on Particles, Strings and Cosmology
  (PASCOS 1991) Boston, Massachusetts, March 25-30, 1991}, pp 0648--666,
  \eprint{hep-th/9110014}

\bibitem[{Hawking and Ellis(1973)}]{HE}
Hawking S, Ellis G (1973) Large scale structure of spacetime. Cambridge
  University Press

\bibitem[{Hawking et~al(1976)Hawking, King, and McCarthy}]{hkm}
Hawking S, King A, McCarthy P (1976) A New Topology for Curved Space-Time Which
  Incorporates the Causal, Differential, and Conformal Structures. J Math Phys

\bibitem[{Hemion(1988)}]{hemion}
Hemion G (1988) A Quantum Theory Of Space And Time. Int J Theor Phys 27:1145

\bibitem[{Henson(2005)}]{joecausality}
Henson J (2005) {Comparing causality principles}. Stud Hist Phil Sci B
  36:519--543, \doi{10.1016/j.shpsb.2005.04.003}, \eprint{quant-ph/0410051}

\bibitem[{Henson(2006{\natexlab{a}})}]{joeinterval}
Henson J (2006{\natexlab{a}}) {Constructing an interval of Minkowski space from
  a causal set}. Class Quantum Grav 23:L29--L35,
  \doi{10.1088/0264-9381/23/4/L02}, \eprint{gr-qc/0601069}

\bibitem[{Henson(2006{\natexlab{b}})}]{joereviewone}
Henson J (2006{\natexlab{b}}) {The Causal set approach to quantum gravity}. In:
  Oriti D (ed) Approaches to quantum gravity, Cambridge University Press,
  Cambridge, pp 393--413, \eprint{gr-qc/0601121}

\bibitem[{Henson(2010)}]{joereviewtwo}
Henson J (2010) {Discovering the Discrete Universe}. In: {Proceedings,
  Foundations of Space and Time: Reflections on Quantum Gravity: Cape Town,
  South Africa}, \eprint{1003.5890}

\bibitem[{Henson(2011)}]{joequantumbell}
Henson J (2011) {Causality, Bell's theorem, and Ontic Definiteness}. arXiv
  e-prints \eprint{1102.2855}

\bibitem[{Henson et~al(2017)Henson, Rideout, Sorkin, and Surya}]{onset}
Henson J, Rideout D, Sorkin RD, Surya S (2017) Onset of the Asymptotic Regime
  for (Uniformly Random) Finite Orders. Experimental Mathematics
  26(3):253--266, \doi{10.1080/10586458.2016.1158134}

\bibitem[{Johnston(2008)}]{johnston}
Johnston S (2008) {Particle propagators on discrete spacetime}. Class Quantum
  Grav 25:202001, \doi{10.1088/0264-9381/25/20/202001}, \eprint{0806.3083}

\bibitem[{Johnston(2009)}]{johnstontwo}
Johnston S (2009) {Feynman Propagator for a Free Scalar Field on a Causal Set}.
  Phys Rev Lett 103:180401, \doi{10.1103/PhysRevLett.103.180401},
  \eprint{0909.0944}

\bibitem[{Johnston(2010)}]{johnstonthesis}
Johnston SP (2010) {Quantum Fields on Causal Sets}. PhD thesis, Imperial Coll.,
  London, \eprint{1010.5514}

\bibitem[{Jubb(2017)}]{jubb}
Jubb I (2017) {The Geometry of Small Causal Cones}. Class Quantum Grav
  34:094005, \doi{10.1088/1361-6382/aa68b7}, \eprint{1611.00785}

\bibitem[{Jubb et~al(2017)Jubb, Samuel, Sorkin, and Surya}]{jsss}
Jubb I, Samuel J, Sorkin R, Surya S (2017) {Boundary and Corner Terms in the
  Action for General Relativity}. Class Quantum Grav 34:065006,
  \doi{10.1088/1361-6382/aa6014}, \eprint{1612.00149}

\bibitem[{Kaloper and Mattingly(2006)}]{swerveconstraint}
Kaloper N, Mattingly D (2006) {Low energy bounds on Poincare violation in
  causal set theory}. Phys Rev D 74:106001, \doi{10.1103/PhysRevD.74.106001},
  \eprint{astro-ph/0607485}

\bibitem[{Khetrapal and Surya(2013)}]{ks}
Khetrapal S, Surya S (2013) {Boundary Term Contribution to the Volume of a
  Small Causal Diamond}. Class Quantum Grav 30:065005,
  \doi{10.1088/0264-9381/30/6/065005}, \eprint{1212.0629}

\bibitem[{Kleitman and Rothschild(1975)}]{kr}
Kleitman DJ, Rothschild BL (1975) Asymptotic enumeration of partial orders on a
  finite set. Transactions of the American Mathematical Society 205:205--220

\bibitem[{Kronheimer and Penrose(1967)}]{kp}
Kronheimer E, Penrose R (1967) On the Structure of causal spaces. Proc Camb
  Phil Soc 63:481

\bibitem[{Lehner et~al(2016)Lehner, Myers, Poisson, and Sorkin}]{lmps}
Lehner L, Myers RC, Poisson E, Sorkin RD (2016) {Gravitational action with null
  boundaries}. Phys Rev D 94:084046, \doi{10.1103/PhysRevD.94.084046},
  \eprint{1609.00207}

\bibitem[{Levichev(1987)}]{levichev}
Levichev AV (1987) Prescribing the conformal geometry of a Lorentz manifold by
  means of its causal structure. Sov Math Dokl 35:452--455

\bibitem[{Liberati and Mattingly(2016)}]{li}
Liberati S, Mattingly D (2016) {Lorentz breaking effective field theory models
  for matter and gravity: theory and observational constraints}. In: Peron R,
  Colpi M, Gorini V, Moschella U (eds) Gravity: Where Do We Stand?, Springer,
  pp 367--417, \doi{10.1007/978-3-319-20224-2_11}, \eprint{1208.1071}

\bibitem[{Loomis and Carlip(2018)}]{carliploomis}
Loomis SP, Carlip S (2018) {Suppression of non-manifold-like sets in the causal
  set path integral}. Class Quantum Grav 35:024002,
  \doi{10.1088/1361-6382/aa980b}, \eprint{1709.00064}

\bibitem[{Louko and Sorkin(1997)}]{sorkinlouko}
Louko J, Sorkin RD (1997) Complex actions in two-dimensional topology change.
  Class Quantum Grav 14:179--204, \doi{10.1088/0264-9381/14/1/018}

\bibitem[{Major et~al(2007)Major, Rideout, and Surya}]{homology}
Major S, Rideout D, Surya S (2007) {On Recovering continuum topology from a
  causal set}. J Math Phys 48:032501, \doi{10.1063/1.2435599},
  \eprint{gr-qc/0604124}

\bibitem[{Major et~al(2009)Major, Rideout, and Surya}]{stablehomology}
Major S, Rideout D, Surya S (2009) {Stable Homology as an Indicator of
  Manifoldlikeness in Causal Set Theory}. Class Quantum Grav 26:175008,
  \doi{10.1088/0264-9381/26/17/175008}, \eprint{0902.0434}

\bibitem[{Major et~al(2006)Major, Rideout, and Surya}]{antichain}
Major SA, Rideout D, Surya S (2006) Spatial hypersurfaces in causal set
  cosmology. Class Quantum Grav 23:4743, \doi{10.1088/0264-9381/23/14/011}

\bibitem[{Malament(1977)}]{malament}
Malament DB (1977) The class of continuous timelike curves determines the
  topology of spacetime. J Math Phys 18:1399--1404, \doi{10.1063/1.523436}

\bibitem[{Marr(2007)}]{sarahthesis}
Marr S (2007) Black Hole entropy from causal sets. PhD thesis, Imperial College

\bibitem[{Martin et~al(2001)Martin, O'Connor, Rideout, and Sorkin}]{csgrg}
Martin X, O'Connor D, Rideout DP, Sorkin RD (2001) {On the `renormalization'
  transformations induced by cycles of expansion and contraction in causal set
  cosmology}. Phys Rev D 63:084026, \doi{10.1103/PhysRevD.63.084026},
  \eprint{gr-qc/0009063}

\bibitem[{Mathur and Surya(2019)}]{mathursurya}
Mathur A, Surya S (2019) {Sorkin-Johnston vacuum for a massive scalar field in
  the 2D causal diamond}. Phys Rev D100(4):045007,
  \doi{10.1103/PhysRevD.100.045007}, \eprint{1906.07952}

\bibitem[{Meyer(1988)}]{meyer}
Meyer D (1988) The Dimension of Causal Sets,. PhD thesis, M.I.T.

\bibitem[{Munkres(1984)}]{munkres}
Munkres JR (1984) Elements of algebraic topology. Addison-Wesley

\bibitem[{Myrheim(1978)}]{myrheim}
Myrheim J (1978) Statistical Geometry. Tech. Rep. CERN-TH-2538, CERN

\bibitem[{Noldus(2002)}]{noldustwo}
Noldus J (2002) {A new topology on the space of Lorentzian metrics on a fixed
  manifold}. Class Quantum Grav 19:6075--6107,
  \doi{10.1088/0264-9381/19/23/313}, \eprint{1104.1811}

\bibitem[{Noldus(2004)}]{noldusone}
Noldus J (2004) {A Lorentzian Lipschitz, Gromov-Hausdoff notion of distance}.
  Class Quantum Grav 21:839--850, \doi{10.1088/0264-9381/21/4/007},
  \eprint{gr-qc/0308074}

\bibitem[{Parrikar and Surya(2011)}]{ps}
Parrikar O, Surya S (2011) {Causal Topology in Future and Past Distinguishing
  Spacetimes}. Class Quantum Grav 28:155020,
  \doi{10.1088/0264-9381/28/15/155020}, \eprint{1102.0936}

\bibitem[{Penrose(1972)}]{penrose}
Penrose R (1972) Techniques of Differential Topology in Relativity. SIAM

\bibitem[{Petersen(2006)}]{petersen}
Petersen P (2006) Riemannian Geometry, 2nd edn. Springer,
  \doi{10.1007/978-0-387-29403-2}

\bibitem[{Philpott et~al(2009)Philpott, Dowker, and Sorkin}]{swervestwo}
Philpott L, Dowker F, Sorkin RD (2009) {Energy-momentum diffusion from
  spacetime discreteness}. Phys Rev D 79:124047,
  \doi{10.1103/PhysRevD.79.124047}, \eprint{0810.5591}

\bibitem[{Promel et~al(2001)Promel, Steger, and Taraz}]{pst}
Promel H, Steger A, Taraz A (2001) {Phase Transitions in the Evolution of
  Partial Orders}. J Combin Theory, Series A 94:230

\bibitem[{Reid(2003)}]{reid}
Reid DD (2003) {The Manifold dimension of a causal set: Tests in conformally
  flat space-times}. Phys Rev D 67:024034, \doi{10.1103/PhysRevD.67.024034},
  \eprint{gr-qc/0207103}

\bibitem[{Rideout and Sorkin(2000{\natexlab{a}})}]{csg}
Rideout D, Sorkin R (2000{\natexlab{a}}) A Classical sequential growth dynamics
  for causal sets. PhysRev D61:024002, \doi{10.1103/PhysRevD.61.024002}

\bibitem[{Rideout and Wallden(2009)}]{rw}
Rideout D, Wallden P (2009) {Spacelike distance from discrete causal order}.
  Class Quantum Grav 26:155013, \doi{10.1088/0264-9381/26/15/155013},
  \eprint{0810.1768}

\bibitem[{Rideout and Zohren(2006)}]{rz}
Rideout D, Zohren S (2006) {Evidence for an entropy bound from fundamentally
  discrete gravity}. Class Quantum Grav 23:6195--6213,
  \doi{10.1088/0264-9381/23/22/008}, \eprint{gr-qc/0606065}

\bibitem[{Rideout(2001)}]{davidthesis}
Rideout DP (2001) {Dynamics of causal sets}. PhD thesis, Syracuse U.

\bibitem[{Rideout and Sorkin(2000{\natexlab{b}})}]{csgone}
Rideout DP, Sorkin RD (2000{\natexlab{b}}) {A Classical sequential growth
  dynamics for causal sets}. Phys Rev D 61:024002,
  \doi{10.1103/PhysRevD.61.024002}, \eprint{gr-qc/9904062}

\bibitem[{Rideout and Sorkin(2001)}]{csgtwo}
Rideout DP, Sorkin RD (2001) {Evidence for a continuum limit in causal set
  dynamics}. Phys Rev D 63:104011, \doi{10.1103/PhysRevD.63.104011},
  \eprint{gr-qc/0003117}

\bibitem[{Riemann(1873)}]{Riemann}
Riemann B (1873) On the Hypotheses Which Lie at the Bases of Geometry. Nature
  VIII(183, 184):14--17, 36, 37, \doi{10.1038/008036a0}, translated by W.~K.
  Clifford from Vol.~VIII of the G\"ottingen Abhandlungen

\bibitem[{Robb(1914)}]{robbone}
Robb A (1914) A Theory of Time and Space. Cambridge University Press

\bibitem[{Robb(1936)}]{robbtwo}
Robb A (1936) Geometry Of Time And Space. At The University Press

\bibitem[{Roy et~al(2013)Roy, Sinha, and Surya}]{rss}
Roy M, Sinha D, Surya S (2013) {Discrete geometry of a small causal diamond}.
  Phys Rev D 87:044046, \doi{10.1103/PhysRevD.87.044046}, \eprint{1212.0631}

\bibitem[{Salgado(2002)}]{salgadoqm}
Salgado RB (2002) {Some identities for the quantum measure and its
  generalizations}. Mod Phys Lett A 17:711--728,
  \doi{10.1142/S0217732302007041}, \eprint{gr-qc/9903015}

\bibitem[{Salgado(2008)}]{salgado}
Salgado RB (2008) Toward a Quantum Dynamics for Causal Sets. PhD thesis,
  Syracuse University

\bibitem[{Samuel and Sinha(2006)}]{samsupurna}
Samuel J, Sinha S (2006) {Surface tension and the cosmological constant}. Phys
  Rev Lett 97:161302, \doi{10.1103/PhysRevLett.97.161302},
  \eprint{cond-mat/0603804}

\bibitem[{Saravani and Afshordi(2017)}]{darkmatter}
Saravani M, Afshordi N (2017) Off-shell dark matter: {A} cosmological relic of
  quantum gravity. Phys Rev D 95:043514, \doi{10.1103/PhysRevD.95.043514}

\bibitem[{Saravani and Aslanbeigi(2014)}]{aspoisson}
Saravani M, Aslanbeigi S (2014) {On the Causal Set-Continuum Correspondence}.
  Class Quantum Grav 31:205013, \doi{10.1088/0264-9381/31/20/205013},
  \eprint{1403.6429}

\bibitem[{Saravani et~al(2014)Saravani, Sorkin, and Yazdi}]{yasamaneecont}
Saravani M, Sorkin RD, Yazdi YK (2014) {Spacetime entanglement entropy in 1 + 1
  dimensions}. Class Quantum Grav 31:214006,
  \doi{10.1088/0264-9381/31/21/214006}, \eprint{1311.7146}

\bibitem[{Sorkin(1991)}]{lambdaone}
Sorkin RD (1991) Spacetime and Causal Sets. In: D'Olivo JC (ed) Relativity and
  Gravitation: Classical and Quantum, World Scientific, Singapore, pp 150--173,
  proceedings of the SILARG VII Conference, Cocoyocan, Mexico

\bibitem[{Sorkin(1994)}]{qmeasureone}
Sorkin RD (1994) {Quantum mechanics as quantum measure theory}. Mod Phys Lett A
  9:3119--3128, \doi{10.1142/S021773239400294X}, \eprint{gr-qc/9401003}

\bibitem[{Sorkin(1995)}]{qmeasuretwo}
Sorkin RD (1995) {Quantum measure theory and its interpretation}. In: {Physics
  and experiments with linear colliders. Proceedings, 3rd Workshop,
  Morioka-Appi, Japan, September 8-12, 1995. Vol. 1, 2}, \eprint{gr-qc/9507057}

\bibitem[{Sorkin(1997)}]{lambdatwo}
Sorkin RD (1997) {Forks in the road, on the way to quantum gravity}. Int J
  Theor Phys 36:2759--2781, \doi{10.1007/BF02435709}, \eprint{gr-qc/9706002}

\bibitem[{Sorkin(2005{\natexlab{a}})}]{sorkinledl}
Sorkin RD (2005{\natexlab{a}}) {Big extra dimensions make lambda too small}.
  Braz J Phys 35:280--283, \doi{10.1590/S0103-97332005000200012},
  \eprint{gr-qc/0503057}

\bibitem[{Sorkin(2005{\natexlab{b}})}]{valdivia}
Sorkin RD (2005{\natexlab{b}}) Causal sets: Discrete gravity. In: Gomberoff A,
  Marolf D (eds) Proceedings of the Valdivia Summer School, New York, Springer,
  Series of the Centro de Estudios Cientificos de Santiago),
  \eprint{gr-qc/0309009}

\bibitem[{Sorkin(2007{\natexlab{a}})}]{sorkinalogic}
Sorkin RD (2007{\natexlab{a}}) {An Exercise in 'anhomomorphic logic'}. J Phys
  Conf Ser 67:012018, \doi{10.1088/1742-6596/67/1/012018},
  \eprint{quant-ph/0703276}

\bibitem[{Sorkin(2007{\natexlab{b}})}]{sorkinnonlocal}
Sorkin RD (2007{\natexlab{b}}) Does Locality Fail at Intermediate
  Length-Scales. In: Oriti D (ed) Approaches to quantum gravity, Cambridge
  University Press, pp 26--43, \eprint{gr-qc/0703099}

\bibitem[{Sorkin(2007{\natexlab{c}})}]{coevents}
Sorkin RD (2007{\natexlab{c}}) An exercise in ``anhomomorphic logic''. J Phys:
  Conf Ser 67:012018, \doi{10.1088/1742-6596/67/1/012018}

\bibitem[{Sorkin(2007{\natexlab{d}})}]{sorkinqmeasure}
Sorkin RD (2007{\natexlab{d}}) {Quantum dynamics without the wave function}. J
  Phys A 40:3207--3222, \doi{10.1088/1751-8113/40/12/S20},
  \eprint{quant-ph/0610204}

\bibitem[{Sorkin(2009)}]{lightlinks}
Sorkin RD (2009) {Light, Links and Causal Sets}. J Phys Conf Ser 174:012018,
  \doi{10.1088/1742-6596/174/1/012018}, \eprint{0910.0673}

\bibitem[{Sorkin(2011{\natexlab{a}})}]{sorkinsj}
Sorkin RD (2011{\natexlab{a}}) {Scalar Field Theory on a Causal Set in
  Histories Form}. J Phys Conf Ser 306:012017,
  \doi{10.1088/1742-6596/306/1/012017}, \eprint{1107.0698}

\bibitem[{Sorkin(2011{\natexlab{b}})}]{ec}
Sorkin RD (2011{\natexlab{b}}) {Toward a `fundamental theorem of quantal
  measure theory'}. arXiv e-prints \eprint{1104.0997}

\bibitem[{Sorkin(2014)}]{sorkinEE}
Sorkin RD (2014) {Expressing entropy globally in terms of (4D)
  field-correlations}. J Phys: Conf Ser 484:012004,
  \doi{10.1088/1742-6596/484/1/012004}, \eprint{1205.2953}

\bibitem[{Sorkin and Yazdi(2018)}]{causetee}
Sorkin RD, Yazdi YK (2018) {Entanglement Entropy in Causal Set Theory}. Class
  Quantum Grav 35:074004, \doi{10.1088/1361-6382/aab06f}, \eprint{1611.10281}

\bibitem[{Stachel(1986)}]{Einstein}
Stachel J (1986) Einstein and the quantum: Fifty years of struggle. In: Colodny
  R (ed) From Quarks to Quasars, Philosophical Problems of Modern Physics, U.
  Pittsburgh Press,, p 379

\bibitem[{Stanley(2011)}]{stanley}
Stanley RP (2011) Enumerative Combinatorics, Volume I, 2nd edn. Cambridge
  University Press

\bibitem[{Stoyan et~al(1995)Stoyan, Kendall, and Mecke}]{stoyan}
Stoyan D, Kendall W, Mecke J (1995) Stochastic geometry and its applications,.
  Wiley

\bibitem[{Surya(2008)}]{suryatop}
Surya S (2008) {Causal set topology}. Theor Comput Sci 405:188--197,
  \doi{10.1016/j.tcs.2008.06.033}, \eprint{0712.1648}

\bibitem[{Surya(2012)}]{2dqg}
Surya S (2012) Evidence for the continuum in 2D causal set quantum gravity.
  Class Quantum Grav 29:132001, \doi{10.1088/0264-9381/29/13/132001}

\bibitem[{Surya et~al(2018)Surya, X, and Yazdi}]{sxy}
Surya S, X N, Yazdi YK (2018) {Studies on the SJ Vacuum in de Sitter
  Spacetime}. arXiv e-prints \eprint{1812.10228}

\bibitem[{Sverdlov and Bombelli(2009)}]{bomsver}
Sverdlov R, Bombelli L (2009) {Gravity and Matter in Causal Set Theory}. Class
  Quantum Grav 26:075011, \doi{10.1088/0264-9381/26/7/075011},
  \eprint{0801.0240}

\bibitem[{Unruh and Wald(1989)}]{unruhwald}
Unruh WG, Wald RM (1989) {Time and the Interpretation of Canonical Quantum
  Gravity}. Phys Rev D 40:2598, \doi{10.1103/PhysRevD.40.2598}

\bibitem[{Varadarajan and Rideout(2006)}]{rv}
Varadarajan M, Rideout D (2006) {A General solution for classical sequential
  growth dynamics of causal sets}. Phys Rev D 73:104021,
  \doi{10.1103/PhysRevD.73.104021}, \eprint{gr-qc/0504066}

\bibitem[{Wald(1984)}]{Wald}
Wald R (1984) General relativity. University of Chicago Press

\bibitem[{Wald(1994)}]{waldqft}
Wald RM (1994) Quantum Field Theory in Curved Spacetime and Black Hole
  Thermodynamics. University of Chicago Press

\bibitem[{Wallden(2013)}]{walldenreview}
Wallden P (2013) {Causal Sets Dynamics: Review \& Outlook}. J Phys Conf Ser
  453:012023, \doi{10.1088/1742-6596/453/1/012023}

\bibitem[{Winkler(1991)}]{winkler}
Winkler P (1991) Random orders of dimension 2. Order 7:329

\bibitem[{Yazdi and Kempf(2017)}]{yasamanspectral}
Yazdi YK, Kempf A (2017) {Towards Spectral Geometry for Causal Sets}. Class
  Quantum Grav 34:094001, \doi{10.1088/1361-6382/aa663f}, \eprint{1611.09947}

\bibitem[{Zeeman(1964)}]{zeeman}
Zeeman EC (1964) Causality Implies the Lorentz Group. J Math Phys
  5(4):490--493, \doi{10.1063/1.1704140}

\bibitem[{Zuntz(2008)}]{zuntz}
Zuntz JA (2008) {The cosmic microwave background in a causal set universe}.
  Phys Rev D 77:043002, \doi{10.1103/PhysRevD.77.043002}, \eprint{0711.2904}

\bibitem[{Zwane et~al(2018)Zwane, Afshordi, and Sorkin}]{recentlambda}
Zwane N, Afshordi N, Sorkin RD (2018) {Cosmological tests of Everpresent
  $\Lambda$}. Class Quantum Grav 35:194002, \doi{10.1088/1361-6382/aadc36},
  \eprint{1703.06265}

\end{thebibliography}



\end{document}